\documentclass[fleqn,usenatbib]{rasti}

\usepackage{newtxtext,newtxmath}

\usepackage[T1]{fontenc}

\DeclareRobustCommand{\VAN}[3]{#2}
\let\VANthebibliography\thebibliography
\def\thebibliography{\DeclareRobustCommand{\VAN}[3]{##3}\VANthebibliography}

\usepackage{graphicx}	
\usepackage{amsmath}	
\usepackage{xcolor}  

\newcommand{\Gant}{\Gamma_{\text{ant}}}

\title[EIGSEP]{The Electromagnetically Isolated Global Signal Estimation Platform (EIGSEP)}

\author[C. H. Bye et al.]{
Christian H. Bye,$^{1,2}$\thanks{E-mail: chbye@berkeley.edu}
Aaron R. Parsons,$^{1,2}$
David R. DeBoer,$^{2,3}$
Matt Dexter,$^2$
Sean Donelan,$^{2,4}$
\newauthor{Aaron Ewall-Wice,$^{1,2}$
Adam Fahs,$^{5,6}$
Pranav Karthik,$^{1,2}$
Komal Kaur,$^{1,2}$
Bahram Khalichi,$^{1,2}$
Wei Liu,$^{1,2}$
}
\newauthor{Raul A. Monsalve,$^{7,8}$
Reid Parsons,$^9$
Richard R. Rodriguez,$^{2,10}$
Richard J. Saeed,$^2$
Charlie G. Tolley,$^2$
}
\newauthor{
Dominic Vazquez,$^{2,10}$
and Dirk Wright$^{1,2}$
}
\\
$^{1}$Department of Astronomy, University of California Berkeley, Berkeley, CA 94720, USA\\
$^{2}$Radio Astronomy Laboratory, University of California Berkeley, Berkeley, CA 94720, USA\\
$^{3}$Sub-department of Astrophysics, University of Oxford, Oxford, OX1 3RH, UK\\
$^{4}$Department of Electrical Engineering and Computer Sciences, University of California Berkeley, Berkeley, CA 94720, USA\\
$^{5}$Department of Physics, University of California Berkeley, Berkeley, CA 94720, USA\\
$^{6}$Space Sciences Laboratory, University of California Berkeley, Berkeley, CA 94720, USA\\
$^{7}$Instituto de Astrofísica, Universidad Andrés Bello, Las Condes, 7591538, Chile\\
$^{8}$School of Earth and Space Exploration, Arizona State University, Tempe, AZ 85287, USA\\
$^{9}$Environmental, Geographic \& Public Health Sciences Department, Fitchburg State University, Fitchburg, MA 01420, USA\\
$^{10}$Department of Physics \& Astronomy, San Francisco State University, San Francisco, CA 94132, USA
}

\date{Accepted XXX. Received YYY; in original form ZZZ}

\pubyear{\the\year{}}

\begin{document}
\label{firstpage}
\pagerange{\pageref{firstpage}--\pageref{lastpage}}
\maketitle

\begin{abstract}
The Electromagnetically Isolated Global Signal Estimation Platform (EIGSEP) is a new instrument designed to measure the global 21-cm signal from Cosmic Dawn and the Epoch of Reionization, redshifted to frequencies below 250\,MHz. 
To reduce spectral structure in the antenna beam associated with ground scattering, EIGSEP uses a shaped bowtie antenna suspended in a canyon, designed to keep terrain and conducting structures about 100\,m away.
We describe the current system design of EIGSEP, including the rotating antenna platform, a transmitter antenna to characterize the beam of the bowtie antenna, and auxiliary ground antennas.
We then discuss the EIGSEP calibration scheme, which incorporates traditional Dicke switching in the receiver, and novel approaches that include beam mapping, beam modulation, and interferometric cross-correlation.
The instrument has been deployed near Marjum Pass, Utah, for testing and initial data collection. 
We discuss the site characteristics and present the measured response of the suspended and rotating antenna to an external transmitter.
We also present an in situ measurement of the antenna reflection coefficient, whose delay spectrum shows no detectable terrestrial scattering at delays below 590\,ns, the round-trip light travel time to the nearest canyon terrain, demonstrating the electromagnetic isolation of the antenna.
Absolutely calibrated sky spectra and the characterization of the antenna beam are deferred to forthcoming work.
\end{abstract}

\begin{keywords}
cosmology: dark ages, reionization, first stars -- instrumentation -- telescopes
\end{keywords}



\section{Introduction} \label{sec:intro}
During the first billion years of cosmic time, the universe evolved from a relatively smooth expanse of hydrogen and helium to the complex, chemically enriched web we observe today \citep{2006PhR...433..181F}. This transformation was driven by the first generations of stars, galaxies, and black holes, whose radiation heated and ionized the intergalactic medium (IGM), while locally seeding stellar growth. Measurements of the cosmic microwave background \citep[CMB;][]{2020A&A...641A...6P, 2021ApJ...908..199R}, quasar absorption spectra \citep{2006AJ....132..117F, 2015MNRAS.447.3402B, 2018MNRAS.479.1055B, 2022MNRAS.514...55B, 2018ApJ...864...53E}, deep galaxy observations \citep{2015MNRAS.450.3032M, 2022ApJ...938L..15C, 2022ApJ...940L..14N, 2023ApJ...946L..13F, 2025MNRAS.542.1952M, 2025ApJ...992...63W}, and Ly$\alpha$ emitter surveys \citep{2007MNRAS.381...75M, 2017ApJ...844...85O, 2024ApJ...975..208T, 2025ApJS..277...37U, 2025Natur.639..897W} can constrain the patchy end of reionization. However, the early stages of our Cosmic Dawn ($50 \gtrsim z \gtrsim 12$), when the first stars and X-ray sources appeared, will remain beyond the reach of optical and infrared probes for the foreseeable future \citep{2015ApJ...813...21M, 2016MNRAS.462..235L}.

The Cosmic Dawn is none the less accessible through observations of the 21-cm hyperfine line of hydrogen \citep{1997ApJ...475..429M, 1999A&A...345..380S, 2000ApJ...528..597T}. 
Interferometers, including HERA \citep{2017PASP..129d5001D, 2024PASP..136d5002B}, LOFAR \citep{2013A&A...556A...2V}, the LWA \citep{2019AJ....158...84E}, the MWA \citep{2013PASA...30....7T, 2018PASA...35...33W}, NenuFAR \citep{2024A&A...681A..62M}, PAPER \citep{2010AJ....139.1468P}, and SKA-Low \citep{2009IEEEP..97.1482D}, aim to measure the spatial anisotropies of the 21-cm signal, to probe inhomogeneities in density, heating, and ionization of the IGM.
Upper limits on the 21-cm power spectrum reported by these experiments are beginning to constrain models of reionization \citep{2020MNRAS.493.4728G, 2025A&A...699A.109G, 2020MNRAS.498.4178M, 2021MNRAS.501....1G, 2022ApJ...924...51A, 2023ApJ...945..124H}.

Information in the global 21-cm signal, i.e. the monopole component, complements measurements of the spatial 21-cm fluctuations \citep{2016MNRAS.457.1864L}.
At high redshifts, the global signal encodes information about the first luminous sources \citep{2017MNRAS.464.1365M}. Ly$\alpha$ photons from the first generations of stars couple the spin-temperature of hydrogen to the kinetic gas temperature, causing an absorption dip relative to the CMB at $z \approx 25$. Compact objects left over from the first stars emit X-rays that heat the gas, possibly driving it into emission at $z \approx 9$, before UV emission from widespread star formation ionizes the IGM and effectively eliminates the signal by $z \approx 5$ \citep{2019MNRAS.485L..24K, 2022MNRAS.514...55B}. The timing and depth of the absorption feature are sensitive to the properties of the first galaxies, including their mass distribution, star formation efficiency, X-ray, Ly$\alpha$, and UV production rates, and the escape fraction of ionizing radiation into the IGM \citep{2016PhR...645....1B, 2020PASP..132f2001L}.
By independently constraining the optical depth of Thomson scattering in the CMB, the global 21-cm signal also helps to improve constraints on the sum of neutrino masses and on the amplitude of primordial fluctuations seeding large-scale structure \citep{2016ApJ...821...59F, 2016PhRvD..93d3013L}.

Some of the ongoing experiments targeting the global signal from the ground include EDGES \citep{2017ApJ...847...64M, 2018Natur.555...67B, 2025PASP..137l5002C}, GINAN \citep{2025arXiv250911846M}, LEDA \citep{2018MNRAS.478.4193P}, MIST \citep{2024MNRAS.530.4125M}, PRI\textsuperscript{Z}M \citep{2019JAI.....850004P}, REACH \citep{2022NatAs...6..984D}, RHINO \citep{2025RASTI...4af046B}, and SARAS \citep{2021ExA....51..193T, 2022NatAs...6..607S}. Space-based missions are also proposed or underway, including CosmoCube \citep{2025RASTI...4...61A}, DSL/Hongmeng \citep{2021RSPTA.37990566C}, LuSEE-Night \citep{2023arXiv230110345B}, and PRATUSH \citep{2023ExA....56..741S}.

The EDGES experiment reported a detection of an unexpectedly large ($\sim500\,\text{mK}$) 21-cm absorption trough in \citet{2018Natur.555...67B}. If of cosmological origin, the measurement could imply new interactions between baryons, photons, and dark matter \citep{2018Natur.555...71B, 2018Natur.557..684M, 2018PhRvL.121c1103P,
2018PhRvD..98b3013S, 2019PDU....24..289S} or an excess radio background early in the Cosmic Dawn \citep{2018ApJ...858L..17F} from previously undetected progenitors of super-massive black holes \citep{2018ApJ...868...63E, 2020MNRAS.492.6086E}, star-forming galaxies \citep{2019MNRAS.483.1980M}, or dark matter annihilation \citep{2018PhLB..785..159F, 2018PhRvD..98j3503Y}. 

Alternatively, the absorption feature might be a systematic artefact of antenna interactions with the environment, cables, or ground planes \citep{2018Natur.564E..32H, 2019ApJ...874..153B, 2019ApJ...880...26S, 2020MNRAS.492...22S}. Such interactions complicate the removal of galactic foreground emission, which is more than four orders of magnitude brighter than the target signal \citep{2002ApJ...564..576D, 2003MNRAS.346..871O} and pose the primary challenge for high-redshift 21-cm experiments. In principle, the spectral smoothness and spatial variability of the foregrounds allow them to be separated from the 21-cm global signal \citep{2010PhRvD..82b3006P, 2013PhRvD..87d3002L, 2017ApJ...840...33S}. However, effects such as antenna beam chromaticity \citep{2015ApJ...799...90B, 2016MNRAS.455.3890M}, ionospheric scattering \citep{2014MNRAS.437.1056V, 2016ApJ...831....6D, 2021MNRAS.503..344S}, and reflections off conductive elements \citep{2022MNRAS.515.1580S, 2024ApJ...974..137A, 2024ApJ...961...56M} introduce significant spectral structure into the measured antenna temperature, thus reducing the efficacy of foreground subtraction or filtering. Even low-level ($\sim10^{-4}$) spectral structure in the antenna beam can result in features that mimic a cosmological signal \citep{2014ApJ...793..102S}. 

In 2022, SARAS reported a non-detection of the EDGES feature at 95.3 per cent confidence \citep{2022NatAs...6..607S}.
SARAS uses an antenna on a raft to set a minimum delay of reflections, corresponding to the light travel time to the bottom of the lake. 
This delay ensured that spectral ripples in SARAS data associated with reflections would have periods less than 1.5\,MHz, enabling them to be distinguished from the 21-cm signal, which is expected to vary more slowly with frequency \citep{2021ITAP...69.6209R, 2022NatAs...6..607S}.
The tension between the EDGES and SARAS results emphasizes the need for experiments that are electromagnetically isolated from their environment at scales relevant for measuring the 21-cm signal.

Here, we introduce the Electromagnetically Isolated Global Signal Estimation Platform (EIGSEP), a new experiment that aims to measure the 21-cm global signal from the Cosmic Dawn and Epoch of Reionization.
Crucial to the EIGSEP design is the use of an antenna suspended in a canyon, about $100\,\textrm{m}$ above the ground and from the nearest electric conductors.
This distance suppresses ground coupling and confines spectral structure induced by reflections to long delays.
Moreover, the electromagnetic isolation may enable accurate electromagnetic modelling of the antenna without detailed knowledge of the electrical properties of the ground. This isolation would greatly simplify the electromagnetic simulations of the antenna, and facilitate direct comparisons between simulations and measurements of the antenna beam pattern and impedance.
EIGSEP also furnishes a rotating antenna platform, allowing us to change the projection of the antenna beam pattern on the sky during measurements.
Rotations of the antenna produce different beam-weighted foregrounds that can be used to separate instrumental systematics from sky emission.
Additionally, the rotating antenna enables the measurement of the beam pattern with a single stationary transmitter.

This paper presents the concept, design, and calibration methodology of EIGSEP, the characterization of the observing site, and the current status of the instrument in the field. We defer absolutely calibrated sky spectra and a measured antenna beam pattern to forthcoming work.
The paper is outlined as follows: We describe the design principles in section~\ref{sec:philosophy} followed by the system design in section~\ref{sec:system} and the calibration scheme in section~\ref{sec:calibration}. We then give an overview of EIGSEP deployments to date, including discussion about the site and radio-frequency interference (RFI), in section~\ref{sec:deployments}.
Finally, we summarize the status of the experiment and future work in section~\ref{sec:conclusion}.

\section{Design Principles} \label{sec:philosophy}
The goal of the EIGSEP experiment is to measure the redshifted global 21-cm signal from the Cosmic Dawn and Epoch of Reionization.
Distinguishing between the possible 21-cm models in the presence of foregrounds and instrumental systematics requires careful foreground removal to minimize signal loss and avoid the introduction of spurious features that mimic the real cosmological signal.
Here, we outline the principles motivating the EIGSEP design, to enable this signal separation.

EIGSEP is taking a conservative approach to foreground removal, where spectral modes contaminated by instrument-weighted foregrounds are all explicitly removed from the data and not used to model the cosmological signal.
Mathematically, we obtain the modes to remove from the data with a singular value decomposition of the instrument-weighted foregrounds, which could be either measured or simulated.
Projecting such foreground eigenmodes out of measured data can suppress systematics to the sub-mK level.
The difficulty lies in ensuring that residuals still retain a measurable contribution from cosmological 21-cm emission.
As noted above, the intrinsic foreground sky is spectrally smooth and may be described by low-order polynomials or maximally smooth functions \citep{2010PhRvD..82b3006P, 2017ApJ...840...33S, 2021MNRAS.502.4405B}. If an antenna or its environment can introduce spectral structure that spans the space of global 21-cm signal models, filtering off bright foreground modes leaves no cosmological signal behind.
This issue might be alleviated by taking advantage of the spatial anisotropies of the foregrounds, to separate them from the monopole cosmological signal \citep{2014ApJ...793..102S}, or by observing with different instruments to separate systematics from sky emission \citep{2023MNRAS.522.1022S}.

This approach leads to three specific design principles for EIGSEP:
\begin{enumerate}
    \item The instrument spectral response can be modelled with a limited number of spectral modes, that have minimal overlap with the 21-cm signal modes.
    \item The instrument can be modified between or during measurements to produce different weightings of systematics.
    \item The systematics of the instrument can be modelled and measured in the field.
\end{enumerate}

In the rest of this section, we detail each principle and explain how they inform the EIGSEP design. 
We focus on the systematics associated with the chromatic antenna beam, which is typically the most challenging systematic to characterize for global signal experiments and is usually accounted for in numerical simulations \citep{2021AJ....162...38M, 2022JAI....1150001C, 2024MNRAS.530.4125M}. Frequency-dependent systematics introduced by the receiver are calibrated with lab measurements and in situ Dicke switching, as outlined in section~\ref{sec:calibration}.
We define in this section the spectral modes of the antenna as the eigenmodes of the spectral covariance matrix of an antenna beam averaged over pointings.

\subsection{Minimizing covariance with the 21-cm signal} \label{subsec:covariance}
If the antenna beam and the foregrounds are described by $N_{\text{ant}}$ and $N_{\text{fg}}$ independent spectral modes respectively, then the beam-weighted foregrounds require of the order $N_{\text{modes}} \sim N_{\text{ant}} N_{\text{fg}}$ modes. The product $N_{\text{modes}}$ is the number of degrees of freedom in the model used to fit and remove foregrounds from EIGSEP data.
If this number is too large, the model becomes so flexible it can fit any feature in the data, including the cosmological signal. 
Furthermore, the task of characterizing the antenna beam pattern through both measurements and simulations -- and comparing the two -- is simplified by limiting the number of modes needed to describe the beam pattern.
Several global signal experiments, including EDGES \citep{2025PASP..137l5002C}, MIST \citep{2024MNRAS.530.4125M}, REACH \citep{2022JAI....1150001C}, and SARAS \citep{2022NatAs...6..607S}, use small antennas with simple geometries to ensure that the beam pattern varies smoothly with frequency. Using the terminology introduced here: their responses may be described by few eigenmodes.
Another benefit of a simple geometry is that it streamlines electromagnetic simulations and reduces discrepancies between the simulated structure and the fabricated one. This simplicity improves the fidelity of the simulations and enables more reliable comparisons between simulations and measurements.
We therefore require a geometrically simple antenna for EIGSEP that can easily be manufactured and replicated, with a beam pattern that can be described by a limited number of spectral eigenmodes, which in turn have minimal overlap with the 21-cm signal modes.
We address the two requirements on the eigenmodes of the antenna beam pattern one at a time, by first assessing how many modes are required to describe the beam-weighted foregrounds, and then how much of the 21-cm signal survives filtering them.

Using representative beams from our electromagnetic simulations of the EIGSEP antenna (section~\ref{subsec:eig_antenna}), a model of the horizon profile at the EIGSEP site (Fig.~\ref{fig:horizon}), and the Global Sky Model \citep{2017MNRAS.464.3486Z}, we simulated one sidereal day of observation by EIGSEP over 50--250\,MHz with the forward model described in section~\ref{subsec:fwd_modelling}. In the simulations, we assumed a constant ground temperature of 300\,K. This beam-weighted ground pickup is included in the antenna temperature and not corrected here.

We repeated the simulation for three beams observing the same sky to separate the effect of the antenna from that of the sky: an achromatic and isotropic beam, the optimized EIGSEP bowtie, and a Vivaldi feed used in isolation, without a dish, as for the ground antennas (section~\ref{sec:vivaldis}).
The Vivaldi feed was used for the October 2024 EIGSEP suspension (section~\ref{subsec:field_meas}) as an interim antenna, before the design of the bowtie antenna was finalized. The bowtie geometry was selected from a range of antenna geometries and optimized afterwards in a broader eigenmode-based analysis of beam chromaticity, foreground complexity, and overlap with global 21-cm signal modes. This design process will be presented in Khalichi et al. 2026 (in prep.).

\begin{figure*}
    \centering
    \includegraphics[width=\linewidth]{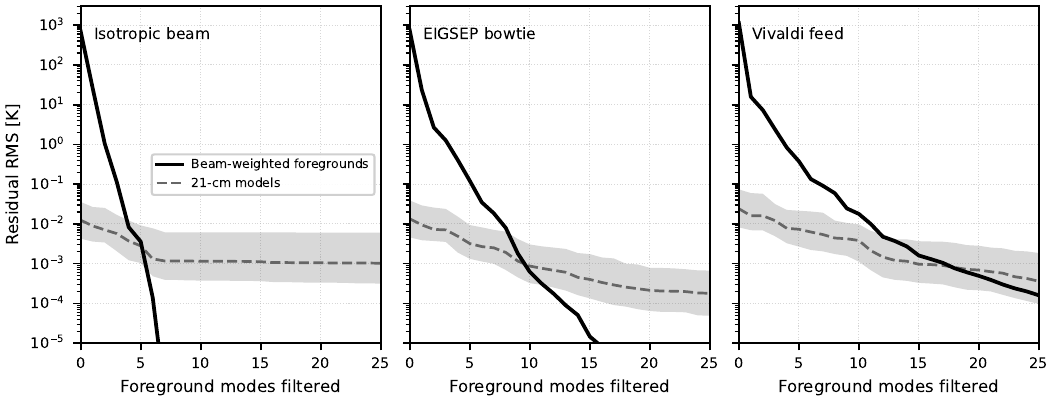}
    \caption{Residual RMS of the simulated antenna temperature (black) and of the 1769 global 21-cm models filtered with the same modes (5--95 per cent of ensemble in grey band, median dashed), as a function of the number of eigenmodes filtered ($N$), for three antennas observing the same sky behind the same horizon. \textit{Left:} an isotropic beam, which has no chromaticity and so isolates the sky contribution. 
    \textit{Centre:} the EIGSEP bowtie.
    \textit{Right:} a Vivaldi feed of the type used for the ground antennas.
    Each antenna is filtered in its own eigenbasis and its models are scaled by its own beam-weighted open-sky fraction, so the grey bands differ between panels. The isotropic residual falls below the axis near $N=6$, where it reaches the numerical noise floor of the simulation.}
    \label{fig:singular_values}
\end{figure*}

We then computed the singular value decomposition of each antenna temperature, producing its orthogonal spectral eigenmodes, and used those eigenmodes to filter the corresponding beam-weighted foregrounds.
In Fig.~\ref{fig:singular_values}, we present the residual RMS after filtering as a function of the number of eigenmodes filtered, $N$, for each of the three beams. The first point -- zero modes filtered -- is the RMS of the raw waterfall pixels about zero, which for the bowtie is approximately 730\,K. Because the decomposition is not centred, the first mode captures the mean antenna temperature spectrum.
The isotropic beam represents an ideal case without a chromatic pattern coupling to the foregrounds, thus producing the beam-weighted foregrounds with the least possible spectral structure -- it takes 6 modes to filter the foregrounds to below 1\,mK in this case.
The realistic beam patterns of the EIGSEP bowtie and the Vivaldi feed are chromatic, thus increasing the number of spectral modes needed to filter the beam-weighted foregrounds to the same level: 10 modes for the bowtie and 18 modes for the Vivaldi feed. The excess over the isotropic case -- 4 modes for the bowtie and 12 for the Vivaldi feed -- counts the modes each beam's own chromaticity adds to what this sky and horizon alone require.
The ordering follows from the measurement being a convolution of the beam and the sky, with chromatic beams coupling their own frequency-dependent angular structure to the foregrounds. In the case of the EIGSEP bowtie, filtering the 10 modes needed to reduce residuals below 1\,mK describes the antenna temperature to one part in $1.2\times10^{6}$.

To assess the spectral overlap between the beam-weighted foregrounds and the expected cosmological signal, we applied each antenna's filter to an ensemble of 1769 global 21-cm models generated with the Zeus21 code\footnote{\url{https://github.com/ZeusCosmo/Zeus21}} \citep{2023MNRAS.523.2587M, 2025PhRvD.111h3503C}.
As for the foregrounds, the beam-weighted ground pickup is left uncorrected for the cosmological signal. The retained amplitude of the 21-cm signal models is therefore scaled by the fraction of the beam that sees the open sky, which across the band is 0.36--0.55 for the bowtie, 0.59--0.80 for the Vivaldi feed, and 0.35 for the isotropic beam.
For the bowtie, the filter explains foreground variance more effectively than the cosmological signal: after $N=10$ modes are removed, the RMS of the median retained 21-cm signal is larger (0.87\,mK) than the RMS of the residual foregrounds (0.62\,mK).
Compared at the same level of foreground suppression, the bowtie retains a median 7.7 per cent of each model's unfiltered 21-cm RMS, against 3.5 per cent for the Vivaldi feed and 13.3 per cent for the isotropic beam.

While Fig.~\ref{fig:singular_values} is a useful heuristic probing the covariance between the beam-weighted foregrounds and the 21-cm signal, it does not represent the complete analysis needed for foreground separation.
The calculation includes the antenna beam, the horizon profile, and the sky, but assumes a perfectly calibrated receiver gain and antenna reflection coefficient, which are treated separately in sections~\ref{subsec:rec_cal} and \ref{subsec:vna}, respectively.
Uncertainty in that calibration, or systematics left unmodelled, can increase the number of modes required to explain the data.
On the other hand, informative priors on the sky or the beam -- e.g. obtained through beam mapping (section~\ref{subsec:beam_mapping}) -- may allow chromatic structure to be modelled rather than filtered.
In practice, EIGSEP will use a differentiable forward model to evaluate systematics and separate the foregrounds, as described in section~\ref{subsec:fwd_modelling}.

Beyond the antenna geometry itself, the conductive environment around an antenna influences the beam chromaticity by creating multiple paths for a received wave to follow. The differential path lengths introduce a frequency-dependent interference ripple whose spectral scale is set by
\begin{equation}
    \tau=\Delta\ell/c=1/\Delta\nu,
    \label{eq:delay}
\end{equation}
where $\tau$ is the signal delay, $\Delta\ell$ is the differential path length between the direct and scattered rays, $c$ is the speed of light, and $\Delta\nu$ is the period of the resulting spectral ripple. For a structure at a distance $d$ from the antenna, $\Delta\ell\leq2d$, with the maximum corresponding to a round trip between the antenna and the structure.
Because global 21-cm signal models vary on scales of 10--100\,MHz, equation~\eqref{eq:delay} suggests that conductive structures within $\sim15$\,m of the antenna require careful scrutiny.
Notably for global signal experiments, the presence of a ground plane below the antenna may potentially introduce spectral artefacts that are degenerate with global signal features, leading to spurious or hard-to-interpret results \citep{2014MNRAS.437.1056V, 2018Natur.564E..32H, 2019ApJ...874..153B, 2019ApJ...880...26S, 2020MNRAS.492...22S}.
On the other hand, experiments operating directly on the soil without a ground plane are finding that interpretations of their results depend crucially on soil modelling \citep{2022MNRAS.515.1580S, 2024ApJ...961...56M}.

\begin{figure*}
    \centering
    \includegraphics[width=\linewidth]{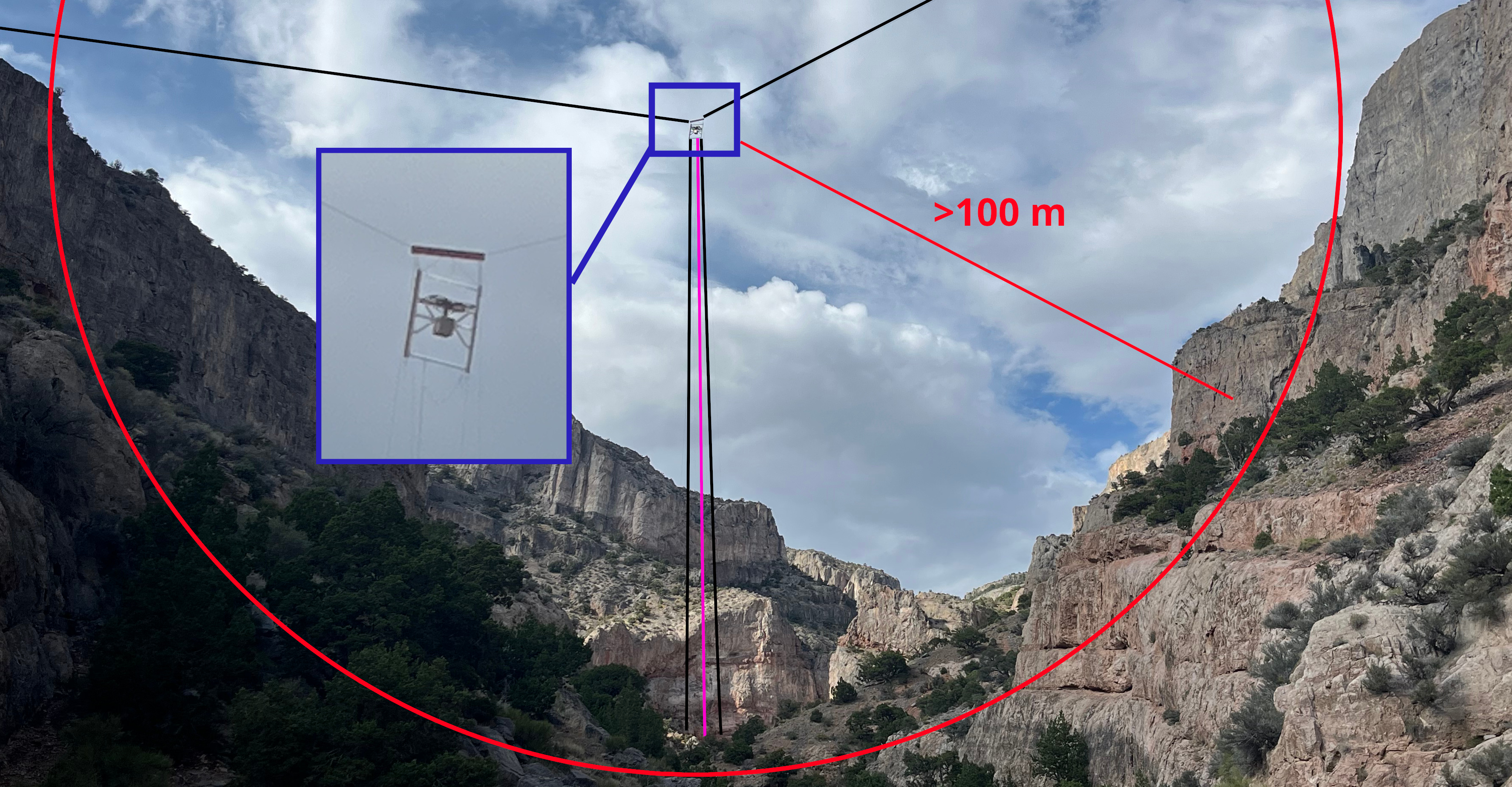}
    \caption{The EIGSEP concept: An antenna suspended in a canyon, with a 100\,m zone free of electrical conductors.
    The inset shows a zoomed in view of the antenna.
    The conductor-free zone is shown in red, with the antenna in the middle of the canyon suspended from Kevlar ropes (black). The signal is transmitted to the backend electronics on the ground via non-conducting fibre-optic (pink).}
    \label{fig:concept}
\end{figure*}

The nominal requirement for EIGSEP is electromagnetic isolation from the surrounding terrain and structures within 100\,m of the antenna. According to equation~\eqref{eq:delay}, this isolation will confine the dominant environmental reflections to delays $\tau \gtrsim670\,\text{ns}$, suppressing spectral ripples with periods of $\Delta\nu \gtrsim 1.5\,\text{MHz}$.
EIGSEP achieves this electromagnetic isolation by using an antenna suspended $\sim100\,\textrm{m}$ above the ground in the canyon.
This concept is shown in Fig.~\ref{fig:concept}. The only conductive structure within the 100\,m air bubble is the antenna platform that holds the front-end electronics. This platform is in the near-field of the antenna and is included in electromagnetic simulations. We require that the metal platform and electronics be contained within 2\,m of the antenna.
The analogue signal from the antenna is transmitted over fibre-optic cables to the ground and the antenna is suspended with non-conducting Kevlar ropes. 

\begin{figure}
    \centering
    \includegraphics[width=\linewidth]{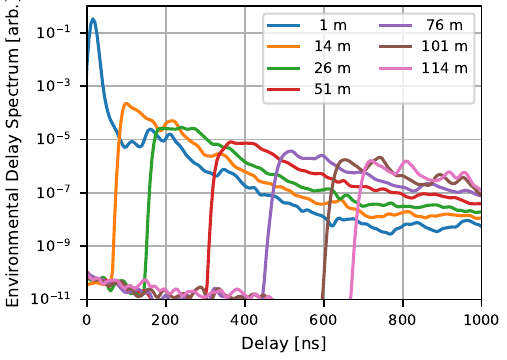}
    \caption{Delay--power spectra of simulated environmental scattering for the EIGSEP bowtie antenna suspended at different heights. The spectra show the environment-induced contribution to the antenna reflection coefficient. Environmental reflections appear at delays set by the two-way travel time of light between the antenna and the surrounding terrain, producing a height-dependent gap in delay space. 
    As the height of the antenna increases, the gap in signal delay increases, due to increased light travel time to the environment. 
    At the highest height above the canyon floor (114\,m, pink), the nearest terrain is 102\,m away and environmental reflections are confined to delays above 670\,ns. The 101\,m curve (brown) is closest to the deployed configuration of section~\ref{subsec:eig_antenna}.
    The curves are normalized by a common reference to highlight relative delay structure.}
    \label{fig:reflections}
\end{figure}

To illustrate the effects of electromagnetic isolation, we modelled the environmental contributions to the EIGSEP antenna reflection coefficient at different heights above the canyon floor. The modelled reflection coefficient has two parts: an intrinsic impedance mismatch between the antenna and the receiver and scattering from the canyon terrain.
The reflection due to impedance mismatch was simulated in Ansys HFSS\footnote{\url{https://www.ansys.com/products/electronics/ansys-hfss}} using a model of the antenna in free space, as outlined in section~\ref{subsec:eig_antenna}.
For the terrain component, we computed the amplitude and phase of the reflected wave analytically -- assuming uniform permittivity and conductivity -- and used a digital elevation model to compute the propagation delay.
The results are presented in Fig.~\ref{fig:reflections}, up to an undetermined amplitude scale factor. This prediction is tested against a field measurement of the suspended antenna in section~\ref{subsec:eig_antenna}. Crucially, increasing the suspension height shifts reflections to longer delays. While these results are computed for a signal transmitted from the antenna, the nature of the spectral ripples is set by the canyon geometry and will also be present in sky measurements.


\subsection{Making independent measurements}
Aside from spectral separation, another way to break degeneracies between beam-weighted foregrounds and the cosmological signal is to leverage the variability of the beam-weighted foregrounds. This is usually proposed in the context of using data from different local sidereal time (LST) bins, where the foregrounds change with the Earth rotation while the (monopole) cosmological signal stays constant \citep{2014ApJ...793..102S, 2020ApJ...897..175T, 2023MNRAS.520..850A}.
A second possibility is to alter the instrument. For example, REACH plans to observe with two different antennas \citep{2022NatAs...6..984D, 2023MNRAS.522.1022S}, and the EDGES \citep{2025PASP..137l5002C}, MIST \citep{2024MNRAS.530.4125M}, and SARAS \citep{2022NatAs...6..607S} experiments have been conducting observations with different instruments and at different sites. 

\begin{figure}
    \centering
    \includegraphics[width=\linewidth]{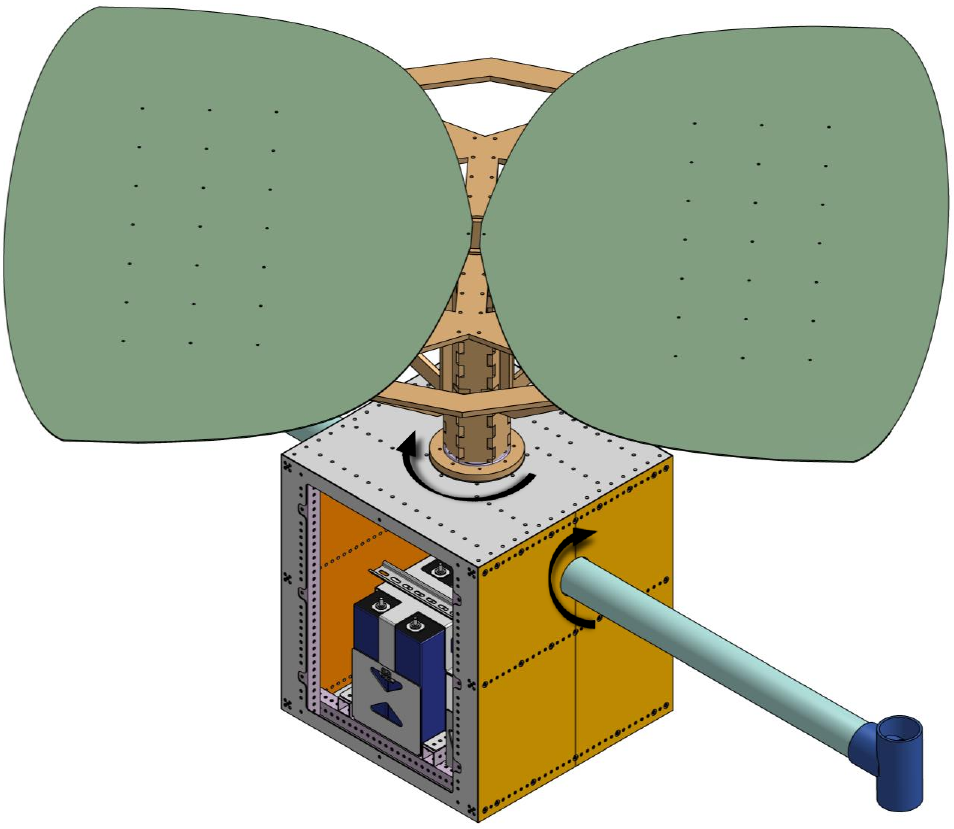}
    \caption{A CAD model of the EIGSEP antenna (green) on top of the antenna platform. The drawing does not include the copper plates that shield the sides of the box. 
    The two black arrows represent the two types of rotations possible with EIGSEP.}
    \label{fig:rotations}
\end{figure}

Making independent measurements with different weighting of instrumental systematics is crucial to achieve a confident detection of the 21-cm signal.
EIGSEP is a portable experiment that can observe from different sites, and we plan on making measurements with different antennas.
Moreover, EIGSEP uses a rotating platform to change the instrumental systematics in a controlled way during observations. Two types of rotations are supported: azimuthal rotations of the antenna (similar to the LuSEE-Night turntable \citep{2023arXiv230110345B}) and axial rolls of the platform itself. A diagram illustrating the rotations is shown in Fig.~\ref{fig:rotations}. 
Measurements done at different antenna orientations change the beam weighting of the emission from the sky and the ground.
In a manner similar to observations at different LST bins, the antenna rotations allow EIGSEP to project out shapes that vary, as the global signal stays constant.
In addition to the rotations, the platform height can be varied in the field to change the horizon profile or evaluate the coupling to the ground, which may impact signal recovery \citep{2021ApJ...923...33B, 2024MNRAS.527.2413P}.

\subsection{Modelling and measuring systematics}
The third design principle for EIGSEP is the ability to model systematics and measure them in the field, during observations.
The relevant systematics associated with the antenna are the beam pattern and impedance. Both quantities are multiplicative and frequency-dependent, and must therefore be known to a precision set by the dynamic range between foregrounds and signal, about one part in $10^4$ \citep{2024MNRAS.531.4734C}.
While many experiments targeting the global 21-cm signal are measuring the impedance with a vector network analyser (VNA), they often use commercial electromagnetic solvers to simulate the antenna beam pattern for calibration and chromatic corrections \citep{2018Natur.555...67B, 2021MNRAS.506.2041A, 2024MNRAS.530.4125M}.
Validations of simulations are often indirect or focused on impedance measurements, due to challenges of measuring the antenna beam directly \citep{2021AJ....162...38M, 2025PASP..137h5002A, 2025GL116618}.


For EIGSEP, we require that these quantities are measured in the field directly to increase our confidence in the calibration and validate simulations.
The design of EIGSEP must be conducive to accurate electromagnetic simulations, and enable direct measurements in the field. The electromagnetic isolation of the EIGSEP antenna is again crucial, as realistic simulations of the antenna can be done without including models of the geometry or electrical properties of the soil below the antenna.
The EIGSEP receiver has a VNA and a switch network to autonomously measure the antenna impedance in the field.
In order to measure the antenna beam, EIGSEP uses a transmitter antenna placed on the ground directly below the suspended antenna. It is important that the transmitter be in the far-field of the suspended antenna. This specification is satisfied by the 100\,m suspension height, a distance of more than 15 wavelengths at the lowest frequencies.
To probe the direction-dependence of the beam, we rotate the antenna platform, thereby shifting the part of the beam sensitive to the transmitter.

\section{System Design}
\label{sec:system}
\begin{figure*}
    \centering
    \includegraphics[width=\linewidth]{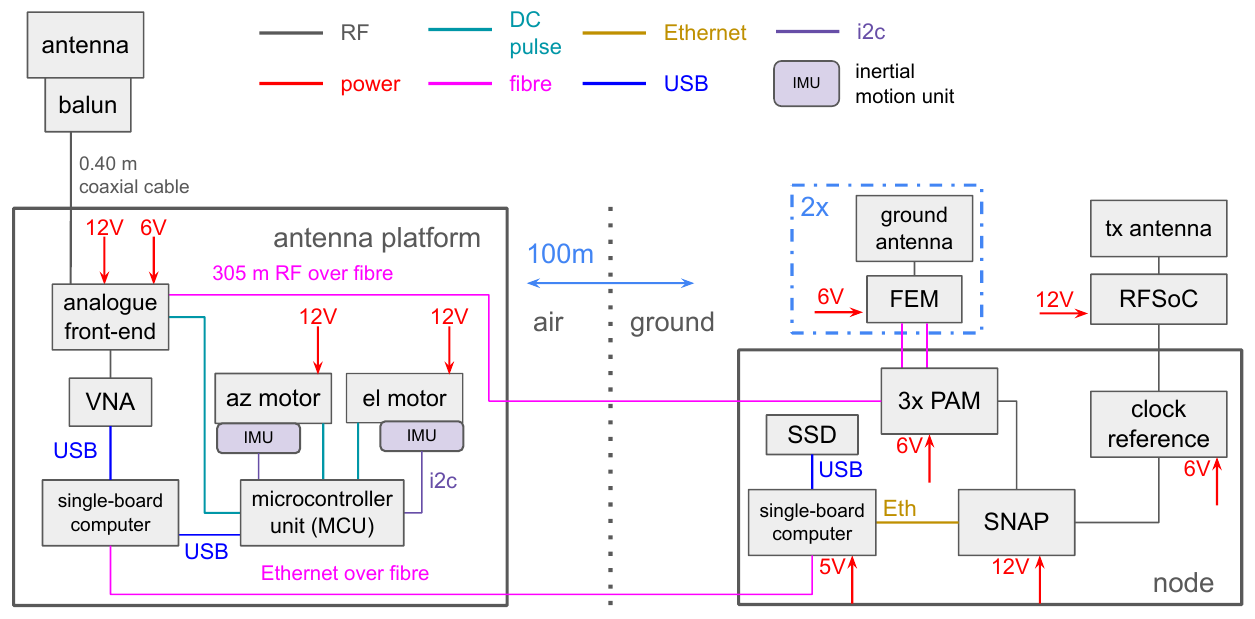}
    \caption{High-level block diagram of the EIGSEP instrument. The diagram shows the configuration used through the July 2025 deployment.
    The signal path for sky measurements goes from the antenna and ground antennas to the analogue front-end (the custom EIGSEP front-end for the suspended antenna, see Fig.~\ref{fig:fe_block}, and the HERA FEM in the case of the ground antennas), through RF over fibre to the PAMs. Coaxial cables connect the PAM to the SNAP, where the signals are digitized and correlated. The SNAP uses Ethernet to communicate with the single-board computer, which saves the data to an external solid-state drive.}
    \label{fig:system_diagram}
\end{figure*}

The EIGSEP instrument is an interferometer, with the main antenna suspended within the canyon and one or two auxiliary antennas on the ground beneath it. Front-end electronics, including a balun, low-noise amplifiers (LNAs) and RF filters are located directly next to each antenna. The analogue signals from all antennas are digitized and correlated at a central node on the ground. 
In this section, we describe the instrument as built and deployed in the field. A summary of the system characteristics is shown in Table~\ref{tab:system}, and a block diagram is shown in Fig.~\ref{fig:system_diagram}.

\begin{table}
    \centering
    \caption{EIGSEP system characteristics}
    \begin{tabular}{cc}
        \hline
        Redshift Range & 27.4--4.7 \\
         Nominal Band & 50--250\,MHz  \\
         Digitized Band & 0--250\,MHz \\
         Frequency Channels & 1024 \\
         Frequency Resolution & 244\,kHz \\
         Integration Time & 0.54--1.07\,s \\
         Antennas & 2--3 \\
         Correlations & 3--11 \\
         Suspension Height & 100\,m \\
         Fibre Length & 305\,m suspended, 152\,m ground \\
         Kevlar Length & 305\,m \\
         Platform Weight & 35\,kg \\
         Tether Tension & 100\,kg \\
        \hline
    \end{tabular}
    \label{tab:system}
\end{table}

\subsection{Suspended antenna}
\label{subsec:eig_antenna}
Following the requirements outlined in section~\ref{sec:philosophy}, a shaped bowtie antenna has been developed as the main antenna for EIGSEP. The bowtie antenna has a simple geometry, an inherently wide bandwidth, and an approximately omnidirectional radiation pattern.
In advanced bowtie antenna designs, conformal analysis has led to modifications of the traditional triangular geometry into elliptical or half-elliptical structures. These shapes offer smoother current distribution and impedance closer to $50\,\Omega$ or $75\,\Omega$, traditionally used as characteristic impedance, thereby enhancing broadband performance. 
The flare angle is an important geometric parameter governing the input impedance of a bowtie antenna, with the impedance generally decreasing as the flare angle increases \citep{lestari2005adaptive, lestari2010modified}.
Elliptical or smoothly varying profiles can provide a smoother path for the surface current and facilitate broadband impedance matching \citep{yang2020radiation}.
Based on these principles, the EIGSEP antenna uses a half-circular baseline geometry that is subsequently refined using exponential parametric equations. The geometrical parameters are optimized in HFSS to improve the impedance bandwidth over 50--250\,MHz while maintaining compact physical dimensions.

The antenna was simulated using the frequency-domain finite-element solver in Ansys HFSS. The electromagnetic model included the shaped bowtie conductors, a simplified feed in which the balun is represented as an ideal transformer, and the rectangular aluminium enclosure representing the antenna platform (section~\ref{subsec:box}). Material properties corresponding to the fabricated structure were assigned to the respective components, while small mechanical features that were not expected to have a significant electromagnetic effect were omitted. The antenna and nearby platform were surrounded by an air region terminated using perfectly matched layers to approximate an open radiating domain. To improve the numerical accuracy and reliability of the solution across the operating band, multi-frequency adaptive meshing was performed at select frequencies between 50 and 250\,MHz. A maximum convergence criterion of 0.005 was specified at each adaptive frequency, with a maximum of 30 adaptive passes. Following convergence, a discrete frequency sweep was performed from 50 to 250\,MHz with a frequency resolution of 1\,MHz.

The final geometry was selected based on the spectral smoothness of the antenna reflection coefficient and the frequency evolution of its radiation pattern over the EIGSEP operating band (section~\ref{subsec:covariance}), following the analysis of Khalichi et al. 2026 (in prep.).

\begin{figure}
    \centering
    \includegraphics[width=\linewidth]{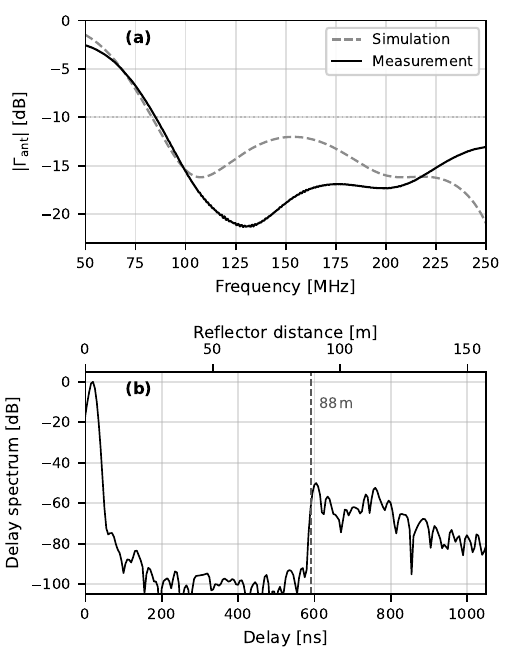}
    \caption{\textit{(a)} Magnitude in dB of the antenna reflection coefficient for the bowtie antenna, measured in situ during the July 2025 deployment (solid) and simulated in free space with HFSS (dashed). \textit{(b)} Delay spectrum of the measurement, obtained by Fourier transforming $\Gant$ over 50--250\,MHz with a Blackman--Harris window and normalized to the peak of the delay spectrum, which is the antenna--receiver impedance mismatch near zero delay. The simulated counterpart is Fig.~\ref{fig:reflections}, which models the canyon terrain explicitly; the free-space simulation of panel (a) contains no terrain. The structure beyond 590\,ns (dashed line, 88\,m) is scattering from the canyon, and between $\sim100\,\text{ns}$ and 590\,ns there is no detectable reflection.}
    \label{fig:bowtie_s11}
\end{figure}

We present the simulated antenna reflection coefficient, $\Gant$, in Fig.~\ref{fig:bowtie_s11}(a) alongside an in situ measurement of the suspended bowtie antenna from the deployment in July 2025 (section~\ref{subsec:field_meas}).
The result demonstrates a smooth reflection coefficient across the desired bandwidth and a good impedance match between the antenna and a $50\,\Omega$ reference impedance, with $|\Gant| < -10\,\text{dB}$ above 85\,MHz for both the simulated and measured reflection coefficient. Below 100\,MHz, the measurement agrees closely with the simulation, with a median difference of 0.5\,dB. The measured coefficient stays below $-15\,\text{dB}$ across 100--220\,MHz. It departs from the simulation over the same range, by up to 8\,dB near 134\,MHz, while above 220\,MHz the sign reverses and the measurement is up to 8\,dB higher. The remaining discrepancy between the simulation and measurements in Fig.~\ref{fig:bowtie_s11} is likely due to simplifications made in the simulations, in particular the idealized representation of the balun at the feed. Laboratory characterization of the balun S-parameters (section~\ref{subsec:rec_cal}) will allow the feed model to be refined, which should reduce the discrepancy and further improve the forward model described in section~\ref{subsec:fwd_modelling}.

In Fig.~\ref{fig:bowtie_s11}(b), we show the delay transform of the measured antenna reflection coefficient. The antenna was suspended close to the nominal 100\,m above the canyon floor, where the nearest canyon terrain is 88\,m away. Equation~\eqref{eq:delay} and Fig.~\ref{fig:reflections} then predict environmental reflections only at delays beyond $\sim590\,\text{ns}$, which is what the measurement shows: between $\sim100\,\text{ns}$ and 590\,ns the delay spectrum sits at the noise floor, and beyond 590\,ns it rises more than 40\,dB above it. The structure below $\sim100\,\text{ns}$ is the antenna--receiver impedance mismatch rather than an environmental reflection.

\begin{figure}
    \centering
    \includegraphics[width=\columnwidth]{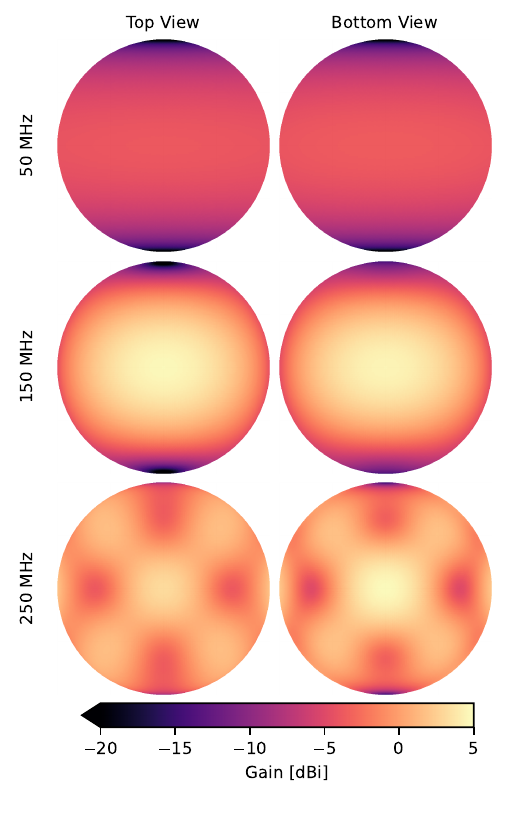}
    \caption{
    Simulated realized gain in dBi of the bowtie antenna at 50\,MHz (top row), 150\,MHz (middle row), and 250\,MHz (bottom row). The columns show the beam projected from above the antenna (top view, left column) and below the antenna (bottom view, right column). The low end of the colour scale is clipped at $-20\,\text{dBi}$ for visual clarity; the gain at 50\,MHz has an actual minimum of $\approx -35\,\text{dBi}$.}
    \label{fig:bowtie_beam}
\end{figure}

Simulations of the antenna beam pattern are shown in Fig.~\ref{fig:bowtie_beam}, at 50\,MHz, 150\,MHz, and 250\,MHz.
The simulations show that the beam pattern is nearly omnidirectional at 50\,MHz and becomes more directive with increasing frequency. In addition, the gain increases with frequency. The simulations show the smooth evolution of the beam pattern from 50\,MHz to 150\,MHz -- the side lobes appear only above 150\,MHz and can be seen in the simulation at 250\,MHz.
Due to the presence of the metallic box, the gain is slightly higher below the antenna (zenith angles greater than $90^\circ$) than above, most prominently seen in the simulation at 250\,MHz.

\subsection{Antenna platform} \label{subsec:box}
The antenna is mounted on an aluminium box that serves as the antenna platform. The box has a square cross-section of side length 14\,in (approximately 36\,cm) and a height of 16\,in (approximately 41\,cm). Copper panels are attached to the outside of the box to provide additional shielding against RFI generated by electronics within the box.
The dimensions of the box and the distance to the antenna were optimized in electromagnetic simulations to keep the reflection coefficient smooth across the frequency band of interest.
The antenna is held 40\,cm above the box, and is mechanically attached to the box using a wooden structure. Wood was chosen because it is lightweight, strong, and non-conductive. During rotations of the antenna relative to the box, this structure rotates with the antenna.

As Fig.~\ref{fig:concept} shows, the box is suspended in the canyon using a frame big enough to support rotations of the antenna around the horizontal axis through the box. The frame is made of hollow polyvinyl chloride (PVC) pipes and suspended with $3/16\,\text{in}$ Kevlar rope. 
Two lines of rope are strung across the canyon and tied to trees on top of each side. This double rope holds a pulley plate with two nylon pulleys. Two suspension ropes and two safety lines go through the pulleys, tie to each side of the frame, and are anchored on the ground. Pulling on these ropes allows the antenna to be raised or lowered from the ground. 
The anchoring to the ground also prevents the antenna from moving significantly side to side.

The box can rotate around two axes, as described above and illustrated in Fig.~\ref{fig:rotations}. A stepper motor driving a slew drive is used for each rotation axis.
This solution maintains the position of the box when the motors are not powered.
By keeping track of the starting position of the box and the steps taken by each motor, the orientation of the box can be derived.
Each axis is also equipped with an inertial measurement unit (IMU), giving independent verification of the box orientation.

The IMUs also record any tilts of the box due to wind during observations. They do not constrain the absolute horizontal position of the antenna, but that motion is small in practice, $\sim 1\,\text{m}$, and shifts the delays of reflected waves by only a few nanoseconds -- negligible for the delay-based filtering used to remove those reflections. It matters more for the horizon: a metre-scale displacement changes the fraction of the sky blocked by the canyon walls, causing kelvin-level changes in the antenna temperature (Fig.~\ref{fig:horizon_shift}). While the changes due to metre-scale displacements are absorbed by the foreground filter (section~\ref{subsec:fwd_modelling}), a lidar monitoring the antenna position was added for the July 2026 deployment to enhance the forward modelling.
A potentiometer on the azimuth axis provides a further independent readout of the platform orientation, complementing the motor step count and the IMU.

In addition to the motors, the box houses the analogue front-end electronics for the signal chain and the digital electronics that control the box. 
The analogue chain is powered by 6\,V batteries.
The motors and the digital electronics are powered by 12\,V batteries; separate 12\,V batteries power the RF switches and noise diode (see section~\ref{subsec:front_end}).
Using separate batteries increases the isolation between the analogue signal chain and noisy digital electronics.

\subsection{Front-end electronics} \label{subsec:front_end}
\begin{figure*}
    \centering
    \includegraphics[width=\linewidth]{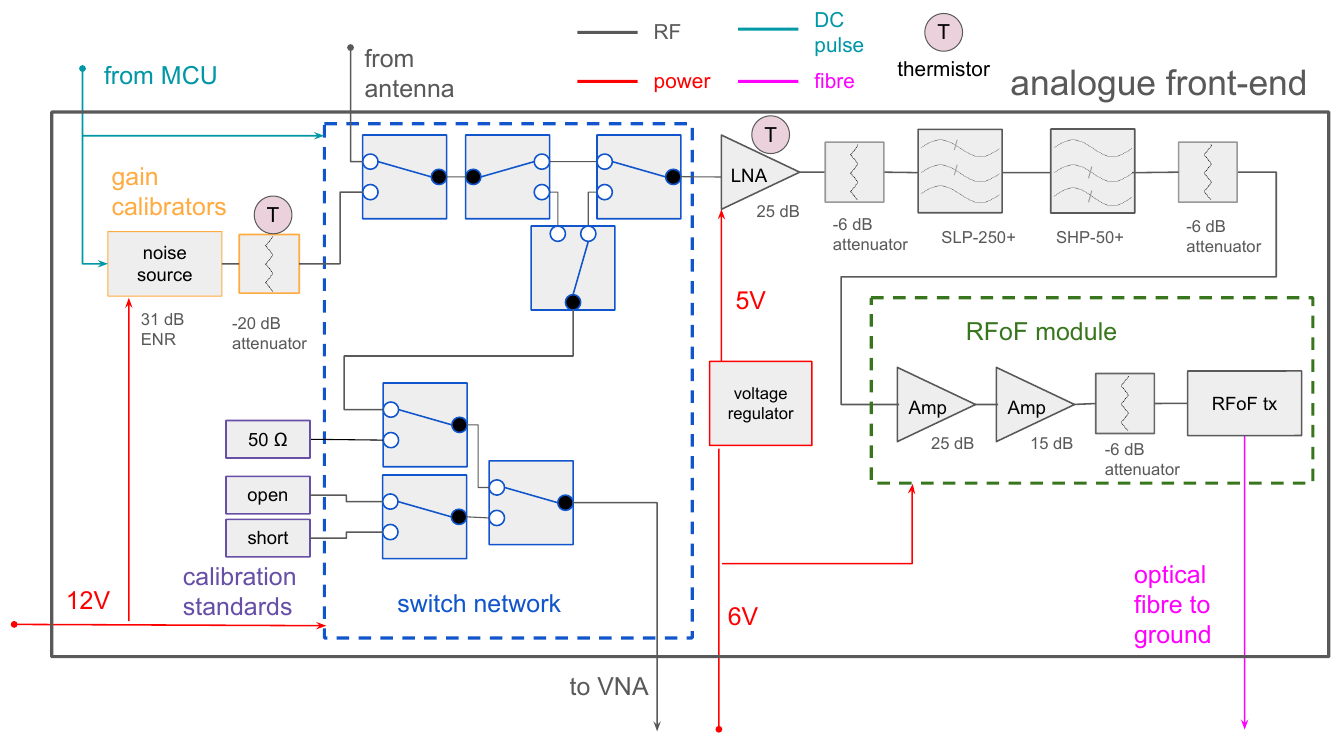}
    \caption{Block diagram of the front-end connected to the suspended antenna.}
    \label{fig:fe_block}
\end{figure*}

A prototype of the EIGSEP signal chain has been developed and is used for observations. The suspended bowtie antenna has a custom-made front-end specifically developed for EIGSEP. This front-end is described in this section and shown schematically in Fig.~\ref{fig:fe_block}.
A simplified version of this front-end was also used for a ground antenna in the July 2026 deployment.
Through the July 2025 deployment, the ground antennas used front-end modules (FEMs) developed for the Hydrogen Epoch of Reionization Array \citep[HERA;][]{2017PASP..129d5001D, 2024PASP..136d5002B}. Details about the ground antennas and the FEMs are provided in section~\ref{sec:vivaldis}.

After the balun, which is placed at the feed of the bowtie antenna, a 40\,cm coaxial SMA cable leads through the central column of the wooden platform and connects to the receiver in the box. The first component inside the box is the RF switch network. There are seven electromechanical switches in the network, of the type Teledyne CCR-33S. They are fail-safe, single pole, double throw (SPDT) switches and the network follows the design used by MIST \citep[see fig. 17 in][]{2024MNRAS.530.4125M}. According to the datasheet, the switches provide 70\,dB of isolation and have a maximum insertion loss of 0.2\,dB. 
We confirmed these specifications in our laboratory, measuring $\leq0.1\,\text{dB}$ insertion loss and $\geq90\,\text{dB}$ isolation over 30--300\,MHz.
The switches require continuous 200\,mA to be switched to the high-power state and remain in the low-power state if no power is supplied.

The main signal path through the switches connects the antenna to the receiver for sky observations. The other switch positions are used for calibration, including reflection coefficient measurements with a VNA and noise source measurements to set the temperature scale. A relay switch is used to turn the noise diode on and off. 
The calibration subsystems are described in section~\ref{sec:calibration}. Here we focus on the amplification chain used for sky observations.
The simplified front-end used for the ground antenna in July 2026 has no switch network, noise source, or VNA: its signal path is the amplification chain described below.

\begin{figure}
    \centering
    \includegraphics[width=\linewidth]{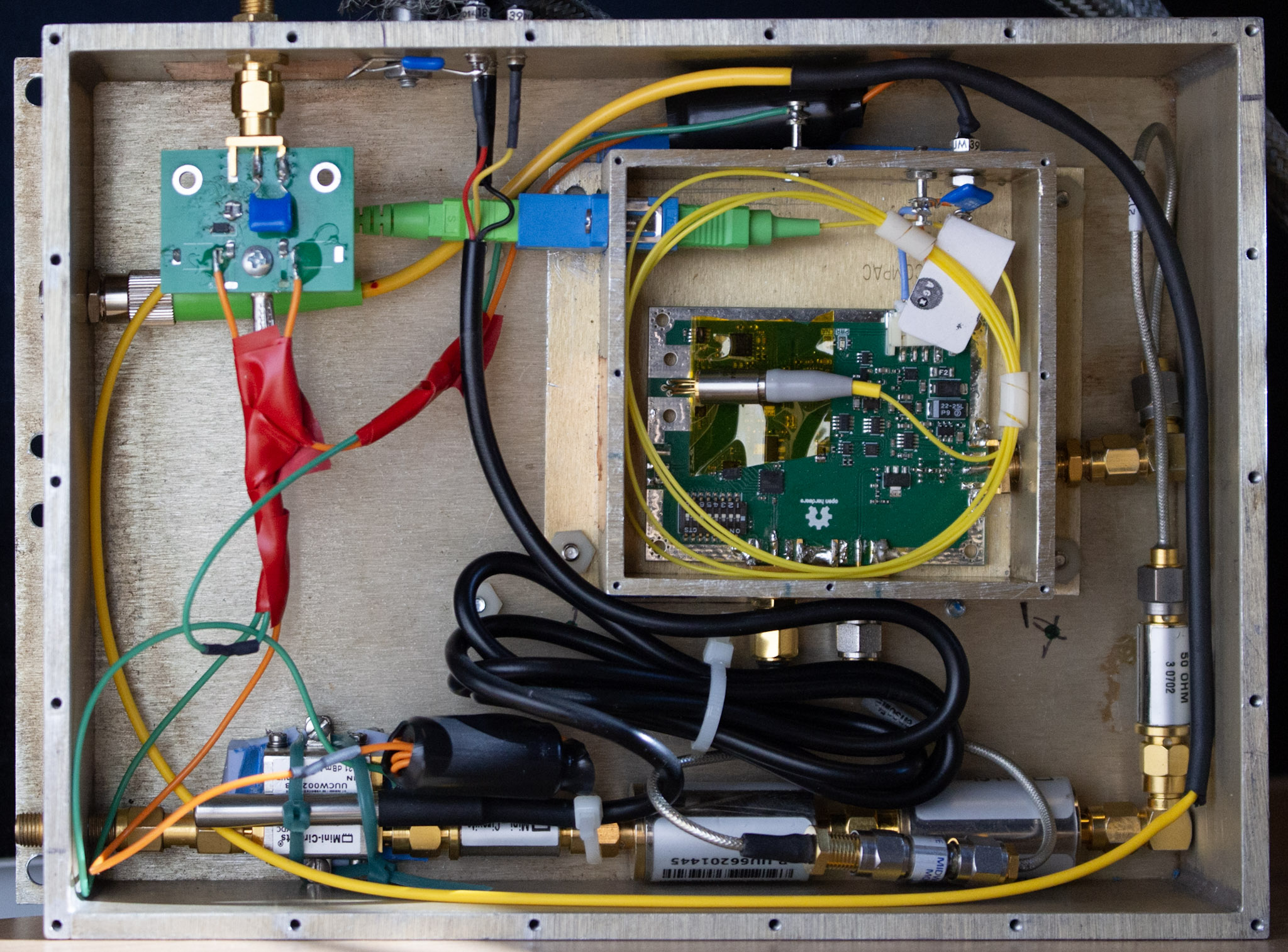}
    \caption{The amplification chain for the bowtie antenna. The RF signal input is in the bottom left corner. The signal goes anticlockwise through the amplifiers and filters on the bottom and right-hand side of the picture, through the RFoF module inside the inner box, and out through the optical output on the top left.
    }
    \label{fig:signal_chain}
\end{figure}

EIGSEP's amplification chain uses a Mini-Circuits ZX60-P103LN+ as the first LNA. At 50\,MHz, this LNA has a gain of 25.3\,dB and noise figure of 1.2\,dB. The ambient temperature of the LNA is monitored with a thermistor.
As shown in the system diagram in Fig.~\ref{fig:fe_block}, the LNA is followed by a 6\,dB attenuator, filters providing a bandpass of 50--250\,MHz, a second attenuator, and an RF-over-fibre (RFoF) module.
The RFoF module is a custom-printed circuit board derived from a design developed for the DSA telescope \citep{2019BAAS...51g.255H}. This module features two onboard amplifiers and a laser diode that modulates the RF signal to optical wavelengths for transmission over fibre-optic cable. The total gain of the onboard amplifiers is $\approx40\,\text{dB}$.
We made small modifications to the original board, including changing the trace width for impedance match and changing the onboard filters to match the EIGSEP frequency range. We also added a 6\,dB external attenuator between the second amplifier and the diode to reduce reflections.
Both the original design and the modified version of the RFoF board are open sourced under the CERN OHL S-V2 licence. They are available at \url{https://gitlab.com/dsa-2000/asp/rfof/ftx} and \url{https://github.com/liuweiseu/rfof-ftx/tree/cefad78}, respectively.
The amplification chain as implemented in 2025 is illustrated in Fig.~\ref{fig:signal_chain}.

A single-mode fibre-optic cable connects the front-end in the box to the post-amplification module on the ground. The cable is 1000\,ft (305\,m) long and is not conductive. The long cable ensures that any potential cable reflections have delays of at least 3000\,ns -- long enough that they can be filtered from the measured spectra without signal loss \citep{2016MNRAS.460.4320E, 2019ApJ...884..105K}.

\subsection{Digital electronics}
The LattePanda Mu\footnote{\url{https://www.lattepanda.com/lattepanda-mu}} single-board computer is used in the box to control the motors, the switches, and the VNA. It has an x86 processor that allows it to run the VNA software.
The computer is connected to the microcontroller unit (MCU), which consists of six Raspberry Pi Pico microcontrollers that interface with the switches, the motors, the IMUs, and the temperature sensors.
The computer connects to a separate single-board computer on the ground through Ethernet over a digital fibre-optic link.
The computer, MCU, and VNA form a unit placed inside a separate metal box within the antenna platform box to minimize self-generated interference.

\subsection{Ground antennas}
\label{sec:vivaldis}
Two Vivaldi feeds of the type developed for HERA Phase II were used as ground antennas through the July 2025 deployment. Each Vivaldi feed features two orthogonal blades, enabling dual-polarization measurements. The antenna is optimized over 50--250\,MHz, where it is impedance matched to the HERA FEM with a reflection coefficient of about $-10\,\text{dB}$. 
Like the custom EIGSEP front-end, the HERA FEM has an RFoF module, and connects to fibre-optic cables at the output. The same type of fibre-optic cables are used for the ground antennas as for the suspended antenna, in 500\,ft (152\,m) lengths.
The Vivaldi feed and HERA FEM are described in detail in \citet{2021ITAP...69.8143F} and \citet{2024PASP..136d5002B}.
While HERA uses the Vivaldi feeds to illuminate a 14-m dish, EIGSEP uses the feeds in isolation without a dish.
This choice keeps the instrument portable and reduces cost \citep{2022JAI....1150001C}.

The FEMs feature an internal load and active noise source, but there is no autonomous switching between sky measurements and these devices in the current EIGSEP design. Only the receiver used for the suspended antenna will be absolutely calibrated. 
In future versions of the system, the same level of calibration may be attempted for the ground antennas.

For the July 2026 deployment, the Vivaldi feeds were replaced by a single ground copy of the suspended system: the same shaped bowtie with a simplified version of the front-end electronics (section~\ref{subsec:front_end}), so that the suspended and ground antennas share the same design and signal path.
The ground-antenna measurements presented in Figs.~\ref{fig:rfi} and \ref{fig:point_source} predate the change and use the Vivaldi feeds, while the new ground copy was used alongside the suspended antenna in the rotation measurements of section~\ref{subsec:beam_mapping}.

\subsection{Post-amplifiers and correlator}
Through the July 2025 deployment, the fibre-optic cables from each antenna connected on the ground to a HERA post-amplifier module (PAM). The HERA PAM was co-designed with the HERA FEM and is also detailed in \citet{2021ITAP...69.8143F} and \citet{2024PASP..136d5002B}.
In each PAM, the optical signal is converted back to radio frequencies and amplified to levels appropriate for digitization.
For the July 2026 deployment, the PAMs were replaced by a custom post-amplifier based on the receiving counterpart of the RFoF module described in section~\ref{subsec:front_end}, followed by custom LNAs.
Both the original HERA PAMs and the custom replacements are housed in an RF-shielded and watertight enclosure.

The final stage of the signal chain is the Smart Networked ADC Processor (SNAP). The SNAP is a digital signal processing module based on a field-programmable gate array (FPGA), co-designed by HERA and the Collaboration for Astronomy Signal Processing and Electronics Research \citep[CASPER;][]{2016JAI.....541001H, 2017PASP..129d5001D}. It provides six inputs, digitized with 8-bit Hittite HMCAD1511 analogue-to-digital converters (ADCs) at 500\,Msps and channelized with a 2048-point polyphase filter bank. Integrated correlations are computed at an accumulation length set in software, 0.54--1.07\,s in the configurations deployed to date, and output via 1\,Gb Ethernet to a Raspberry Pi.
The July 2025 configuration used five inputs: one from the single-polarization suspended antenna and two from each of the two dual-polarization ground antennas. This configuration yielded 11 correlations, including 5 autocorrelations and 6 cross-correlations. The cross-correlations consist of the 4 products formed between the suspended antenna and the ground antennas, in addition to the 2 co-polarized products (N--N and E--E) between the ground antennas. 
The ground copy used from the July 2026 deployment (section~\ref{sec:vivaldis}) is single-polarization, so that configuration uses two inputs, giving two autocorrelations and one cross-correlation.

\begin{figure*}
    \centering
    \includegraphics[width=\linewidth]{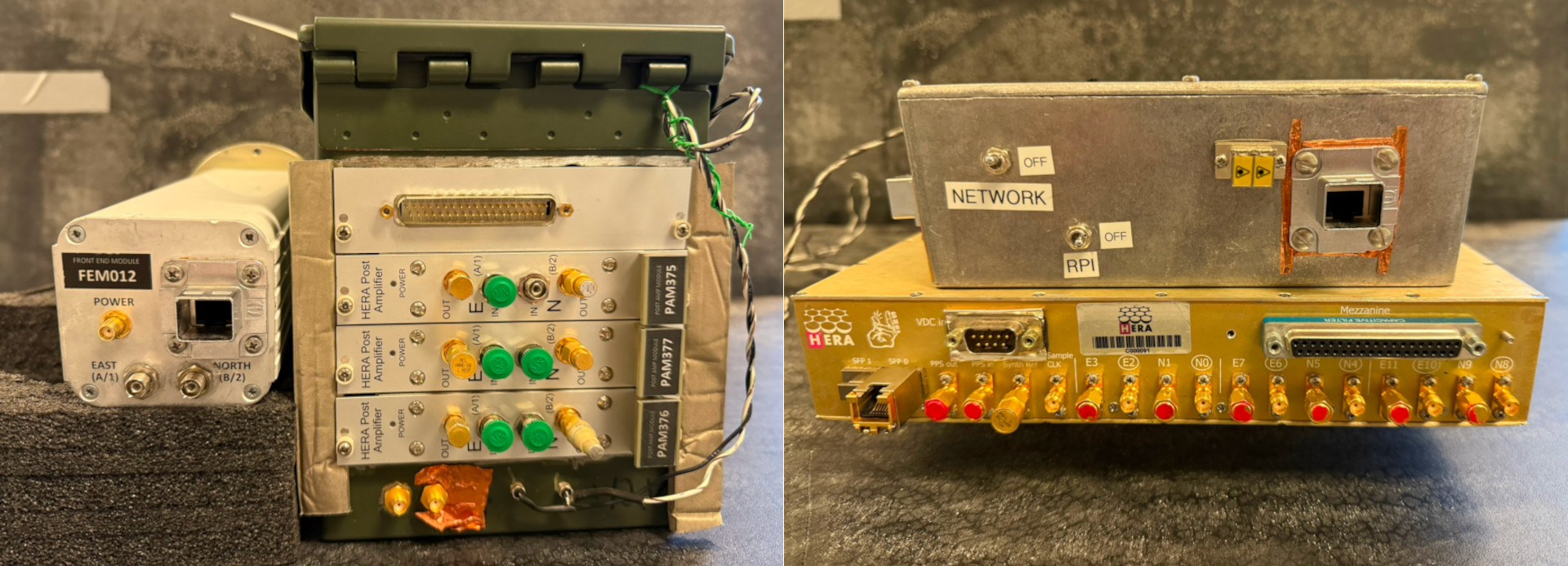}
    \caption{Components of the EIGSEP receiver as deployed in the field through the July 2025 campaign. \textit{From left to right}: The FEM receives the analogue signal at each ground antenna, amplifies it, and transmits it with RFoF to the node. Each FEM connects to a PAM via optical fibre, in the same way the custom front-end at the suspended antenna connects to a PAM. 
    The three PAMs are housed together. Analogue signals from the three antennas are digitized in the SNAP and correlated. On top of the SNAP is the enclosed Raspberry Pi, which programs the SNAP and saves data to disk. Each component is placed inside connectorized RF-tight enclosures.}
    \label{fig:receiver}
\end{figure*}

A low-power single-board computer (Raspberry Pi) writes correlation data to USB storage, configures the SNAP, and records sensor data. 
The Raspberry Pi is housed inside a metal box.
The total data volume is approximately 5\,GB for 24 hours of observation.
Fig.~\ref{fig:receiver} shows the FEM, PAM, SNAP, and Raspberry Pi.

\section{Calibration Methodology} \label{sec:calibration}
Measurements of the global 21-cm signal require exquisite calibration due to the large dynamic range between the foregrounds and the signal. An important part of this requirement is to establish an absolute calibration scale, which converts measured voltages to sky temperature, taking all gains, losses, and reflections in the signal path into account. In general, these are frequency-dependent quantities, meaning that imprecise calibration -- even at a low level -- may artificially introduce spectral shapes that resemble the global 21-cm signal.

Instrumental systematics come in three distinct flavours:
\begin{enumerate}
    \item Additive bias such as amplifier noise and impedance mismatch, which adds frequency-dependent signal power.
    \item Direction-independent multiplicative bias from, for instance, amplifiers, filters, and cable reflections, or due to impedance mismatch, which modulate foreground-dominated signals with a multiplicative bandpass.
    \item Direction-dependent multiplicative bias from the antenna beam and surrounding environment, which imprints incoming radiation with a multiplicative distortion as a function of direction.
\end{enumerate}

Due to the stringent calibration requirements, a key aspect of the EIGSEP approach is using independent measurements to calibrate the same features. Independent measurements reduce the reliance on any one method, and the measurements can be used to validate each other.
Below, we describe the mathematical model of our measurements and outline the methodology used to calibrate the instrumental systematics.

\subsection{Model of measurement}
The suspended antenna measures a frequency-dependent temperature $T_{\text{ant}}(\nu)$ that can be written as
\begin{equation} \label{eq:convolution}
    T_{\text{ant}}(\nu) = \int_0^{2\pi} \int_{0}^{\pi} A(\nu, \theta, \phi) T_{\text{emit}}(\nu, \theta, \phi) \sin{\theta} \,\mathrm{d}\theta \,\mathrm{d}\phi,
\end{equation}
where $A$ is the normalized antenna beam pattern and $T_{\text{emit}}$ is the temperature associated with any emission from the sky or from the ground. Both quantities are functions of frequency $\nu$ and the angles $\theta$ and $\phi$ on the sky. In equation~\eqref{eq:convolution}, the antenna beam pattern $A$ is normalized over the sphere such that
\begin{equation} \label{eq:beam_norm}
    \int_0^{2\pi} \int_{0}^{\pi} A(\nu, \theta, \phi) \sin{\theta} \,\mathrm{d}\theta \,\mathrm{d}\phi = 1.
\end{equation}
In this formalism, $A$ does not include losses associated with the finite antenna conductivity or impedance mismatches. Instead, we incorporate these into the receiver gain $g_{\text{rx}}$, defined below. 

Aside from RFI, which we assume is already removed in this model, $T_{\text{emit}}$ is primarily a sum of sky emission $T_{\text{sky}}$ and ground emission $T_{\text{gnd}}$. EIGSEP aims to measure $T_{\text{sky}}$, but, as seen in Fig.~\ref{fig:bowtie_beam}, the antenna is very sensitive to ground radiation. This radiation includes thermal radiation and reflected sky emission, delayed by the light travel time from the ground to the antenna.
Due to the electromagnetic isolation of the EIGSEP antenna, we anticipate that most of the reflected sky emission can be filtered in delay space, and the remaining reflected emission is heavily attenuated due to the long distance.
Thus, we assume that the residual $T_{\text{gnd}}$ is dominated by thermal emission associated with the physical temperature of the environment.

For brevity, we omit the explicit dependencies on $\nu$, $\theta$, and $\phi$ in the notation.
We may rewrite equation~\eqref{eq:convolution} by separating $T_{\text{emit}}$ into $T_{\text{sky}}$ and $T_{\text{gnd}}$.
We denote the part of the sky that is not blocked by the canyon walls by $\Omega_{\text{above}}$, and the ground and the canyon walls by $\Omega_{\text{below}}$. We get
\begin{equation} \label{eq:conv_split}
    T_{\text{ant}} = \int_{\Omega_{\text{above}}} AT_{\text{sky}} \sin{\theta} \,\mathrm{d} \theta\,\mathrm{d} \phi + \int_{\Omega_{\text{below}}} AT_{\text{gnd}} \sin{\theta} \,\mathrm{d} \theta\,\mathrm{d} \phi.
\end{equation}

The power $P_{\text{ant}}$ received from the suspended antenna, measured as the autocorrelation spectrum, is related to the antenna temperature $T_{\text{ant}}$ and the gain $\left(g_{\text{rx}}\right)$ and the temperature $\left(T_{\text{rx}}\right)$ of the receiver by
\begin{equation} \label{eq:gain_scale}
    P_{\text{ant}} (\nu) = g_{\text{rx}} (\nu) \left(T_{\text{ant}} (\nu) + T_{\text{rx}} (\nu)\right).
\end{equation}

For every measurement of $P_{\text{ant}}$, the EIGSEP calibration first involves solving for $T_{\text{ant}}$ in equation~\eqref{eq:gain_scale}.
We then require accurate models and measurements of the antenna beam $A$ and of the horizon profile to estimate $T_{\text{sky}}$ from equation~\eqref{eq:conv_split}.
A direct inversion of equation~\eqref{eq:conv_split} may not be necessary to extract the 21-cm global signal, but we likely need accurate understanding of the spectral modes of $A$ and $T_{\text{sky}}$. The details of the signal extraction and estimates of the constraining power of EIGSEP are left for future work. In the remainder of this section, we focus on the calibration needed to obtain $T_{\text{ant}}$ and $A$.

\subsection{Receiver calibration} \label{subsec:rec_cal}
To measure system gain and characterize the bandpass, EIGSEP is working to implement a version of the receiver calibration formalism pioneered by EDGES. The method is detailed in \citet{2012RaSc...47.0K06R} and \citet{2017ApJ...835...49M}; here, we give a summary of the implementation for EIGSEP.

We need to determine the receiver gain and temperature to invert equation~\eqref{eq:gain_scale} and solve for $T_{\text{ant}}$. In the field, the EIGSEP receiver switches between measuring the antenna and the internal noise source. When the noise source is off, the measured noise temperature is the physical temperature of the $-20\,\text{dB}$ attenuator. The off condition represents the ambient load, and we monitor its physical temperature with a thermistor. When the noise source is on, the injected noise temperature increases from that of the ambient load by a factor given by the excess noise ratio (ENR) of the noise source.
These two calibration measurements can be combined to obtain a first-order estimate of the antenna temperature \citep[equation 1 in][]{2017ApJ...835...49M}. This estimate does not account for reflections at the interface between the antenna and the receiver due to impedance mismatch or that the noise is injected at a different switch position from the receiver input.
The calibration of these effects requires measuring the reflection coefficients of the antenna and the receiver, and the receiver noise wave parameters \citep{1978ITMTT..26...34M}. 
The noise wave parameters are five frequency-dependent parameters that represent noise from the LNA reflected at the interface between the antenna and the receiver, and corrections to the assumed injected noise temperature of the ambient load and noise source.
We will determine these by measuring four external calibration standards in the laboratory, following the procedure in \citet{2017ApJ...835...49M}.

In our calibration formalism, we have implicitly absorbed into $g_{\text{rx}}$ and $T_{\text {rx}}$ the reflection coefficients of the antenna and receiver, the antenna radiation efficiency, and the loss from the balun. While the reflection coefficients are measured in the field, we rely on electromagnetic simulations of the antenna to determine the radiation efficiency. The efficiency of the balun will be determined through a combination of simulations and laboratory characterization of the S-parameters, following \citet{2024MNRAS.530.4125M}.

\subsection{Reflection coefficient measurements} \label{subsec:vna}
EIGSEP employs a Copper Mountain Technologies R60 1-port VNA for measurements of the antenna and receiver reflection coefficients.
These coefficients are periodically measured in situ. The VNA is calibrated following the procedure used by EDGES and MIST \citep{2017ApJ...835...49M, 2024MNRAS.530.4125M, 2018Natur.555...67B}. Immediately before each measurement of the antenna reflection coefficient, the VNA is used to measure internal open, short, and load (OSL) calibration standards. This set of measurements provides a first-order calibration with the reference plane internal to the receiver.
The reflection coefficient of the receiver is measured with a lower power level, to avoid saturation of the LNA. We repeat the OSL calibration of the VNA at this power level prior to the receiver reflection coefficient measurement.
To shift the reference plane of the measurements to the receiver input, the S-parameters associated with cables and switches inside the receiver are de-embedded. These S-parameters are measured in the laboratory using the S911T calibration standards from Copper Mountain Technologies.

\subsection{Interferometric measurements}
As the SNAP correlates all its inputs, EIGSEP has cross-correlation spectra available alongside autocorrelations. The auxiliary ground antennas are therefore correlated with the suspended antenna and, when more than one is deployed, with each other.
The ground antennas are portable and are placed where the terrain allows. Two ground antennas unspooled in opposite directions give a baseline of up to 300\,m between them. In the July 2025 configuration, the baseline to the suspended antenna was about 150\,m horizontally and 100\,m vertically, or 180\,m in total.

The cross-correlation spectra serve two practical purposes.
They are used to identify RFI, which is correlated between antennas and distinct from the diffuse sky emission (Fig.~\ref{fig:rfi}).
Moreover, the cross-correlations average down the uncorrelated receiver noise and resolve out diffuse galactic emission thanks to the long interferometric baseline, making them useful for observing bright celestial sources.
For a source near the zenith, the fringe spacing is set by the horizontal component of the baseline: about $2^\circ$ at 50\,MHz and $0.5^\circ$ at 250\,MHz for a 150\,m separation, far finer than the scales over which the diffuse emission varies.
In particular, Cygnus A passes $\sim1.5^\circ$ from the zenith at the site and has a well-characterized spectrum \citep{1977A&A....61...99B, 2017ApJS..230....7P}, making it a convenient external reference source for EIGSEP.
We also use cross-correlations to qualitatively evaluate terrain coupling. 
The ground antennas are not absolutely calibrated and their beam patterns are not measured in the field. Interferometric data are therefore not used for absolute calibration and their use for sky modelling is left to future work.

\subsection{Antenna beam mapping} \label{subsec:beam_mapping}
To calibrate the direction-dependent and frequency-dependent antenna beam, EIGSEP will use a combination of numerical electromagnetic simulations and field measurements.
Simulations are done in HFSS and produce both predictions for the antenna reflection coefficient and antenna beam pattern.

EIGSEP aims to measure the beam pattern of the suspended antenna using a transmitting antenna placed on the ground directly below.
The first implementation used an electrically short dipole, a commercial RTL-SDR Blog Dipole Antenna\footnote{\url{https://www.rtl-sdr.com}} of total length 46\,cm. The commercial dipole is well approximated as a point source in equation~\eqref{eq:beam_mapping} and has a radiation pattern that can be described analytically \citep[e.g.][]{1982nyhr.book.....B}.
For the July 2026 deployment, the transmitter was upgraded to a dual-polarization blade dipole with a custom reflector, to increase directivity and reduce multipath scattering off the canyon walls. The dual-polarization transmitter helps break the degeneracy between the radiation pattern and the polarization loss factor changing with the rotation of the suspended bowtie.

\begin{figure*}
    \centering
    \includegraphics[height=6.2cm]{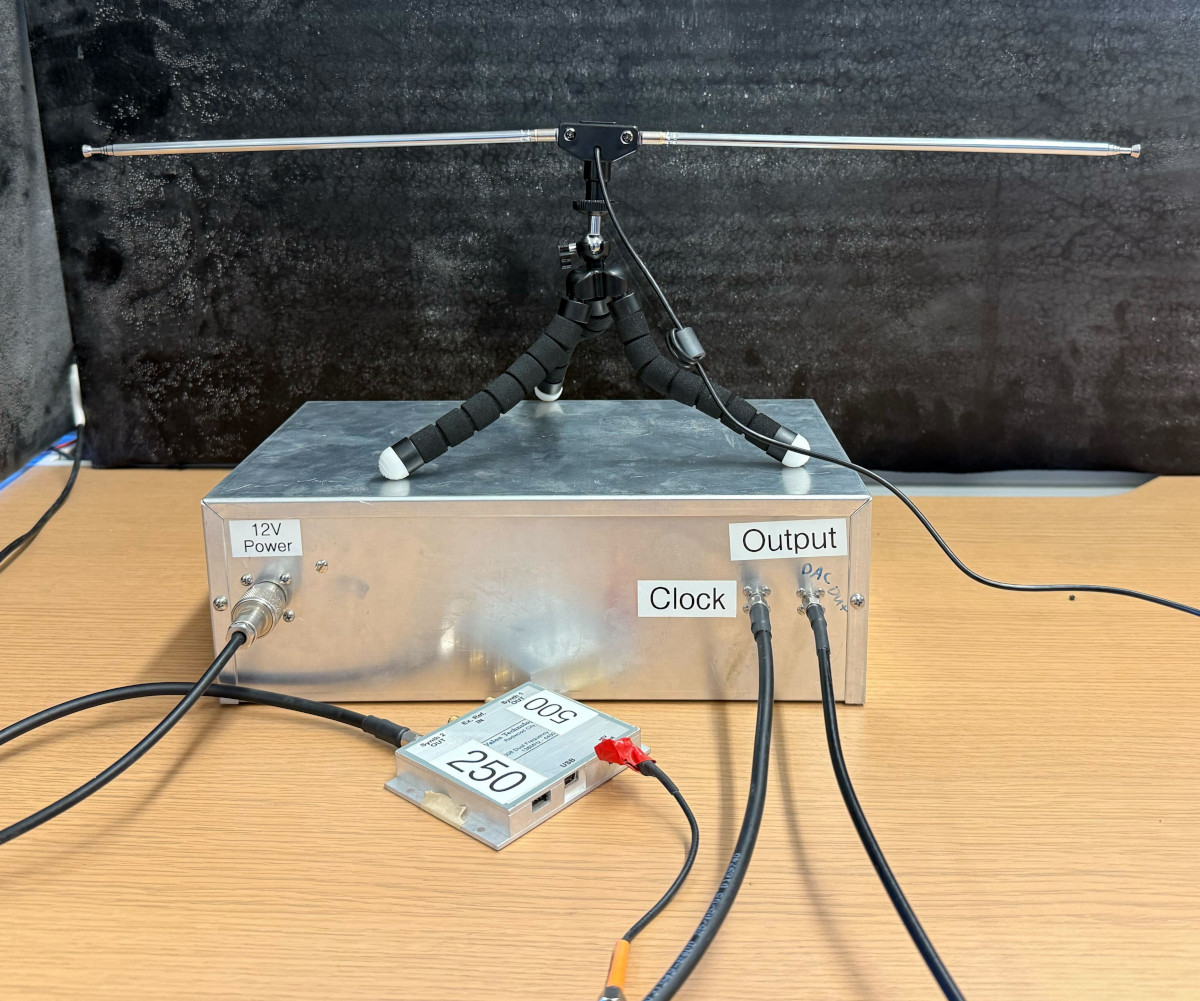}
    \hfill
    \includegraphics[height=6.2cm]{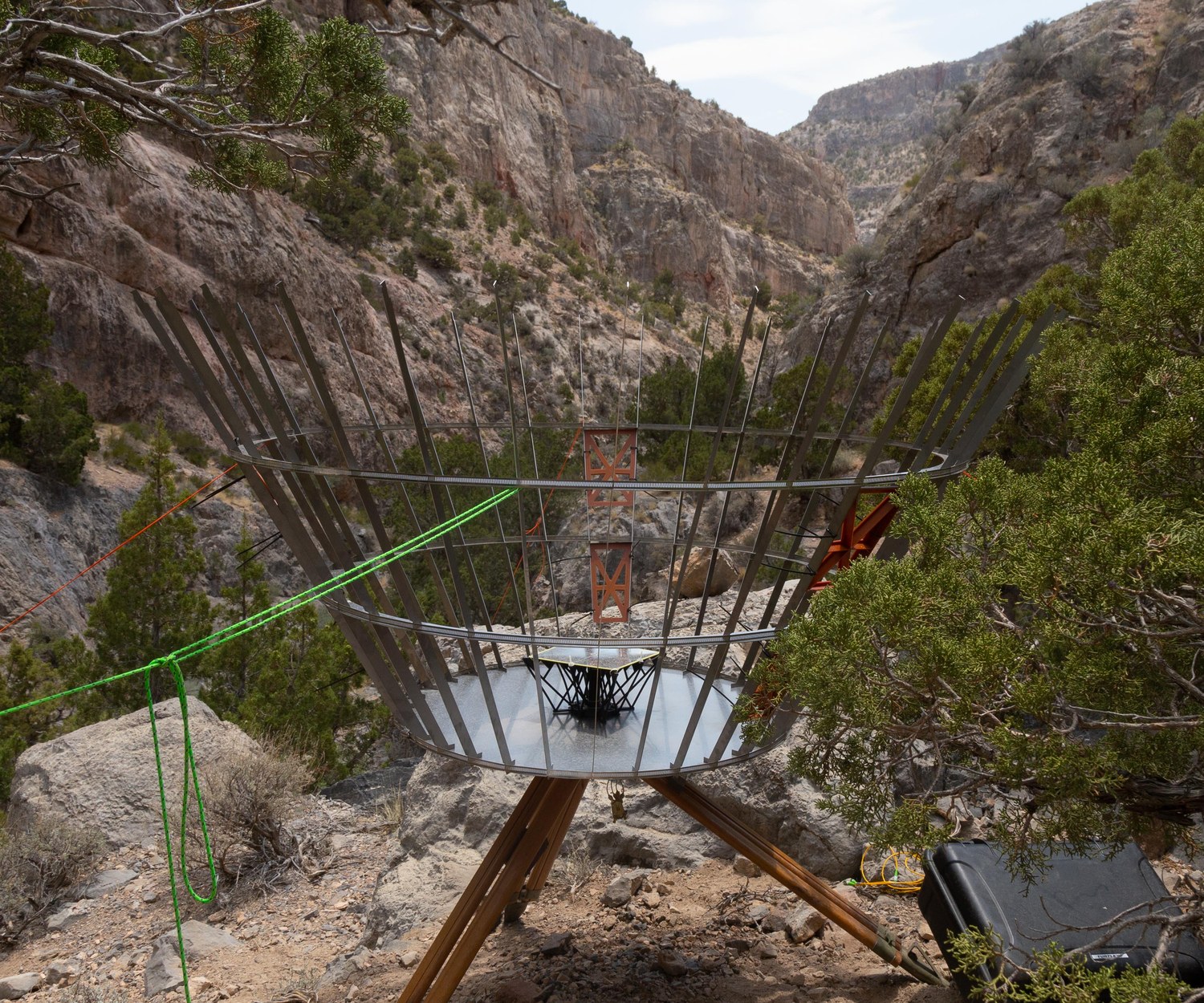}
    \caption{The EIGSEP transmitters. \textit{(a)} The short dipole used in the July 2025 deployment. The antenna is connected to the RFSoC board, placed inside the metal box underneath the antenna, which generates the transmitted Dirac comb. The clock is seen in front of the RFSoC box. During field measurements, the metal box is separated from the antenna to minimize its impact on the antenna performance. \textit{(b)} The transmitter used in July 2026: a dual-polarization blade dipole with a custom reflector, on a tripod on the canyon floor below the suspended antenna. The reflector makes the transmitter more directive, which is intended to reduce multipath scattering off the terrain. The same RFSoC board was used to generate the combs in both cases.}
    \label{fig:transmitter}
\end{figure*}

The short dipole antenna transmitted a sequence of Dirac delta functions on every 16th channel ($\approx 3.9\,\text{MHz}$) across the frequency band, generated by a digital-to-analogue converter (DAC) on a Xilinx Radio Frequency System-on-Chip (RFSoC) 2x2\footnote{\url{https://xilinx.github.io/RFSoC2x2-PYNQ/}}.
In 2026, the dual-polarization antenna was used to transmit in every 8th channel with alternating polarization, producing two interleaved Dirac combs. 
The clock of the RFSoC transmitter board is synchronized with the clock of the SNAP correlator so that, in the ideal case, the transmitted signal is exactly a Dirac comb with no spectral leakage.
In practice, phase noise or imperfect synchronization could broaden the comb slightly.
Laboratory measurements place an upper bound on the leakage into adjacent channels of 61\,dB below the comb power.
If we find evidence of the transmitted Dirac comb leaking to adjacent channels, those channels may be flagged.
Duty-cycling of the transmitter during field measurement could also be considered to further validate that the comb is confined to the transmitter channels.
Both transmitters are shown in Fig.~\ref{fig:transmitter}.

The transmitter antenna is continuously transmitting during the observations. With no spectral leakage of the Dirac comb, the transmitted power is confined to 64 of the 1024 frequency channels for a single polarization. The two polarizations of the 2026 transmitter transmit on alternating channels of the comb, so that together they occupy every 8th channel, 128 channels in total.
In the other channels, equation~\eqref{eq:conv_split} still applies as the model for our measurements without contamination by the transmitted signal.
We tune the transmit power level $P_{\text{tx}}$ in the field to make the tones detectable above the continuum at the bottom of the band, where the transmitter is least efficient and the sky is brightest, while keeping it low enough that the tones stay within the linear range of the correlator at the top of the band.
We remove the residual sky and receiver contribution in the transmitter frequency channels, estimated from the median of the neighbouring non-comb channels, so that the model for our measurement becomes (instead of equations~\eqref{eq:conv_split} and \eqref{eq:gain_scale})
\begin{equation} \label{eq:beam_mapping}
    P_{\text{ant}}(\nu_{\text{tx}}) \propto A(\nu_{\text{tx}}, \theta_{\text{tx}}, \phi_{\text{tx}}) P_{\text{tx}}(\nu_{\text{tx}}),
\end{equation}
where $\nu_{\text{tx}}$ denotes the transmitter frequency channels and the transmitter is at the position ($\theta_{\text{tx}}$, $\phi_{\text{tx}}$) defined in the local antenna coordinate system.
Here, we treat the transmitter as a point source.
The frequency-dependent factor $P_{\text{tx}}$ in equation~\eqref{eq:beam_mapping} depends only on the properties of the transmitter antenna -- its radiation efficiency, gain, and reflection coefficient.
Provided that the position of the suspended antenna is stable relative to the transmitter antenna, we can directly probe the angular dependence of the bowtie beam by rotating it and measuring $P_{\text{ant}}(\nu_{\text{tx}})$ as a function of orientation.
At each frequency, this measurement yields the angular antenna beam pattern up to the single factor $P_{\text{tx}}$; because the beam is normalized over the sphere in the measurement equation~\eqref{eq:beam_norm}, that factor equals the integral of the measured beam pattern over the solid angle.
The frequency-dependent beam shape can therefore be recovered without calibrating the transmitter.
The absolute temperature scale that converts measured power to sky temperature comes separately from the internal receiver calibration (sections~\ref{subsec:rec_cal} and \ref{subsec:vna}), not from the transmitter.

Transmitting continuously during observations also allows us to monitor the temporal stability of the receiver gain.
An absolutely calibrated transmitter could additionally serve as an external gain reference, though this is not required for the beam mapping.

Rather than absolute calibration of the transmitter antenna, the main practical challenge is to separate the directly transmitted signal from multipath reflections off the terrain.
The reflector used in 2026 is intended to reduce this systematic by making the transmitter more directive.
A full description of the beam mapping method, its validation, and the recovery of the antenna beam pattern from the field measurements will be presented in Vazquez et al. 2026 (in prep.).

\begin{figure}
    \centering
    \includegraphics[width=\linewidth]{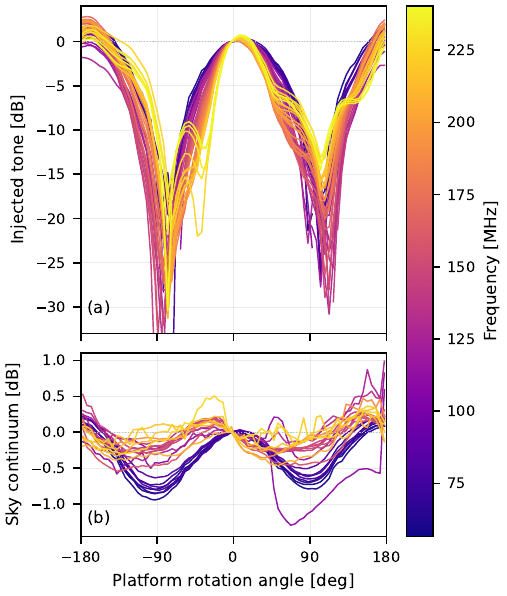}
    \caption{Response of the suspended EIGSEP antenna to platform rotation, measured against an injected signal during the July 2026 deployment. Both panels show power as a function of rotation angle, binned in $5^\circ$ steps, referenced to the power at $0^\circ$ rotation and coloured by frequency.
    \textit{(a)} The injected tones, one curve per transmitted tone, each with the sky continuum subtracted from its channel. Tones contaminated by RFI are excluded, including those inside the FM broadcast band and those whose continuum estimate draws on channels inside it.
    Each curve is drawn wherever the continuum-subtracted power is positive, so a curve breaks where the tone falls to the level of the continuum.
    The vertical axis is clipped at $-33$\,dB, just below the deepest point at which any tone is still significantly detected; tails reaching that limit are noise.
    \textit{(b)} The continuum estimate subtracted from each tone in \textit{(a)}, over the same rotation: the median of the eight neighbouring non-comb channels, which carry the sky and the receiver noise rather than the injected signal.
    It is drawn for the 24 of the 42 tones, spanning 57--225\,MHz, whose neighbouring channels were stable to within 0.35\,dB on the stationary ground copy over the same integrations.}
    \label{fig:beam_modulation}
\end{figure}

The method is demonstrated on field data from the July 2026 deployment in Fig.~\ref{fig:beam_modulation}(a). The suspended platform was rotated about the elevation axis, at a fixed azimuth of $-90^\circ$, while the transmitter injected the calibration comb. The transmitter is dual-polarized and the figure shows the comb corresponding to the transmitter polarization most aligned with the bowtie at this azimuth.
The continuum is subtracted from each tone channel to remove the sky emission and the receiver noise power, which contribute additively to every channel, leaving the transmitted signal.
Over a single turn, the response varies by 25\,dB for the median tone and by up to 33\,dB. The maxima correspond to when the antenna faces the transmitter and the nulls are near $\pm90^\circ$ of rotation, when the antenna is orthogonal to the transmitter. The ground copy of the suspended antenna and receiver, observing the same transmitter over the same integrations, varied by 0.03\,dB, confirming that the modulation is the response of the moving antenna rather than drift in the transmitter or the observing conditions.
Above 158\,MHz the tone stays above the noise floor through the null in 21 of the 22 tones, but at lower frequencies the tone reaches the noise before the null does and the depths are therefore upper limits on the power received in the null. The depths quoted use only the bins in which a tone stays above its own noise level, which differs from tone to tone.
Of the 51 comb channels of this polarization that fall in the observing band, 8 are not used due to significant RFI contamination: 6 in or beside the FM band, and 2 at 244 and 248\,MHz, which coincide with an allocated UHF satellite communications band. The lowest frequency comb in the band, at 52.7\,MHz, is not detected above the noise in a single turn.

\subsection{Antenna beam modulation} \label{subsec:beam_modulation}
\begin{figure}
    \centering
    \includegraphics[width=\linewidth]{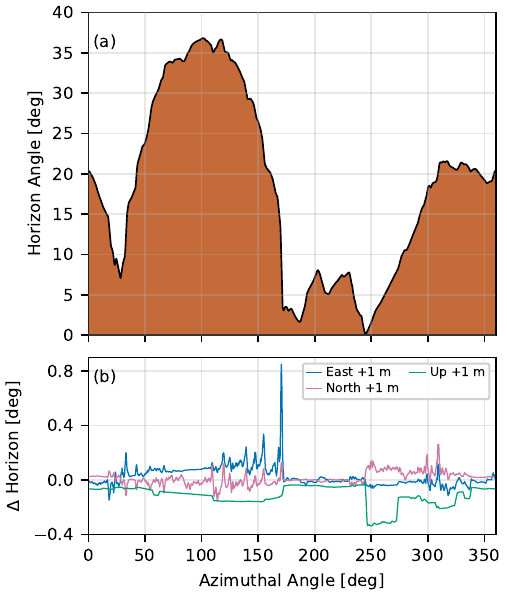}
    \caption{Horizon profile at the nominal antenna site as a function of azimuth (panel a), and the change in horizon angle produced by 1\,m antenna displacements along east, north, and up (panel b). In (a) the brown region represents the canyon terrain $\left(\Omega_{\text{below}}\right)$ and the white region is the view of the sky $\left(\Omega_{\text{above}}\right)$; azimuth is measured from north towards east. The horizon profiles in both panels are smoothed in azimuth to $1.4^{\circ}$ resolution. The horizon shifts in (b) are of order $0.1^{\circ}$ (RMS) for all three directions. The horizontal displacements (east and north) concentrate at the steep cliffs surrounding the antenna, reaching $0.85^{\circ}$ to the east and $0.26^{\circ}$ to the north.
    The vertical displacement lowers the horizon by $0.12^\circ$ on average ($0.14^\circ$ RMS), an angle comparable to the horizontal shifts, but at all azimuths, making the change in solid angle blocked by the canyon much larger: the vertical displacement uncovers $39\,\text{deg}^2$ of sky, while north and east displacements increases solid angle blocked by the terrain by $5\,\text{deg}^2$ and $8\,\text{deg}^2$, respectively.}
    \label{fig:horizon}
\end{figure}

By rotating the antenna platform, EIGSEP can change the projection of the antenna beam pattern on the sky and terrain during observations.
The canyon in which the antenna is suspended has a horizon profile that varies significantly with azimuth angle, as shown in Fig.~\ref{fig:horizon}(a). The profile represents the integration domain $\Omega_{\text{below}}$ in equation~\eqref{eq:conv_split}, and is computed from a nominal antenna position suspended in the canyon.
As the antenna rotates, the weighting of the emission coming from the sky and from the ground changes, helping us to separate the effects of the antenna beam from the intrinsic sky emission.

The rotation follows a scan strategy in which a grid in azimuth and elevation is traced continuously, with one motor moving at a time.
In the set-up, the box completes one full revolution around one axis (the fast axis) for each motor step around the other axis (the slow axis).
The motors move slowly: completing a $360^{\circ}$ rotation around both axes takes about 1.5\,h in the nominal set-up.
Sky data are collected continuously, independent of the rotations.
We model the smearing of the beam caused by the rotations explicitly, but we expect this effect to be small given the wide beam pattern of the bowtie and the short integration time relative to the motor motion.

In our nominal scanning strategy, the fast axis completes a full sweep in a few minutes, much faster than the sidereal drift of the sky.
Therefore, the orientation of the beam $A$ in equation~\eqref{eq:conv_split} is changing, while $T_{\text{sky}}$ and $T_{\text{gnd}}$ remain effectively constant over a single sweep.
Along the slow axis the sidereal drift is significant and is included in the model.
In practice, we take measurements with both the elevation axis and the azimuth axis as the fast axis.
The rotating platform thus provides flexibility in how the beam and sky are weighted.
Each configuration yields independent information, which, when combined, helps us better separate the antenna beam from the intrinsic sky emission.

In Fig.~\ref{fig:beam_modulation}(b), the channels between the transmitted tones, which carry the sky continuum rather than the injected signal, modulate by a median of 0.8\,dB over a turn -- a peak-to-trough swing of about 20 per cent in power.
The continuum is shown only for those tones whose neighbouring channels were steady on the stationary ground copy over the same integrations, in order to isolate the beam modulation of the moving antenna from other variations, such as intermittent interference.
The sky modulation is far shallower than that of the injected tone because the sky is dominated by diffuse emission which the wide beam averages over, while the injected tone is a point source that directly probes the beam pattern.
The cut is not applied to the injected tones in Fig.~\ref{fig:beam_modulation}(a), because the decibel-level perturbations in the neighbouring channels are small compared to the 25\,dB modulation of the injected signal, except close to the nulls, where the depths are in any case reported as upper limits.

Furthermore, we aim to use beam modulation to measure and separate spatial foreground modes.
When the antenna rotates, it will be sensitive to different parts of the sky. Because the 21-cm monopole is constant across the sky, we can use this modulation to constrain foregrounds.
We note that some change in the spectrum is due to different weighting of the sky and the ground (equation~\eqref{eq:conv_split}), but this effect can be estimated with models and measurements of the antenna beam pattern $A$ and of the ground temperature $T_{\text{gnd}}$, which we assume to be close to a blackbody after filtering reflections.
This technique is similar to fitting spectra from different LST bins to improve foreground modelling and subtraction \citep{2013PhRvD..87d3002L, 2020ApJ...897..175T, 2023MNRAS.520..850A}, but with the added benefit that the direction and timing of the rotation are not fixed to the Earth rotation. For EIGSEP, we aim to use both to improve constraints.

\subsection{Forward modelling} \label{subsec:fwd_modelling}
A complementary approach to the calibration described above is to compare measured spectra directly against forward models of the antenna temperature.
With an increasing focus on accurate modelling of diffuse synchrotron and point-source emission at low frequencies \citep{2008MNRAS.388..247D, 2017AJ....153...26S, 2017MNRAS.464.3486Z}, comparing data to direct sky simulations is becoming a viable approach for validating calibrated data.
EIGSEP uses the CROISSANT simulation package\footnote{\url{https://github.com/christianhbye/croissant}} to generate synthetic datasets given models of the foregrounds, antenna beam, horizon profile, and antenna position and orientation.
Written in JAX \citep{jax2018github}, the EIGSEP forward models are automatically differentiable, thereby providing both simulated antenna temperature and its gradients, allowing us to efficiently explore the dependence of the antenna temperature on the input astrophysical and instrumental parameters.
While current models do not have the accuracy to calibrate the instrument alone, they none the less present an additional cross-check on calibration that can be used for verification.
In particular, given a good enough beam model from measurements (section~\ref{subsec:beam_mapping}), it is possible to use EIGSEP data to solve for the sky emission and compare with sky models. 

\begin{figure*}
    \centering
    \includegraphics[width=\linewidth]{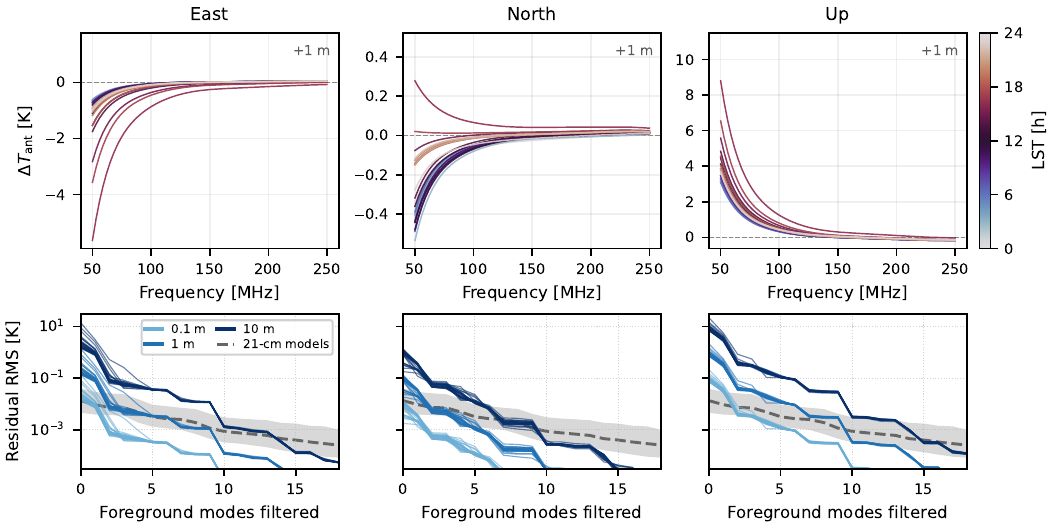}
    \caption{Change in the simulated antenna temperature, $\Delta T_{\text{ant}}$, when the suspended antenna is displaced to the east (left column), north (middle column), and up (right column). The displacement shifts the local horizon, shown in Fig.~\ref{fig:horizon}(b), thereby changing the fraction of sky occulted by the canyon walls and hence $T_{\text{ant}}$.
    \textit{Top row:} $\Delta T_{\text{ant}}$ for a $+1$\,m displacement. Each curve corresponds to a different LST (one per sidereal hour), coloured as indicated by the colour bar. \textit{Bottom row:} the RMS over frequency of $\Delta T_{\text{ant}}$ after filtering the leading eigenmodes of the unperturbed antenna temperature -- the same modes as the central panel of Fig.~\ref{fig:singular_values} -- for displacements of 0.1, 1, and 10\,m, coloured light to dark with increasing displacement magnitude, with all 24 LSTs drawn for each. Grey shows the 21-cm models of Fig.~\ref{fig:singular_values} retained under the same filter (5--95 per cent of ensemble, median dashed). A 1\,m displacement changes $T_{\text{ant}}$ by up to 9.5\,K, but the change lies largely in the leading eigenmodes: by $N=10$, the worst LST of a 1\,m shift has fallen to 0.40 of the median retained 21-cm signal for the vertical displacement, 0.16 for the eastward, and 0.044 for the northward. The vertical shift leaves the largest residual at every magnitude and every LST drawn: for the 1\,m shift its worst LST is 2.6 times the eastward and 9.1 times the northward.}
    \label{fig:horizon_shift}
\end{figure*}

As an example, we used CROISSANT to evaluate how sensitive the antenna temperature is to movements of the antenna position. Due to the non-trivial horizon profile in the canyon (Fig.~\ref{fig:horizon}(a)), changes in the antenna position will change what part of the sky is blocked by the canyon wall.
Fig.~\ref{fig:horizon}(b) shows how the horizon angle changes relative to the nominal profile when the antenna is displaced by 1\,m.
To quantify the effect of the horizon coupling, we computed the change in antenna temperature, $\Delta T_{\text{ant}}(\nu)$, caused by displacing the antenna to the east, north, and up, relative to the nominal position used for the horizon profile of Fig.~\ref{fig:horizon}. Each axis was displaced by 0.1, 1, and 10\,m, to assess the dependence of the antenna temperature on the magnitude of the displacement. 
The simulation uses the same models of the sky, ground, and beam as the bowtie simulations of section~\ref{subsec:covariance}. The results are shown as a function of LST and frequency in Fig.~\ref{fig:horizon_shift}.

A 1\,m displacement changes the antenna temperature by up to 6.3\,K to the east, 0.6\,K to the north, and 9.5\,K upward, in each case peaking at the bottom of the band. These maxima are over all simulated LSTs, not only the 24 drawn in Fig.~\ref{fig:horizon_shift}.
When filtered with the eigenmodes of the antenna temperature of the unperturbed antenna -- i.e. the same as in Fig.~\ref{fig:singular_values} -- the residual at $N=10$ rises from 0.62\,mK to 0.70\,mK for a vertical shift and to 0.63\,mK for an eastward one, and is unchanged at 0.62\,mK for a northward one, thus staying below the median retained 21-cm signal of 0.87\,mK.
The residual floor here is the spectral structure a displacement leaves in a perfectly calibrated receiver (section~\ref{subsec:covariance}).
Note that this increased residual is purely due to ignorance of the position, not the absolute position in the canyon.
If the eigenmode basis is built using the correct antenna position, the residual floor is comparable to that of the nominal position: even a 10\,m displacement leaves it between 0.53 and 0.73\,mK, against 0.62\,mK unperturbed. The only absolute position requirement is the $\gtrsim100\,\text{m}$ isolation from the canyon.

The vertical direction differs in kind from the other two, because a vertical upward shift lowers the local horizon at every azimuth at once. The effect of a vertical displacement on the occulted sky fraction therefore accumulates instead of cancelling (Fig.~\ref{fig:horizon}), and the residual it leaves grows roughly in proportion to the displacement. A horizontal shift moves the antenna towards some canyon walls and away from others, so the horizon rises at some azimuths and falls at others and the two largely cancel. The cancellation is less complete to the east. Because the canyon runs north--south (section~\ref{subsec:geology}), a northward shift moves the antenna along its axis and barely changes its distance to either wall, whereas an eastward shift moves it towards one wall and away from the other.

The cost of a position error is thus strongly direction-dependent, and grows quickly beyond the metre scale. At 10\,m the induced floor is 3.3\,mK vertically, 3.8 times the retained signal, against 1.4\,mK eastward and 0.68\,mK northward. Since the antenna is suspended with a highline going east/west and with tethers to the ground, its motion is the most constrained in the east/west and up/down directions, where an unmodelled error is most problematic for signal recovery.
For the more likely northward displacements, the unmodelled error has the smallest effect on the residual floor.

Fig.~\ref{fig:horizon_shift} shows the structure left behind by a displacement the model does not know about. The ignorance of the antenna position will be marginalized over in the analysis pipeline, and the quoted sensitivities set how wide the prior may be: wide horizontally at almost no cost, and, vertically, wide enough to cover the metre-scale motion the suspension permits (section~\ref{subsec:box}), with the induced floor overtaking the retained signal only between 1 and 10\,m. The lidar monitoring of the antenna position and the IMU measurements of its orientation will be folded into the forward model used for the analysis pipeline.

\section{Deployments}
\label{sec:deployments}
Early versions of the EIGSEP instrument were deployed in the field in 2023 and 2024, primarily to characterize the instrument and the site while the instrument was under development. The version of the EIGSEP instrument described in this work was deployed in July 2025 and July 2026.
The deployment site is Marjum Pass, Utah (39.2477$^\circ$ N, 113.4033$^\circ$ W) in the western United States.
The site is remote, with steep topographical relief far from radio transmitters. It was selected from among a dozen candidate sites on the basis of broadband surveys showing very low levels of RFI. The deployments to date have established the site as a base for development and testing of the instrument, and the in situ measurement of section~\ref{subsec:eig_antenna} shows the canyon providing the electromagnetic isolation from the terrain that the concept requires.
This section gives an overview of the site and the deployments to date.

\subsection{Site selection}
\begin{figure}
    \centering
    \includegraphics[width=\linewidth]{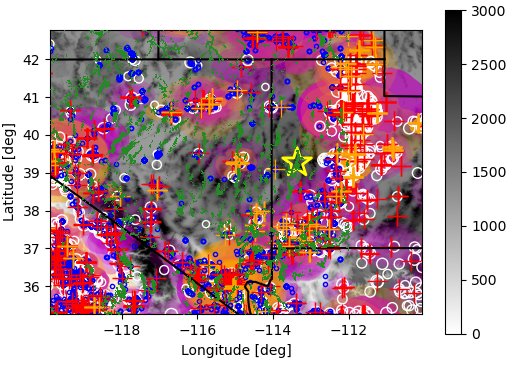}
    \caption{Digital elevation map (greyscale, m) of the western US with areas on public land with topographic features $>100\,\text{m}$ on either side in green, and the deepest $>95$ percentile in pink. FCC-registered FM and TV stations are shown in magenta and orange, respectively, with crosses at the transmitter and shading inside modelled 60\,dBu contours. Blue circles indicate registered, but possibly inactive, FM and TV antennas. White circles indicate city centres. The yellow star marks the selected EIGSEP site.
    The canyon at the EIGSEP site runs north--south, providing shielding from RFI from population centres to the east.}
    \label{fig:site_selection}
\end{figure}

A suitable site to deploy EIGSEP needs to have low levels of RFI and feature a canyon large enough to provide a 100\,m air gap for the suspended antenna.
It is also advantageous that the site is easily accessible from Berkeley, California, while the instrument is still under active development at the University of California Berkeley. The search for a site was therefore restricted to the western United States.
To find candidate sites, we combined maps of radio transmitters from the Federal Communications Commission\footnote{\url{https://www.fcc.gov/media/radio/fm-query} for FM radio transmitters and \url{https://www.fcc.gov/media/television/tv-query} for digital TV transmitters.} with digital elevation maps from the US Geological Survey\footnote{\url{https://www.usgs.gov/3d-elevation-program}}.
The combined map is presented in Fig.~\ref{fig:site_selection}. Using this map, the site was selected from the areas outside the 60\,dBu contours of the registered transmitters that also featured canyons more than 100\,m deep.

With a dozen candidate sites selected, we conducted RFI surveys and a terrain inspection in 2022.
The RFI level was measured with a discone antenna and a Keysight FieldFox handheld RF analyzer\footnote{\url{https://www.keysight.com/us/en/products/network-analyzers/fieldfox-handheld-rf-microwave-analyzers.html}}.
From these field measurements, Marjum Pass was selected as a promising candidate site. It also featured several possible suspension points for the antenna.
Additional measurements at Marjum Pass in 2023, with a prototype of the EIGSEP instrument, confirmed that the site provides a good compromise between RFI, terrain, and accessibility for EIGSEP.

\subsection{RFI}
\begin{figure}
    \centering
    \includegraphics[width=\linewidth]{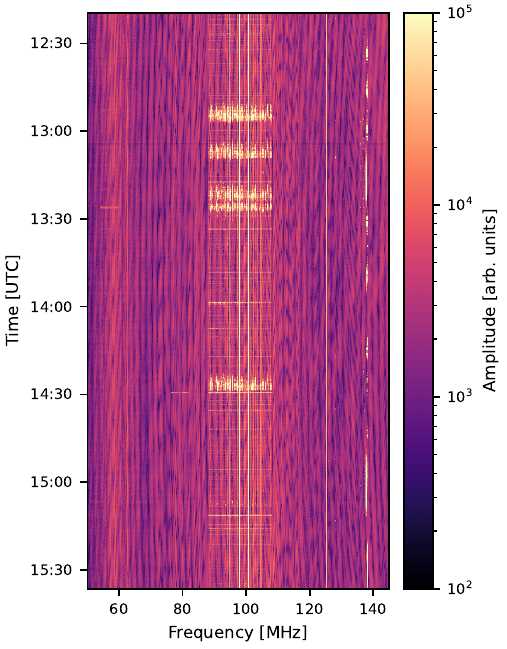}
    \caption{
    The measurements represent raw cross-correlation data from the correlator in logarithmic units taken on 2024 July 21.
    The integration time of each spectrum is 1.07\,s.
    When the sky is free of reflectors, the spectra show low levels of RFI, even in the FM band. With airplanes or micrometeors overhead, significant RFI is present. Longer periods of RFI in the FM band (e.g. the feature right before 13:30) are associated with aircraft, whereas short blips (e.g. at 14:00) are associated with micrometeors. The transmission at 137\,MHz is from Orbcomm satellites.}
    \label{fig:rfi}
\end{figure}

Field measurements at Marjum Pass reveal reflected FM radio as the main source of RFI, primarily reflected off airplanes and, to a lesser extent, off meteor trails.
Fig.~\ref{fig:rfi} shows cross-correlation measurements from the two ground antennas at the site as a function of time and frequency.
RFI is almost always present in five frequency channels in the FM band (88--108\,MHz).

When an airplane is above the antenna, it reflects RFI from nearby transmitters, causing interference across the FM radio band.
The signature of this airplane-reflected RFI is seen for example right before 13:30 in Fig.~\ref{fig:rfi}, where the power increases for about two minutes in the FM band.
A similar effect, but at shorter time-scales -- generally less than five seconds -- is seen, for example at 14:00. We believe this is FM transmission reflected from meteor trails of electrons in the ionosphere, an effect that has been observed for decades \citep[e.g.][]{1947MNRAS.107..176H, 1948Natur.161..596E} and more recently in \citet{2014RaSc...49..157H}.

\begin{figure}
    \centering
    \includegraphics[width=\linewidth]{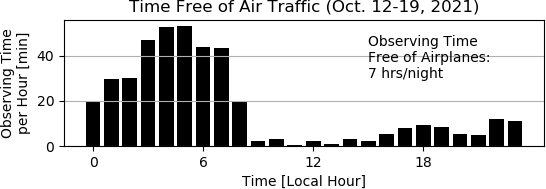}
    \caption{Observing time free of airplane traffic, computed from a horizon model and weekly flight information from \url{flightradar24.com}. This site conservatively yields seven hours of airplane-free observing per night.}
    \label{fig:airplanes}
\end{figure}

Because of low-duty-cycle RFI on site associated with aircraft, we used flight data from \url{flightradar24.com} in combination with our horizon model (Fig.~\ref{fig:horizon}) to model airplane visibility above the horizon seen from the antenna. The results, displayed in Fig.~\ref{fig:airplanes}, suggest that nighttime is relatively flight-free, yielding conservatively $\approx 7$\,hours of observing per night with no aircraft above the horizon.

In addition to RFI associated with FM radio, we see RFI from digital TV transmission at a lower level. This is not apparent in Fig.~\ref{fig:rfi}, but is seen in longer integrations at frequencies below FM. 

\subsection{Terrain and geology of site}
\label{subsec:geology}
The Dome Canyon runs north--south through the House Range of west-central Utah (Millard County). At the antenna site, the eastern rim stands about 200\,m above the canyon floor and the western rim about 380\,m. The bedrock consists predominantly of cliff-forming limestone with interbedded shale \citep{Hintze_1974}.
The sedimentary units in this sequence dip gently ($\approx15^\circ$) towards the south-east at the location of the canyon, but dip orientation and magnitude vary regionally.

We were unable to find any published studies of ground-penetrating radar surveys at our site that directly measure the dielectric properties of these specific units.
For limestone, the dielectric constant and conductivity generally lie within 7--9 \citep{Parasnis_1956} and $0.003\text{--}0.3\, \mathrm{mS\,m^{-1}}$ \citep{2004NSGeo...2..151F}, respectively, over 0.3--1000\,MHz.
For shale, the dielectric constant is typically 5--8 \citep{Josh_2016}.
We note that the electrical properties depend on impurities and water content \citep{2004NSGeo...2..151F}.
They are also frequency-dependent: for example, the conductivity of shale increases from $10\,\mathrm{mS\,m^{-1}}$ at 50\,MHz to $13\,\mathrm{mS\,m^{-1}}$ at 100\,MHz \citep{Josh_2016}.
The complex spatial and spectral dependence of the soil electrical parameters again illustrates the importance of separating the antenna from the ground for EIGSEP.
While there are observing sites that can be characterized with simpler soil models \citep{2015ITAP...63.5433S, 2021AJ....162...38M, 2022MNRAS.515.1580S, 2025GL116618},
residual soil-dependent structure in the antenna beam must be carefully understood and calibrated.

\subsection{Suspensions}
The antenna is suspended using a highline between the eastern and western canyon rims. The highline is anchored to trees on each side.
The choice of anchoring points defines the position of the elevated antenna.
Using digital elevation maps of the site, the anchor points were decided based on accessibility and the horizon profile seen by the antenna. The horizon profile changes substantially between anchor points.
As the canyon walls provide significant shielding from RFI, and the dominant source of RFI is reflections off airplanes, it is preferable for EIGSEP to have as small a field of view to the sky as possible despite the cost of a higher fraction of ground radiation in the total antenna temperature.
Additionally, we favour anchor points well above the target antenna suspension height, keeping the suspension angle steep and reducing the rope tension required to hold the antenna.

The final anchor points were chosen based on the horizon profiles, tension requirements, and accessibility.
The highline is set up and anchored by a team of professional climbers from Sea to Sky Explorers Association\footnote{\url{https://www.stsea.org/}}.
After initial set-up, the antenna is raised from the ground using the pulley system.
This system gives flexibility to change batteries or make necessary fixes to the instrument without climbing to the top of the canyon. It also allows for changing the height during observation to assess the effect of a changing horizon profile or evaluate terrain coupling.

\subsection{Deployment history} \label{subsec:field_meas}
Three versions of the EIGSEP instrument have been suspended. In July 2024, a prototype platform was suspended about 20\,m above the ground in an adjacent canyon at the pass. This test resulted in the first measurements taken by EIGSEP from above the ground, and the first field operation of a rotating antenna.
The suspension with the highline and pulley system was first used in October 2024, using a Vivaldi feed of the same kind as the ground antennas. This feed was suspended about 95\,m above the ground.

The new bowtie antenna and antenna platform were suspended in July 2025 for four days, reaching about 100\,m above the ground. During this period, the antenna was lowered and raised several times.
The antenna was rotated continuously for approximately five hours each day, and the external calibration signal was continuously transmitted from the short dipole antenna. Reflection coefficient measurements and receiver calibration measurements were also taken for the first time, exercising these subsystems end to end under deployment conditions. The reflection coefficient measured during this deployment is presented in Fig.~\ref{fig:bowtie_s11}. 

Each deployment through July 2025 suspended a different antenna, and each involved substantial changes to the hardware. These campaigns provided valuable testing and commissioning of the instrument, including the first field operation of the rotating platform, the transmitter, and the calibration hardware.
Instrumental effects that we have not fully explained remain in the sky data from these campaigns. Self-generated interference from the instrument is one contributor, but it is not separated from the others in these data, and we do not attempt to apportion them here.

The instrument was redeployed in July 2026 with the same bowtie antenna, at the same suspension point and a comparable height, incorporating fixes and lessons from the earlier campaigns. The receiver electronics were upgraded, the transmitter was replaced with a dual-polarization design (section~\ref{subsec:beam_mapping}), and the antenna position and platform orientation were monitored with a lidar and a potentiometer on the azimuth axis.
On the ground, the Vivaldi feeds were replaced by a single copy of the suspended system (section~\ref{sec:vivaldis}).
Data from this campaign are in active analysis.

\subsection{Terrain scattering from a ground antenna}
\label{subsec:virgo}
\begin{figure}
    \centering
    \includegraphics[width=\linewidth]{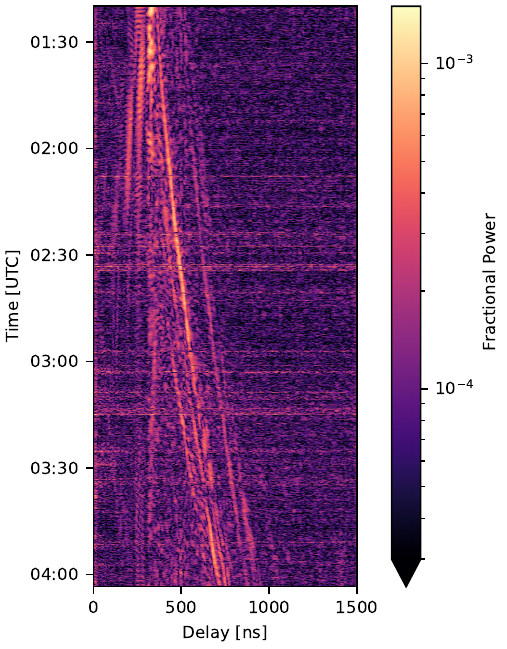}
   \caption{
    Delay spectrum of DPSS-filtered autocorrelation data from a ground-based Vivaldi feed. The data were taken on 2024 October 19. 
    The spectrum is normalized by the total power received at each time.
    The diagonal lines reveal emission from Virgo A, scattering off distinct features in the terrain.
    Note the logarithmic colour scale, clipped at $3\times10^{-5}$ to suppress noise.
    }
    \label{fig:point_source}
\end{figure}

One early result from the October 2024 deployment illustrates the terrain-scattering environment at the site.
It uses delay-transformed autocorrelation data from one of the ground-based Vivaldi feeds, and reveals evidence of multipath scattering from Virgo A.
To isolate the point source, we flag RFI in the raw autocorrelation data and use a filter based on discrete prolate spheroidal sequences \citep[DPSS;][]{1978ATTTJ..57.1371S} to remove spectrally smooth foregrounds.
As noted by \citet{2021MNRAS.500.5195E}, these sequences are optimal for suppressing power within a finite band in Fourier space.
Fig.~\ref{fig:point_source} displays the delay spectrum of the residuals after DPSS-filtering the data, normalized by the total power measured at each time. 
Diagonal tracks appear in these residuals, corresponding to the time-varying geometric delay of a source as it transits.
We attribute them to Virgo A, the only bright point source above the canyon walls throughout the observation, and the multiple tracks to reflections from different parts of the canyon.
The delays are consistent with the known position of Virgo A during the observation, assuming that the dominant scatterer is the eastern canyon wall.

\subsection{Ongoing and future work} \label{subsec:ongoing}
Work is in progress to identify and characterize the instrumental systematics present in the data from the 2025 and 2026 campaigns.
In the laboratory, we are measuring the temporal stability of the receiver and its sensitivity to thermal variations. These tests will inform the strategy for receiver calibration in the field.
A parallel effort addresses the self-generated interference identified in the commissioning data.
We locate the sources of emission, in the laboratory and in the field, by isolating suspected components -- e.g. motors, switching power supplies, unshielded wires -- and measuring their contribution to the system noise. We suppress them with nested RF-tight enclosures, with filtering and shielding of cables, and in some cases by removing the components identified as strong emitters.
These measures were applied to the instrument deployed in July 2026, and their effect on the sky data is being assessed as part of the analysis of that campaign.

We are also implementing the absolute receiver calibration described in section~\ref{subsec:rec_cal} and analysing the beam mapping data described in section~\ref{subsec:beam_mapping}.
Together with the systematics work above, these are the steps towards the absolutely calibrated sky spectra and the measured beam pattern that we defer to forthcoming work.

\section{Conclusions} \label{sec:conclusion}
We have presented the EIGSEP experiment to measure the global 21-cm signal from Cosmic Dawn. EIGSEP uses one suspended antenna and one or two auxiliary antennas on the ground, observing over 50--250\,MHz.
The key design consideration for EIGSEP is minimizing instrumental systematics with spectral structure that, when coupled to bright foregrounds, may mimic the cosmological signal.
To minimize the problematic systematics, EIGSEP employs a rotating bowtie antenna suspended in a canyon, with an $\sim100\,\text{m}$ air bubble surrounding the antenna.
We presented the bowtie antenna design and our approach for characterizing the antenna beam with a ground-based transmitter.
We then presented our plans for calibration of the instrument, including field measurements of internal calibrators, external signal injection, reflection coefficient measurements, interferometric measurements, and lab characterization.


The EIGSEP experiment has been deployed at Marjum Pass five times, most recently in July 2026 with the custom bowtie antenna on a rotating antenna platform.
The site exhibits a clean radio frequency environment, with transient reflections of transmitters off airplanes representing the primary source of interference.
Initial analysis of data from a ground-based Vivaldi feed shows terrain scattering of emission from Virgo A.
The in situ measurement of the antenna reflection coefficient agrees with the free-space simulation to a median of 0.5\,dB below 100\,MHz, with the residual departure at higher frequencies likely due to the idealized balun in the simulated feed.
In delay space, the measurement confines terrain scattering to delays beyond 590\,ns, the round-trip light travel time to the nearest canyon terrain, with no detectable scattering at shorter delays, consistent with the delay gap simulated for this suspension height.
Measurements of an injected calibration signal through a full rotation of the platform show the antenna response modulating by up to 33\,dB, while the ground copy of the antenna and receiver varies by 0.03\,dB.


Future work is focused on data analysis, laboratory characterization of the receiver, and mitigation of self-generated interference. The results of these efforts will be used to identify and remove instrumental systematics.
In parallel, analysis of transmitter data is underway to characterize the antenna beam and compare it with numerical simulations.
With the better characterized instrument, we plan extended sky measurements.
Absolutely calibrated sky spectra and the measured antenna beam are the subject of work in preparation.

\section*{Acknowledgements}
We thank Khaled Allen, Sam Cipriani, Chris Gill, Yongsheng Li, Sam Ludwig, Ashlie Reott, Elon Smith, Grant Syllaba, and Sean Taylor from Sea to Sky Explorers Association for help with suspensions of the instrument. We thank Ken Parsons for help with site selection and contributions during deployments. We thank Siddharth Goswami and Judah, Elena, and Greta Spiegel for help with deployments.
We thank Kiran Shila and the DSA collaboration for sharing the design of the RFoF board.
This research is funded by the Gordon and Betty Moore Foundation, Grant GBMF12211.

\section*{Data Availability}

Data, simulation results, and code used to generate the plots are publicly available on Zenodo at \url{https://doi.org/10.5281/zenodo.18383267}. The code is also available on GitHub at \url{https://github.com/christianhbye/eigsep_paper_notebooks}.
The design of the RFoF board is available on GitHub at \url{https://github.com/liuweiseu/rfof-ftx/tree/cefad78}.

\section*{Conflict of Interest}
Authors declare no conflict of interest.



\bibliographystyle{rasti}
\bibliography{export-bibtex, geo, site}

@ARTICLE{2023MNRAS.523.2587M,
       author = {{Mu{\~n}oz}, Julian B.},
        title = "{An effective model for the cosmic-dawn 21-cm signal}",
      journal = {\mnras},
         year = 2023,
        month = aug,
       volume = {523},
       number = {2},
        pages = {2587-2607},
          doi = {10.1093/mnras/stad1512},
archivePrefix = {arXiv},
       eprint = {2302.08506},
 primaryClass = {astro-ph.CO},
       adsurl = {https://ui.adsabs.harvard.edu/abs/2023MNRAS.523.2587M}
}

@ARTICLE{2025PhRvD.111h3503C,
       author = {{Cruz}, Hector Afonso G. and {Mu{\~n}oz}, Julian B. and {Sabti}, Nashwan and {Kamionkowski}, Marc},
        title = "{Effective model for the 21-cm signal with population III stars}",
      journal = {\prd},
         year = 2025,
        month = apr,
       volume = {111},
       number = {8},
          eid = {083503},
        pages = {083503},
          doi = {10.1103/PhysRevD.111.083503},
archivePrefix = {arXiv},
       eprint = {2407.18294},
 primaryClass = {astro-ph.CO},
       adsurl = {https://ui.adsabs.harvard.edu/abs/2025PhRvD.111h3503C}
}

@ARTICLE{2025PASP..137l5002C,
       author = {{Cappallo}, Rigel C. and {Rogers}, Alan E.~E. and {Lonsdale}, Colin J. and {Bowman}, Judd D. and {Barrett}, John P. and {Murray}, Steven G. and {Mahesh}, Nivedita and {Sims}, Peter and {Vydula}, Akshatha K. and {Monsalve}, Raul A. and {Eckert}, Christopher J. and {Steen}, Parker and {Wilson}, Kenneth M.},
        title = "{EDGES-3: Instrument Design and Commissioning}",
      journal = {\pasp},
         year = 2025,
        month = dec,
       volume = {137},
       number = {12},
          eid = {125002},
        pages = {125002},
          doi = {10.1088/1538-3873/ae1bd6},
archivePrefix = {arXiv},
       eprint = {2508.02577},
 primaryClass = {astro-ph.IM},
       adsurl = {https://ui.adsabs.harvard.edu/abs/2025PASP..137l5002C}
}

@ARTICLE{2025ApJ...992...63W,
       author = {{Whitler}, Lily and {Stark}, Daniel P. and {Topping}, Michael W. and {Robertson}, Brant and {Rieke}, Marcia and {Hainline}, Kevin N. and {Endsley}, Ryan and {Chen}, Zuyi and {Baker}, William M. and {Bhatawdekar}, Rachana and {Bunker}, Andrew J. and {Carniani}, Stefano and {Charlot}, St{\'e}phane and {Chevallard}, Jacopo and {Curtis-Lake}, Emma and {Egami}, Eiichi and {Eisenstein}, Daniel J. and {Helton}, Jakob M. and {Ji}, Zhiyuan and {Johnson}, Benjamin D. and {P{\'e}rez-Gonz{\'a}lez}, Pablo G. and {Rinaldi}, Pierluigi and {Tacchella}, Sandro and {Williams}, Christina C. and {Willmer}, Christopher N.~A. and {Willott}, Chris and {Witstok}, Joris},
        title = "{The z {\ensuremath{\gtrsim}} 9 Galaxy UV Luminosity Function from the JWST Advanced Deep Extragalactic Survey: Insights into Early Galaxy Evolution and Reionization}",
      journal = {\apj},
         year = 2025,
        month = oct,
       volume = {992},
       number = {1},
          eid = {63},
        pages = {63},
          doi = {10.3847/1538-4357/adfddc},
archivePrefix = {arXiv},
       eprint = {2501.00984},
 primaryClass = {astro-ph.GA},
       adsurl = {https://ui.adsabs.harvard.edu/abs/2025ApJ...992...63W}
}

@ARTICLE{2025MNRAS.542.1952M,
       author = {{Meyer}, Romain A. and {Roberts-Borsani}, Guido and {Oesch}, Pascal A. and {Ellis}, Richard S.},
        title = "{Probing patchy reionization with JWST: IGM opacity constraints from the Lyman {\ensuremath{\alpha}} forest of galaxies in legacy extragalactic fields}",
      journal = {\mnras},
         year = 2025,
        month = sep,
       volume = {542},
       number = {3},
        pages = {1952-1968},
          doi = {10.1093/mnras/staf1312},
archivePrefix = {arXiv},
       eprint = {2504.02683},
 primaryClass = {astro-ph.GA},
       adsurl = {https://ui.adsabs.harvard.edu/abs/2025MNRAS.542.1952M}
}

@ARTICLE{2025arXiv250911846M,
       author = {{McKay}, Luke and {Subrahmanyan}, Ravi and {Chippendale}, Aaron P. and {Bolli}, Pietro and {Kyriakou}, Georgios and {Dunning}, Alex and {Ekers}, Ronald},
        title = "{Precise measurement of the absolute sky brightness at 60--350 MHz}",
      journal = {Nature Astronomy},
         year = 2026,
        month = jun,
          doi = {10.1038/s41550-026-02888-5},
archivePrefix = {arXiv},
       eprint = {2509.11846},
 primaryClass = {astro-ph.CO},
       adsurl = {https://ui.adsabs.harvard.edu/abs/2025arXiv250911846M}
}

@ARTICLE{2025PASP..137h5002A,
       author = {{Altamirano}, Cinthia and {Bustos}, Ricardo and {Monsalve}, Raul A. and {Restrepo}, Silvia E. and {Bidula}, Vadym and {Bye}, Christian H. and {Chiang}, H. Cynthia and {Guo}, Xinze and {Hendricksen}, Ian and {McGee}, Francis and {Mena}, F. Patricio and {Nasu-Yu}, Lisa and {Sievers}, Jonathan L. and {Thyagarajan}, Nithyanandan},
        title = "{Using the Antenna Impedance to Estimate Soil Electrical Parameters for the MIST Global 21 cm Experiment}",
      journal = {\pasp},
         year = 2025,
        month = aug,
       volume = {137},
       number = {8},
          eid = {085002},
        pages = {085002},
          doi = {10.1088/1538-3873/adfa43},
archivePrefix = {arXiv},
       eprint = {2503.19401},
 primaryClass = {astro-ph.IM},
       adsurl = {https://ui.adsabs.harvard.edu/abs/2025PASP..137h5002A}
}

@ARTICLE{2025A&A...699A.109G,
       author = {{Ghara}, R. and {Zaroubi}, S. and {Ciardi}, B. and {Mellema}, G. and {Giri}, S.~K. and {Mertens}, F.~G. and {Mevius}, M. and {Koopmans}, L.~V.~E. and {Iliev}, I.~T. and {Acharya}, A. and {Brackenhoff}, S.~A. and {Ceccotti}, E. and {Chege}, K. and {Georgiev}, I. and {Ghosh}, S. and {Hothi}, I. and {H{\"o}fer}, C. and {Ma}, Q. and {Munshi}, S. and {Offringa}, A.~R. and {Shaw}, A.~K. and {Pandey}, V.~N. and {Yatawatta}, S. and {Choudhury}, M.},
        title = "{Constraints on the state of the intergalactic medium at z{\ensuremath{\sim}}8 ‑ 10 using redshifted 21 cm observations with LOFAR}",
      journal = {\aap},
         year = 2025,
        month = jul,
       volume = {699},
          eid = {A109},
        pages = {A109},
          doi = {10.1051/0004-6361/202554163},
archivePrefix = {arXiv},
       eprint = {2505.00373},
 primaryClass = {astro-ph.CO},
       adsurl = {https://ui.adsabs.harvard.edu/abs/2025A&A...699A.109G}
}

@ARTICLE{2025ApJS..277...37U,
       author = {{Umeda}, Hiroya and {Ouchi}, Masami and {Kikuta}, Satoshi and {Harikane}, Yuichi and {Ono}, Yoshiaki and {Shibuya}, Takatoshi and {Inoue}, Akio K. and {Shimasaku}, Kazuhiro and {Liang}, Yongming and {Matsumoto}, Akinori and {Saito}, Shun and {Kusakabe}, Haruka and {Kageura}, Yuta and {Nakane}, Minami},
        title = "{SILVERRUSH. XIV. Ly{\ensuremath{\alpha}} Luminosity Functions and Angular Correlation Functions from 20,000 Ly{\ensuremath{\alpha}} Emitters at z {\ensuremath{\sim}} 2.2{\textendash}7.3 from up to 24 deg$^{2}$ HSC-SSP and CHORUS Surveys: Linking the Postreionization Epoch to the Heart of Reionization}",
      journal = {\apjs},
         year = 2025,
        month = apr,
       volume = {277},
       number = {2},
          eid = {37},
        pages = {37},
          doi = {10.3847/1538-4365/adb1c0},
archivePrefix = {arXiv},
       eprint = {2411.15495},
 primaryClass = {astro-ph.GA},
       adsurl = {https://ui.adsabs.harvard.edu/abs/2025ApJS..277...37U}
}

@article{2025GL116618,
author = {Hendricksen, I. and Monsalve, R. A. and Bidula, V. and Altamirano, C. and Bustos, R. and Bye, C. H. and Chiang, H. C. and Guo, X. and McGee, F. and Mena, F. P. and Nasu-Yu, L. and Omelon, C. and Restrepo, S. E. and Sievers, J. L. and Thomson, L. and Thyagarajan, N.},
title = {Estimating Soil Electrical Parameters in the {Canadian} {High} {Arctic} From Impedance Measurements of the {MIST} Antenna Above the Surface},
journal = {Geophysical Research Letters},
year = {2026},
volume = {53},
number = {2},
pages = {e2025GL116618},
doi = {https://doi.org/10.1029/2025GL116618},
url = {https://agupubs.onlinelibrary.wiley.com/doi/abs/10.1029/2025GL116618},
eprint = {https://agupubs.onlinelibrary.wiley.com/doi/pdf/10.1029/2025GL116618},
note = {e2025GL116618 2025GL116618},
}

@ARTICLE{2025Natur.639..897W,
       author = {{Witstok}, Joris and {Jakobsen}, Peter and {Maiolino}, Roberto and {Helton}, Jakob M. and {Johnson}, Benjamin D. and {Robertson}, Brant E. and {Tacchella}, Sandro and {Cameron}, Alex J. and {Smit}, Renske and {Bunker}, Andrew J. and {Saxena}, Aayush and {Sun}, Fengwu and {Alberts}, Stacey and {Arribas}, Santiago and {Baker}, William M. and {Bhatawdekar}, Rachana and {Boyett}, Kristan and {Cargile}, Phillip A. and {Carniani}, Stefano and {Charlot}, St{\'e}phane and {Chevallard}, Jacopo and {Curti}, Mirko and {Curtis-Lake}, Emma and {D'Eugenio}, Francesco and {Eisenstein}, Daniel J. and {Hainline}, Kevin N. and {Jones}, Gareth C. and {Kumari}, Nimisha and {Maseda}, Michael V. and {P{\'e}rez-Gonz{\'a}lez}, Pablo G. and {Rinaldi}, Pierluigi and {Scholtz}, Jan and {{\"U}bler}, Hannah and {Williams}, Christina C. and {Willmer}, Christopher N.~A. and {Willott}, Chris and {Zhu}, Yongda},
        title = "{Witnessing the onset of reionization through Lyman-{\ensuremath{\alpha}} emission at redshift 13}",
      journal = {\nat},
         year = 2025,
        month = mar,
       volume = {639},
       number = {8056},
        pages = {897-901},
          doi = {10.1038/s41586-025-08779-5},
archivePrefix = {arXiv},
       eprint = {2408.16608},
 primaryClass = {astro-ph.GA},
       adsurl = {https://ui.adsabs.harvard.edu/abs/2025Natur.639..897W}
}

@ARTICLE{2025RASTI...4...61A,
       author = {{Artuc}, Kaan and {de Lera Acedo}, Eloy},
        title = "{The spectrometer development of CosmoCube, lunar orbiting satellite to detect 21-cm hydrogen signal from cosmic dark ages}",
      journal = {RAS Techniques and Instruments},
         year = 2025,
        month = jan,
       volume = {4},
          eid = {rzae061},
        pages = {rzae061},
          doi = {10.1093/rasti/rzae061},
archivePrefix = {arXiv},
       eprint = {2406.10096},
 primaryClass = {astro-ph.IM},
       adsurl = {https://ui.adsabs.harvard.edu/abs/2025RASTI...4...61A}
}

@ARTICLE{2025RASTI...4af046B,
       author = {{Bull}, Philip and {El-Makadema}, Ahmed and {Garsden}, Hugh and {Edgley}, John and {Roddis}, Neil and {Chluba}, Jens and {Conselice}, Christopher J. and {Dutta}, Sohini and {Glasscock}, Katrine A. and {Nasirudin}, Ainulnabilah and {Norris}, Jordan and {Wilensky}, Michael J. and {Ye}, Isabelle and {Zhang}, Zheng},
        title = "{RHINO: a large horn antenna for detecting the 21 cm global signal}",
      journal = {RAS Techniques and Instruments},
         year = 2025,
        month = jan,
       volume = {4},
          eid = {rzaf046},
        pages = {rzaf046},
          doi = {10.1093/rasti/rzaf046},
archivePrefix = {arXiv},
       eprint = {2410.00076},
 primaryClass = {astro-ph.IM},
       adsurl = {https://ui.adsabs.harvard.edu/abs/2025RASTI...4af046B}
}

@ARTICLE{2024ApJ...975..208T,
       author = {{Tang}, Mengtao and {Stark}, Daniel P. and {Topping}, Michael W. and {Mason}, Charlotte and {Ellis}, Richard S.},
        title = "{JWST/NIRSpec Observations of Lyman {\ensuremath{\alpha}} Emission in Star-forming Galaxies at 6.5 {\ensuremath{\lesssim}} z {\ensuremath{\lesssim}} 13}",
      journal = {\apj},
         year = 2024,
        month = nov,
       volume = {975},
       number = {2},
          eid = {208},
        pages = {208},
          doi = {10.3847/1538-4357/ad7eb7},
archivePrefix = {arXiv},
       eprint = {2408.01507},
 primaryClass = {astro-ph.GA},
       adsurl = {https://ui.adsabs.harvard.edu/abs/2024ApJ...975..208T}
}

@ARTICLE{2024ApJ...974..137A,
       author = {{Agrawal}, Yash and {Kavitha}, K. and {Singh}, Saurabh},
        title = "{Direction-dependent Effects on Global 21 cm Detection}",
      journal = {\apj},
         year = 2024,
        month = oct,
       volume = {974},
       number = {1},
          eid = {137},
        pages = {137},
          doi = {10.3847/1538-4357/ad6d68},
archivePrefix = {arXiv},
       eprint = {2401.10756},
 primaryClass = {astro-ph.CO},
       adsurl = {https://ui.adsabs.harvard.edu/abs/2024ApJ...974..137A}
}

@ARTICLE{2024MNRAS.531.4734C,
       author = {{Cumner}, John and {Pieterse}, Carla and {de Villiers}, Dirk and {de Lera Acedo}, Eloy},
        title = "{The effects of the antenna power pattern uncertainty within a global 21 cm experiment}",
      journal = {\mnras},
         year = 2024,
        month = jul,
       volume = {531},
       number = {4},
        pages = {4734-4745},
          doi = {10.1093/mnras/stae1475},
archivePrefix = {arXiv},
       eprint = {2311.07392},
 primaryClass = {astro-ph.IM},
       adsurl = {https://ui.adsabs.harvard.edu/abs/2024MNRAS.531.4734C}
}

@ARTICLE{2024MNRAS.530.4125M,
       author = {{Monsalve}, R.~A. and {Altamirano}, C. and {Bidula}, V. and {Bustos}, R. and {Bye}, C.~H. and {Chiang}, H.~C. and {D{\'\i}az}, M. and {Fern{\'a}ndez}, B. and {Guo}, X. and {Hendricksen}, I. and {Hornecker}, E. and {Lucero}, F. and {Mani}, H. and {McGee}, F. and {Mena}, F.~P. and {Pess{\^o}a}, M. and {Prabhakar}, G. and {Restrepo}, O. and {Sievers}, J.~L. and {Thyagarajan}, N.},
        title = "{Mapper of the IGM spin temperature: instrument overview}",
      journal = {\mnras},
         year = 2024,
        month = jun,
       volume = {530},
       number = {4},
        pages = {4125-4147},
          doi = {10.1093/mnras/stae1138},
archivePrefix = {arXiv},
       eprint = {2309.02996},
 primaryClass = {astro-ph.IM},
       adsurl = {https://ui.adsabs.harvard.edu/abs/2024MNRAS.530.4125M}
}

@ARTICLE{2020PASP..132f2001L,
       author = {{Liu}, Adrian and {Shaw}, J. Richard},
        title = "{Data Analysis for Precision 21 cm Cosmology}",
      journal = {\pasp},
         year = 2020,
        month = jun,
       volume = {132},
       number = {1012},
          eid = {062001},
        pages = {062001},
          doi = {10.1088/1538-3873/ab5bfd},
archivePrefix = {arXiv},
       eprint = {1907.08211},
 primaryClass = {astro-ph.IM},
       adsurl = {https://ui.adsabs.harvard.edu/abs/2020PASP..132f2001L}
}

@ARTICLE{2024PASP..136d5002B,
       author = {{Berkhout}, Lindsay M. and {Jacobs}, Daniel C. and {Abdurashidova}, Zuhra and {Adams}, Tyrone and {Aguirre}, James E. and {Alexander}, Paul and {Ali}, Zaki S. and {Baartman}, Rushelle and {Balfour}, Yanga and {Beardsley}, Adam P. and {Bernardi}, Gianni and {Billings}, Tashalee S. and {Bowman}, Judd D. and {Bradley}, Richard F. and {Bull}, Philip and {Burba}, Jacob and {Byrne}, Ruby and {Carey}, Steven and {Carilli}, Chris L. and {Chen}, Kai-Feng and {Cheng}, Carina and {Choudhuri}, Samir and {DeBoer}, David R. and {de Lera Acedo}, Eloy and {Dexter}, Matt and {Dillon}, Joshua S. and {Dynes}, Scott and {Eksteen}, Nico and {Ely}, John and {Ewall-Wice}, Aaron and {Fagnoni}, Nicolas and {Fritz}, Randall and {Furlanetto}, Steven R. and {Gale-Sides}, Kingsley and {Garsden}, Hugh and {Gehlot}, Bharat Kumar and {Ghosh}, Abhik and {Glendenning}, Brian and {Gorce}, Adelie and {Gorthi}, Deepthi and {Greig}, Bradley and {Grobbelaar}, Jasper and {Halday}, Ziyaad and {Hazelton}, Bryna J. and {Hewitt}, Jacqueline N. and {Hickish}, Jack and {Huang}, Tian and {Josaitis}, Alec and {Julius}, Austin and {Kariseb}, MacCalvin and {Kern}, Nicholas S. and {Kerrigan}, Joshua and {Kim}, Honggeun and {Kittiwisit}, Piyanat and {Kohn}, Saul A. and {Kolopanis}, Matthew and {Lanman}, Adam and {La Plante}, Paul and {Liu}, Adrian and {Loots}, Anita and {Ma}, Yin-Zhe and {Edward MacMahon}, David Harold and {Malan}, Lourence and {Malgas}, Cresshim and {Malgas}, Keith and {Marero}, Bradley and {Martinot}, Zachary E. and {Mesinger}, Andrei and {Molewa}, Mathakane and {Morales}, Miguel F. and {Mosiane}, Tshegofalang and {Murray}, Steven G. and {Neben}, Abraham R. and {Nikolic}, Bojan and {Nunhokee}, Chuneeta Devi and {Nuwegeld}, Hans and {Parsons}, Aaron R. and {Pascua}, Robert and {Patra}, Nipanjana and {Pieterse}, Samantha and {Qin}, Yuxiang and {Rath}, Eleanor and {Razavi-Ghods}, Nima and {Riley}, Daniel and {Robnett}, James and {Rosie}, Kathryn and {Santos}, Mario G. and {Sims}, Peter and {Singh}, Saurabh and {Storer}, Dara and {Swarts}, Hilton and {Tan}, Jianrong and {Thyagarajan}, Nithyanandan and {van Wyngaarden}, Pieter and {Williams}, Peter K.~G. and {Zheng}, Haoxuan and {Xu}, Zhilei},
        title = "{Hydrogen Epoch of Reionization Array (HERA) Phase II Deployment and Commissioning}",
      journal = {\pasp},
         year = 2024,
        month = apr,
       volume = {136},
       number = {4},
          eid = {045002},
        pages = {045002},
          doi = {10.1088/1538-3873/ad3122},
archivePrefix = {arXiv},
       eprint = {2401.04304},
 primaryClass = {astro-ph.IM},
       adsurl = {https://ui.adsabs.harvard.edu/abs/2024PASP..136d5002B}
}

@ARTICLE{2024ApJ...961...56M,
       author = {{Monsalve}, Raul A. and {Bye}, Christian H. and {Sievers}, Jonathan L. and {Bidula}, Vadym and {Bustos}, Ricardo and {Chiang}, H. Cynthia and {Guo}, Xinze and {Hendricksen}, Ian and {McGee}, Francis and {Mena}, F. Patricio and {Prabhakar}, Garima and {Restrepo}, Oscar and {Thyagarajan}, Nithyanandan},
        title = "{Simulating the Detection of the Global 21 cm Signal with MIST for Different Models of the Soil and Beam Directivity}",
      journal = {\apj},
         year = 2024,
        month = jan,
       volume = {961},
       number = {1},
          eid = {56},
        pages = {56},
          doi = {10.3847/1538-4357/ad0f1b},
archivePrefix = {arXiv},
       eprint = {2310.07741},
 primaryClass = {astro-ph.IM},
       adsurl = {https://ui.adsabs.harvard.edu/abs/2024ApJ...961...56M}
}

@ARTICLE{2024MNRAS.527.2413P,
       author = {{Pattison}, Joe H.~N. and {Anstey}, Dominic J. and {de Lera Acedo}, Eloy},
        title = "{Modelling a hot horizon in global 21-cm experimental foregrounds}",
      journal = {\mnras},
         year = 2024,
        month = jan,
       volume = {527},
       number = {2},
        pages = {2413-2425},
          doi = {10.1093/mnras/stad3378},
archivePrefix = {arXiv},
       eprint = {2307.02908},
 primaryClass = {astro-ph.CO},
       adsurl = {https://ui.adsabs.harvard.edu/abs/2024MNRAS.527.2413P}
}

@ARTICLE{2024A&A...681A..62M,
       author = {{Munshi}, S. and {Mertens}, F.~G. and {Koopmans}, L.~V.~E. and {Offringa}, A.~R. and {Semelin}, B. and {Aubert}, D. and {Barkana}, R. and {Bracco}, A. and {Brackenhoff}, S.~A. and {Cecconi}, B. and {Ceccotti}, E. and {Corbel}, S. and {Fialkov}, A. and {Gehlot}, B.~K. and {Ghara}, R. and {Girard}, J.~N. and {Grie{\ss}meier}, J.~M. and {H{\"o}fer}, C. and {Hothi}, I. and {M{\'e}riot}, R. and {Mevius}, M. and {Ocvirk}, P. and {Shaw}, A.~K. and {Theureau}, G. and {Yatawatta}, S. and {Zarka}, P. and {Zaroubi}, S.},
        title = "{First upper limits on the 21 cm signal power spectrum from cosmic dawn from one night of observations with NenuFAR}",
      journal = {\aap},
         year = 2024,
        month = jan,
       volume = {681},
          eid = {A62},
        pages = {A62},
          doi = {10.1051/0004-6361/202348329},
archivePrefix = {arXiv},
       eprint = {2311.05364},
 primaryClass = {astro-ph.CO},
       adsurl = {https://ui.adsabs.harvard.edu/abs/2024A&A...681A..62M}
}

@ARTICLE{2023ExA....56..741S,
       author = {{Sathyanarayana Rao}, Mayuri and {Singh}, Saurabh and {K.~S.}, Srivani and {B.~S.}, Girish and {Sathish}, Keerthipriya and {Somashekar}, R. and {Agaram}, Raghunathan and {Kavitha}, K. and {Vishwapriya}, Gautam and {Anand}, Ashish and {Udaya Shankar}, N. and {Seetha}, S.},
        title = "{PRATUSH experiment concept and design overview}",
      journal = {Experimental Astronomy},
         year = 2023,
        month = sep,
       volume = {56},
       number = {2-3},
        pages = {741-778},
          doi = {10.1007/s10686-023-09909-5},
archivePrefix = {arXiv},
       eprint = {2507.05654},
 primaryClass = {astro-ph.IM},
       adsurl = {https://ui.adsabs.harvard.edu/abs/2023ExA....56..741S}
}

@ARTICLE{2023MNRAS.522.1022S,
       author = {{Saxena}, Anchal and {Meerburg}, P. Daniel and {de Lera Acedo}, Eloy and {Handley}, Will and {Koopmans}, L{\'e}on V.~E.},
        title = "{Sky-averaged 21-cm signal extraction using multiple antennas with an SVD framework: the REACH case}",
      journal = {\mnras},
         year = 2023,
        month = jun,
       volume = {522},
       number = {1},
        pages = {1022-1032},
          doi = {10.1093/mnras/stad1047},
archivePrefix = {arXiv},
       eprint = {2212.07415},
 primaryClass = {astro-ph.CO},
       adsurl = {https://ui.adsabs.harvard.edu/abs/2023MNRAS.522.1022S}
}

@ARTICLE{2023MNRAS.520..850A,
       author = {{Anstey}, Dominic and {de Lera Acedo}, Eloy and {Handley}, Will},
        title = "{Use of time dependent data in Bayesian global 21-cm foreground and signal modelling}",
      journal = {\mnras},
         year = 2023,
        month = mar,
       volume = {520},
       number = {1},
        pages = {850-865},
          doi = {10.1093/mnras/stad156},
archivePrefix = {arXiv},
       eprint = {2210.04707},
 primaryClass = {astro-ph.CO},
       adsurl = {https://ui.adsabs.harvard.edu/abs/2023MNRAS.520..850A}
}

@ARTICLE{2023ApJ...946L..13F,
       author = {{Finkelstein}, Steven L. and {Bagley}, Micaela B. and {Ferguson}, Henry C. and {Wilkins}, Stephen M. and {Kartaltepe}, Jeyhan S. and {Papovich}, Casey and {Yung}, L.~Y. Aaron and {Arrabal Haro}, Pablo and {Behroozi}, Peter and {Dickinson}, Mark and {Kocevski}, Dale D. and {Koekemoer}, Anton M. and {Larson}, Rebecca L. and {Le Bail}, Aur{\'e}lien and {Morales}, Alexa M. and {P{\'e}rez-Gonz{\'a}lez}, Pablo G. and {Burgarella}, Denis and {Dav{\'e}}, Romeel and {Hirschmann}, Michaela and {Somerville}, Rachel S. and {Wuyts}, Stijn and {Bromm}, Volker and {Casey}, Caitlin M. and {Fontana}, Adriano and {Fujimoto}, Seiji and {Gardner}, Jonathan P. and {Giavalisco}, Mauro and {Grazian}, Andrea and {Grogin}, Norman A. and {Hathi}, Nimish P. and {Hutchison}, Taylor A. and {Jha}, Saurabh W. and {Jogee}, Shardha and {Kewley}, Lisa J. and {Kirkpatrick}, Allison and {Long}, Arianna S. and {Lotz}, Jennifer M. and {Pentericci}, Laura and {Pierel}, Justin D.~R. and {Pirzkal}, Nor and {Ravindranath}, Swara and {Ryan}, Russell E. and {Trump}, Jonathan R. and {Yang}, Guang and {Bhatawdekar}, Rachana and {Bisigello}, Laura and {Buat}, V{\'e}ronique and {Calabr{\`o}}, Antonello and {Castellano}, Marco and {Cleri}, Nikko J. and {Cooper}, M.~C. and {Croton}, Darren and {Daddi}, Emanuele and {Dekel}, Avishai and {Elbaz}, David and {Franco}, Maximilien and {Gawiser}, Eric and {Holwerda}, Benne W. and {Huertas-Company}, Marc and {Jaskot}, Anne E. and {Leung}, Gene C.~K. and {Lucas}, Ray A. and {Mobasher}, Bahram and {Pandya}, Viraj and {Tacchella}, Sandro and {Weiner}, Benjamin J. and {Zavala}, Jorge A.},
        title = "{CEERS Key Paper. I. An Early Look into the First 500 Myr of Galaxy Formation with JWST}",
      journal = {\apjl},
         year = 2023,
        month = mar,
       volume = {946},
       number = {1},
          eid = {L13},
        pages = {L13},
          doi = {10.3847/2041-8213/acade4},
archivePrefix = {arXiv},
       eprint = {2211.05792},
 primaryClass = {astro-ph.GA},
       adsurl = {https://ui.adsabs.harvard.edu/abs/2023ApJ...946L..13F}
}

@ARTICLE{2023ApJ...945..124H,
       author = {{HERA Collaboration} and {Abdurashidova}, Zara and {Adams}, Tyrone and {Aguirre}, James E. and {Alexander}, Paul and {Ali}, Zaki S. and {Baartman}, Rushelle and {Balfour}, Yanga and {Barkana}, Rennan and {Beardsley}, Adam P. and {Bernardi}, Gianni and {Billings}, Tashalee S. and {Bowman}, Judd D. and {Bradley}, Richard F. and {Breitman}, Daniela and {Bull}, Philip and {Burba}, Jacob and {Carey}, Steve and {Carilli}, Chris L. and {Cheng}, Carina and {Choudhuri}, Samir and {DeBoer}, David R. and {de Lera Acedo}, Eloy and {Dexter}, Matt and {Dillon}, Joshua S. and {Ely}, John and {Ewall-Wice}, Aaron and {Fagnoni}, Nicolas and {Fialkov}, Anastasia and {Fritz}, Randall and {Furlanetto}, Steven R. and {Gale-Sides}, Kingsley and {Garsden}, Hugh and {Glendenning}, Brian and {Gorce}, Ad{\'e}lie and {Gorthi}, Deepthi and {Greig}, Bradley and {Grobbelaar}, Jasper and {Halday}, Ziyaad and {Hazelton}, Bryna J. and {Heimersheim}, Stefan and {Hewitt}, Jacqueline N. and {Hickish}, Jack and {Jacobs}, Daniel C. and {Julius}, Austin and {Kern}, Nicholas S. and {Kerrigan}, Joshua and {Kittiwisit}, Piyanat and {Kohn}, Saul A. and {Kolopanis}, Matthew and {Lanman}, Adam and {La Plante}, Paul and {Lewis}, David and {Liu}, Adrian and {Loots}, Anita and {Ma}, Yin-Zhe and {MacMahon}, David H.~E. and {Malan}, Lourence and {Malgas}, Keith and {Malgas}, Cresshim and {Maree}, Matthys and {Marero}, Bradley and {Martinot}, Zachary E. and {McBride}, Lisa and {Mesinger}, Andrei and {Mirocha}, Jordan and {Molewa}, Mathakane and {Morales}, Miguel F. and {Mosiane}, Tshegofalang and {Mu{\~n}oz}, Julian B. and {Murray}, Steven G. and {Nagpal}, Vighnesh and {Neben}, Abraham R. and {Nikolic}, Bojan and {Nunhokee}, Chuneeta D. and {Nuwegeld}, Hans and {Parsons}, Aaron R. and {Pascua}, Robert and {Patra}, Nipanjana and {Pieterse}, Samantha and {Qin}, Yuxiang and {Razavi-Ghods}, Nima and {Robnett}, James and {Rosie}, Kathryn and {Santos}, Mario G. and {Sims}, Peter and {Singh}, Saurabh and {Smith}, Craig and {Swarts}, Hilton and {Tan}, Jianrong and {Thyagarajan}, Nithyanandan and {Wilensky}, Michael J. and {Williams}, Peter K.~G. and {van Wyngaarden}, Pieter and {Zheng}, Haoxuan},
        title = "{Improved Constraints on the 21 cm EoR Power Spectrum and the X-Ray Heating of the IGM with HERA Phase I Observations}",
      journal = {\apj},
         year = 2023,
        month = mar,
       volume = {945},
       number = {2},
          eid = {124},
        pages = {124},
          doi = {10.3847/1538-4357/acaf50},
archivePrefix = {arXiv},
       eprint = {2210.04912},
 primaryClass = {astro-ph.CO},
       adsurl = {https://ui.adsabs.harvard.edu/abs/2023ApJ...945..124H}
}

@ARTICLE{2023arXiv230110345B,
       author = {{Bale}, Stuart D. and {Bassett}, Neil and {Burns}, Jack O. and {Dorigo Jones}, Johnny and {Goetz}, Keith and {Hellum-Bye}, Christian and {Hermann}, Sven and {Hibbard}, Joshua and {Maksimovic}, Milan and {McLean}, Ryan and {Monsalve}, Raul and {O'Connor}, Paul and {Parsons}, Aaron and {Pulupa}, Marc and {Pund}, Rugved and {Rapetti}, David and {Rotermund}, Kaja M. and {Saliwanchik}, Ben and {Slosar}, Anze and {Sundkvist}, David and {Suzuki}, Aritoki},
        title = "{LuSEE 'Night': The Lunar Surface Electromagnetics Experiment}",
      journal = {arXiv e-prints},
         year = 2023,
        month = jan,
          eid = {arXiv:2301.10345},
        pages = {arXiv:2301.10345},
          doi = {10.48550/arXiv.2301.10345},
archivePrefix = {arXiv},
       eprint = {2301.10345},
 primaryClass = {astro-ph.IM},
       adsurl = {https://ui.adsabs.harvard.edu/abs/2023arXiv230110345B}
}

@ARTICLE{2022ApJ...940L..14N,
       author = {{Naidu}, Rohan P. and {Oesch}, Pascal A. and {van Dokkum}, Pieter and {Nelson}, Erica J. and {Suess}, Katherine A. and {Brammer}, Gabriel and {Whitaker}, Katherine E. and {Illingworth}, Garth and {Bouwens}, Rychard and {Tacchella}, Sandro and {Matthee}, Jorryt and {Allen}, Natalie and {Bezanson}, Rachel and {Conroy}, Charlie and {Labbe}, Ivo and {Leja}, Joel and {Leonova}, Ecaterina and {Magee}, Dan and {Price}, Sedona H. and {Setton}, David J. and {Strait}, Victoria and {Stefanon}, Mauro and {Toft}, Sune and {Weaver}, John R. and {Weibel}, Andrea},
        title = "{Two Remarkably Luminous Galaxy Candidates at z {\ensuremath{\approx}} 10-12 Revealed by JWST}",
      journal = {\apjl},
         year = 2022,
        month = nov,
       volume = {940},
       number = {1},
          eid = {L14},
        pages = {L14},
          doi = {10.3847/2041-8213/ac9b22},
archivePrefix = {arXiv},
       eprint = {2207.09434},
 primaryClass = {astro-ph.GA},
       adsurl = {https://ui.adsabs.harvard.edu/abs/2022ApJ...940L..14N}
}

@ARTICLE{2022ApJ...938L..15C,
       author = {{Castellano}, Marco and {Fontana}, Adriano and {Treu}, Tommaso and {Santini}, Paola and {Merlin}, Emiliano and {Leethochawalit}, Nicha and {Trenti}, Michele and {Vanzella}, Eros and {Mestric}, Uros and {Bonchi}, Andrea and {Belfiori}, Davide and {Nonino}, Mario and {Paris}, Diego and {Polenta}, Gianluca and {Roberts-Borsani}, Guido and {Boyett}, Kristan and {Brada{\v{c}}}, Maru{\v{s}}a and {Calabr{\`o}}, Antonello and {Glazebrook}, Karl and {Grillo}, Claudio and {Mascia}, Sara and {Mason}, Charlotte and {Mercurio}, Amata and {Morishita}, Takahiro and {Nanayakkara}, Themiya and {Pentericci}, Laura and {Rosati}, Piero and {Vulcani}, Benedetta and {Wang}, Xin and {Yang}, Lilan},
        title = "{Early Results from GLASS-JWST. III. Galaxy Candidates at z  9-15}",
      journal = {\apjl},
         year = 2022,
        month = oct,
       volume = {938},
       number = {2},
          eid = {L15},
        pages = {L15},
          doi = {10.3847/2041-8213/ac94d0},
archivePrefix = {arXiv},
       eprint = {2207.09436},
 primaryClass = {astro-ph.GA},
       adsurl = {https://ui.adsabs.harvard.edu/abs/2022ApJ...938L..15C}
}

@ARTICLE{2022MNRAS.515.1580S,
       author = {{Spinelli}, M. and {Kyriakou}, G. and {Bernardi}, G. and {Bolli}, P. and {Greenhill}, L.~J. and {Fialkov}, A. and {Garsden}, H.},
        title = "{Antenna beam characterization for the global 21-cm experiment LEDA and its impact on signal model parameter reconstruction}",
      journal = {\mnras},
         year = 2022,
        month = sep,
       volume = {515},
       number = {2},
        pages = {1580-1597},
          doi = {10.1093/mnras/stac1804},
archivePrefix = {arXiv},
       eprint = {2206.12158},
 primaryClass = {astro-ph.IM},
       adsurl = {https://ui.adsabs.harvard.edu/abs/2022MNRAS.515.1580S}
}

@ARTICLE{2022NatAs...6..984D,
       author = {{de Lera Acedo}, E. and {de Villiers}, D.~I.~L. and {Razavi-Ghods}, N. and {Handley}, W. and {Fialkov}, A. and {Magro}, A. and {Anstey}, D. and {Bevins}, H.~T.~J. and {Chiello}, R. and {Cumner}, J. and {Josaitis}, A.~T. and {Roque}, I.~L.~V. and {Sims}, P.~H. and {Scheutwinkel}, K.~H. and {Alexander}, P. and {Bernardi}, G. and {Carey}, S. and {Cavillot}, J. and {Croukamp}, W. and {Ely}, J.~A. and {Gessey-Jones}, T. and {Gueuning}, Q. and {Hills}, R. and {Kulkarni}, G. and {Maiolino}, R. and {Meerburg}, P.~D. and {Mittal}, S. and {Pritchard}, J.~R. and {Puchwein}, E. and {Saxena}, A. and {Shen}, E. and {Smirnov}, O. and {Spinelli}, M. and {Zarb-Adami}, K.},
        title = "{The REACH radiometer for detecting the 21-cm hydrogen signal from redshift z {\ensuremath{\approx}} 7.5-28}",
      journal = {Nature Astronomy},
         year = 2022,
        month = jul,
       volume = {6},
        pages = {984-998},
          doi = {10.1038/s41550-022-01709-9},
archivePrefix = {arXiv},
       eprint = {2210.07409},
 primaryClass = {astro-ph.CO},
       adsurl = {https://ui.adsabs.harvard.edu/abs/2022NatAs...6..984D}
}

@ARTICLE{2022MNRAS.514...55B,
       author = {{Bosman}, Sarah E.~I. and {Davies}, Frederick B. and {Becker}, George D. and {Keating}, Laura C. and {Davies}, Rebecca L. and {Zhu}, Yongda and {Eilers}, Anna-Christina and {D'Odorico}, Valentina and {Bian}, Fuyan and {Bischetti}, Manuela and {Cristiani}, Stefano V. and {Fan}, Xiaohui and {Farina}, Emanuele P. and {Haehnelt}, Martin G. and {Hennawi}, Joseph F. and {Kulkarni}, Girish and {Mesinger}, Andrei and {Meyer}, Romain A. and {Onoue}, Masafusa and {Pallottini}, Andrea and {Qin}, Yuxiang and {Ryan-Weber}, Emma and {Schindler}, Jan-Torge and {Walter}, Fabian and {Wang}, Feige and {Yang}, Jinyi},
        title = "{Hydrogen reionization ends by z = 5.3: Lyman-{\ensuremath{\alpha}} optical depth measured by the XQR-30 sample}",
      journal = {\mnras},
         year = 2022,
        month = jul,
       volume = {514},
       number = {1},
        pages = {55-76},
          doi = {10.1093/mnras/stac1046},
archivePrefix = {arXiv},
       eprint = {2108.03699},
 primaryClass = {astro-ph.CO},
       adsurl = {https://ui.adsabs.harvard.edu/abs/2022MNRAS.514...55B}
}

@ARTICLE{2022NatAs...6..607S,
       author = {{Singh}, Saurabh and {Jishnu}, Nambissan T. and {Subrahmanyan}, Ravi and {Udaya Shankar}, N. and {Girish}, B.~S. and {Raghunathan}, A. and {Somashekar}, R. and {Srivani}, K.~S. and {Sathyanarayana Rao}, Mayuri},
        title = "{On the detection of a cosmic dawn signal in the radio background}",
      journal = {Nature Astronomy},
         year = 2022,
        month = feb,
       volume = {6},
        pages = {607-617},
          doi = {10.1038/s41550-022-01610-5},
archivePrefix = {arXiv},
       eprint = {2112.06778},
 primaryClass = {astro-ph.CO},
       adsurl = {https://ui.adsabs.harvard.edu/abs/2022NatAs...6..607S}
}

@ARTICLE{2022JAI....1150001C,
       author = {{Cumner}, J. and {de Lera Acedo}, E. and {de Villiers}, D.~I.~L. and {Anstey}, D. and {Kolitsidas}, C.~I. and {Gurdon}, B. and {Fagnoni}, N. and {Alexander}, P. and {Bernardi}, G. and {Bevins}, H.~T.~J. and {Carey}, S. and {Cavillot}, J. and {Chiello}, R. and {Craeye}, C. and {Croukamp}, W. and {Ely}, J.~A. and {Fialkov}, A. and {Gessey-Jones}, T. and {Gueuning}, Q. and {Handley}, W. and {Hills}, R. and {Josaitis}, A.~T. and {Kulkarni}, G. and {Magro}, A. and {Maiolino}, R. and {Meerburg}, P.~D. and {Mittal}, S. and {Pritchard}, J.~R. and {Puchwein}, E. and {Razavi-Ghods}, N. and {Roque}, I.~L.~V. and {Saxena}, A. and {Scheutwinkel}, K.~H. and {Shen}, E. and {Sims}, P.~H. and {Smirnov}, O. and {Spinelli}, M. and {Zarb-Adami}, K.},
        title = "{Radio Antenna Design for Sky-Averaged 21cm Cosmology Experiments: The REACH Case}",
      journal = {Journal of Astronomical Instrumentation},
         year = 2022,
        month = jan,
       volume = {11},
       number = {1},
          eid = {2250001-2058},
        pages = {2250001-2058},
          doi = {10.1142/S2251171722500015},
archivePrefix = {arXiv},
       eprint = {2109.10098},
 primaryClass = {astro-ph.IM},
       adsurl = {https://ui.adsabs.harvard.edu/abs/2022JAI....1150001C}
}

@ARTICLE{2022ApJ...924...51A,
       author = {{Abdurashidova}, Zara and {Aguirre}, James E. and {Alexander}, Paul and {Ali}, Zaki S. and {Balfour}, Yanga and {Barkana}, Rennan and {Beardsley}, Adam P. and {Bernardi}, Gianni and {Billings}, Tashalee S. and {Bowman}, Judd D. and {Bradley}, Richard F. and {Bull}, Philip and {Burba}, Jacob and {Carey}, Steve and {Carilli}, Chris L. and {Cheng}, Carina and {DeBoer}, David R. and {Dexter}, Matt and {de Lera Acedo}, Eloy and {Dillon}, Joshua S. and {Ely}, John and {Ewall-Wice}, Aaron and {Fagnoni}, Nicolas and {Fialkov}, Anastasia and {Fritz}, Randall and {Furlanetto}, Steven R. and {Gale-Sides}, Kingsley and {Glendenning}, Brian and {Gorthi}, Deepthi and {Greig}, Bradley and {Grobbelaar}, Jasper and {Halday}, Ziyaad and {Hazelton}, Bryna J. and {Heimersheim}, Stefan and {Hewitt}, Jacqueline N. and {Hickish}, Jack and {Jacobs}, Daniel C. and {Julius}, Austin and {Kern}, Nicholas S. and {Kerrigan}, Joshua and {Kittiwisit}, Piyanat and {Kohn}, Saul A. and {Kolopanis}, Matthew and {Lanman}, Adam and {La Plante}, Paul and {Lekalake}, Telalo and {Lewis}, David and {Liu}, Adrian and {Ma}, Yin-Zhe and {MacMahon}, David and {Malan}, Lourence and {Malgas}, Cresshim and {Maree}, Matthys and {Martinot}, Zachary E. and {Matsetela}, Eunice and {Mesinger}, Andrei and {Mirocha}, Jordan and {Molewa}, Mathakane and {Morales}, Miguel F. and {Mosiane}, Tshegofalang and {Mu{\~n}oz}, Julian B. and {Murray}, Steven G. and {Neben}, Abraham R. and {Nikolic}, Bojan and {Nunhokee}, Chuneeta D. and {Parsons}, Aaron R. and {Patra}, Nipanjana and {Pieterse}, Samantha and {Pober}, Jonathan C. and {Qin}, Yuxiang and {Razavi-Ghods}, Nima and {Reis}, Itamar and {Ringuette}, Jon and {Robnett}, James and {Rosie}, Kathryn and {Santos}, Mario G. and {Sikder}, Sudipta and {Sims}, Peter and {Smith}, Craig and {Syce}, Angelo and {Thyagarajan}, Nithyanandan and {Williams}, Peter K.~G. and {Zheng}, Haoxuan},
        title = "{HERA Phase I Limits on the Cosmic 21 cm Signal: Constraints on Astrophysics and Cosmology during the Epoch of Reionization}",
      journal = {\apj},
         year = 2022,
        month = jan,
       volume = {924},
       number = {2},
          eid = {51},
        pages = {51},
          doi = {10.3847/1538-4357/ac2ffc},
archivePrefix = {arXiv},
       eprint = {2108.07282},
 primaryClass = {astro-ph.CO},
       adsurl = {https://ui.adsabs.harvard.edu/abs/2022ApJ...924...51A}
}

@ARTICLE{2021ITAP...69.8143F,
       author = {{Fagnoni}, Nicolas and {de Lera Acedo}, Eloy and {Drought}, Nick and {DeBoer}, David R. and {Riley}, Daniel and {Razavi-Ghods}, Nima and {Carey}, Steven and {Parsons}, Aaron R.},
        title = "{Design of the New Wideband Vivaldi Feed for the HERA Radio-Telescope Phase II}",
      journal = {IEEE Transactions on Antennas and Propagation},
         year = 2021,
        month = dec,
       volume = {69},
       number = {12},
        pages = {8143-8157},
          doi = {10.1109/TAP.2021.3083788},
archivePrefix = {arXiv},
       eprint = {2009.07939},
 primaryClass = {astro-ph.IM},
       adsurl = {https://ui.adsabs.harvard.edu/abs/2021ITAP...69.8143F}
}

@ARTICLE{2021ApJ...923...33B,
       author = {{Bassett}, Neil and {Rapetti}, David and {Tauscher}, Keith and {Nhan}, Bang D. and {Bordenave}, David D. and {Hibbard}, Joshua J. and {Burns}, Jack O.},
        title = "{Lost Horizon: Quantifying the Effect of Local Topography on Global 21 cm Cosmology Data Analysis}",
      journal = {\apj},
         year = 2021,
        month = dec,
       volume = {923},
       number = {1},
          eid = {33},
        pages = {33},
          doi = {10.3847/1538-4357/ac1cde},
archivePrefix = {arXiv},
       eprint = {2106.02153},
 primaryClass = {astro-ph.CO},
       adsurl = {https://ui.adsabs.harvard.edu/abs/2021ApJ...923...33B}
}

@ARTICLE{2021ITAP...69.6209R,
       author = {{Raghunathan}, Agaram and {Subrahmanyan}, Ravi and {Shankar}, N. Udaya and {Singh}, Saurabh and {Nambissan}, Jishnu and {Kavitha}, K. and {Mahesh}, Nivedita and {Somashekar}, R. and {Sindhu}, Gaddam and {Girish}, B.~S. and {Srivani}, K.~S. and {Rao}, Mayuri S.},
        title = "{A Floating Octave Bandwidth Cone-Disk Antenna for Detection of Cosmic Dawn}",
      journal = {IEEE Transactions on Antennas and Propagation},
         year = 2021,
        month = oct,
       volume = {69},
       number = {10},
        pages = {6209-6217},
          doi = {10.1109/TAP.2021.3069563},
       adsurl = {https://ui.adsabs.harvard.edu/abs/2021ITAP...69.6209R}
}

@ARTICLE{2021MNRAS.506.2041A,
       author = {{Anstey}, Dominic and {de Lera Acedo}, Eloy and {Handley}, Will},
        title = "{A general Bayesian framework for foreground modelling and chromaticity correction for global 21 cm experiments}",
      journal = {\mnras},
         year = 2021,
        month = sep,
       volume = {506},
       number = {2},
        pages = {2041-2058},
          doi = {10.1093/mnras/stab1765},
archivePrefix = {arXiv},
       eprint = {2010.09644},
 primaryClass = {astro-ph.IM},
       adsurl = {https://ui.adsabs.harvard.edu/abs/2021MNRAS.506.2041A}
}

@ARTICLE{2021AJ....162...38M,
       author = {{Mahesh}, Nivedita and {Bowman}, Judd D. and {Mozdzen}, Thomas J. and {Rogers}, Alan E.~E. and {Monsalve}, Raul A. and {Murray}, Steven G. and {Lewis}, David},
        title = "{Validation of the EDGES Low-band Antenna Beam Model}",
      journal = {\aj},
         year = 2021,
        month = aug,
       volume = {162},
       number = {2},
          eid = {38},
        pages = {38},
          doi = {10.3847/1538-3881/abfdab},
archivePrefix = {arXiv},
       eprint = {2103.00423},
 primaryClass = {astro-ph.IM},
       adsurl = {https://ui.adsabs.harvard.edu/abs/2021AJ....162...38M}
}

@ARTICLE{2021MNRAS.503..344S,
       author = {{Shen}, Emma and {Anstey}, Dominic and {de Lera Acedo}, Eloy and {Fialkov}, Anastasia and {Handley}, Will},
        title = "{Quantifying ionospheric effects on global 21-cm observations}",
      journal = {\mnras},
         year = 2021,
        month = may,
       volume = {503},
       number = {1},
        pages = {344-353},
          doi = {10.1093/mnras/stab429},
archivePrefix = {arXiv},
       eprint = {2011.10517},
 primaryClass = {astro-ph.IM},
       adsurl = {https://ui.adsabs.harvard.edu/abs/2021MNRAS.503..344S}
}

@ARTICLE{2021ExA....51..193T,
       author = {{Jishnu Nambissan}, T. and {Subrahmanyan}, Ravi and {Somashekar}, R. and {Shankar}, N. Udaya and {Singh}, Saurabh and {Raghunathan}, A. and {Girish}, B.~S. and {Srivani}, K.~S. and {Rao}, Mayuri Sathyanarayana},
        title = "{SARAS 3 CD/EoR radiometer: design and performance of the receiver}",
      journal = {Experimental Astronomy},
         year = 2021,
        month = apr,
       volume = {51},
       number = {2},
        pages = {193-234},
          doi = {10.1007/s10686-020-09697-2},
       adsurl = {https://ui.adsabs.harvard.edu/abs/2021ExA....51..193T}
}

@ARTICLE{2021MNRAS.502.4405B,
       author = {{Bevins}, H.~T.~J. and {Handley}, W.~J. and {Fialkov}, A. and {de Lera Acedo}, E. and {Greenhill}, L.~J. and {Price}, D.~C.},
        title = "{MAXSMOOTH: rapid maximally smooth function fitting with applications in Global 21-cm cosmology}",
      journal = {\mnras},
         year = 2021,
        month = apr,
       volume = {502},
       number = {3},
        pages = {4405-4425},
          doi = {10.1093/mnras/stab152},
archivePrefix = {arXiv},
       eprint = {2007.14970},
 primaryClass = {astro-ph.CO},
       adsurl = {https://ui.adsabs.harvard.edu/abs/2021MNRAS.502.4405B}
}

@ARTICLE{2021ApJ...908..199R,
       author = {{Reichardt}, C.~L. and {Patil}, S. and {Ade}, P.~A.~R. and {Anderson}, A.~J. and {Austermann}, J.~E. and {Avva}, J.~S. and {Baxter}, E. and {Beall}, J.~A. and {Bender}, A.~N. and {Benson}, B.~A. and {Bianchini}, F. and {Bleem}, L.~E. and {Carlstrom}, J.~E. and {Chang}, C.~L. and {Chaubal}, P. and {Chiang}, H.~C. and {Chou}, T.~L. and {Citron}, R. and {Moran}, C. Corbett and {Crawford}, T.~M. and {Crites}, A.~T. and {de Haan}, T. and {Dobbs}, M.~A. and {Everett}, W. and {Gallicchio}, J. and {George}, E.~M. and {Gilbert}, A. and {Gupta}, N. and {Halverson}, N.~W. and {Harrington}, N. and {Henning}, J.~W. and {Hilton}, G.~C. and {Holder}, G.~P. and {Holzapfel}, W.~L. and {Hrubes}, J.~D. and {Huang}, N. and {Hubmayr}, J. and {Irwin}, K.~D. and {Knox}, L. and {Lee}, A.~T. and {Li}, D. and {Lowitz}, A. and {Luong-Van}, D. and {McMahon}, J.~J. and {Mehl}, J. and {Meyer}, S.~S. and {Millea}, M. and {Mocanu}, L.~M. and {Mohr}, J.~J. and {Montgomery}, J. and {Nadolski}, A. and {Natoli}, T. and {Nibarger}, J.~P. and {Noble}, G. and {Novosad}, V. and {Omori}, Y. and {Padin}, S. and {Pryke}, C. and {Ruhl}, J.~E. and {Saliwanchik}, B.~R. and {Sayre}, J.~T. and {Schaffer}, K.~K. and {Shirokoff}, E. and {Sievers}, C. and {Smecher}, G. and {Spieler}, H.~G. and {Staniszewski}, Z. and {Stark}, A.~A. and {Tucker}, C. and {Vanderlinde}, K. and {Veach}, T. and {Vieira}, J.~D. and {Wang}, G. and {Whitehorn}, N. and {Williamson}, R. and {Wu}, W.~L.~K. and {Yefremenko}, V.},
        title = "{An Improved Measurement of the Secondary Cosmic Microwave Background Anisotropies from the SPT-SZ + SPTpol Surveys}",
      journal = {\apj},
         year = 2021,
        month = feb,
       volume = {908},
       number = {2},
          eid = {199},
        pages = {199},
          doi = {10.3847/1538-4357/abd407},
archivePrefix = {arXiv},
       eprint = {2002.06197},
 primaryClass = {astro-ph.CO},
       adsurl = {https://ui.adsabs.harvard.edu/abs/2021ApJ...908..199R}
}

@ARTICLE{2021MNRAS.501....1G,
       author = {{Greig}, Bradley and {Mesinger}, Andrei and {Koopmans}, L{\'e}on V.~E. and {Ciardi}, Benedetta and {Mellema}, Garrelt and {Zaroubi}, Saleem and {Giri}, Sambit K. and {Ghara}, Raghunath and {Ghosh}, Abhik and {Iliev}, Ilian T. and {Mertens}, Florent G. and {Mondal}, Rajesh and {Offringa}, Andr{\'e} R. and {Pandey}, Vishambhar N.},
        title = "{Interpreting LOFAR 21-cm signal upper limits at z {\ensuremath{\approx}} 9.1 in the context of high-z galaxy and reionization observations}",
      journal = {\mnras},
         year = 2021,
        month = jan,
       volume = {501},
       number = {1},
        pages = {1-13},
          doi = {10.1093/mnras/staa3593},
archivePrefix = {arXiv},
       eprint = {2006.03203},
 primaryClass = {astro-ph.CO},
       adsurl = {https://ui.adsabs.harvard.edu/abs/2021MNRAS.501....1G}
}

@ARTICLE{2021RSPTA.37990566C,
       author = {{Chen}, Xuelei and {Yan}, Jingye and {Deng}, Li and {Wu}, Fengquan and {Wu}, Lin and {Xu}, Yidong and {Zhou}, Li},
        title = "{Discovering the sky at the longest wavelengths with a lunar orbit array}",
      journal = {Philosophical Transactions of the Royal Society of London Series A},
         year = 2021,
        month = jan,
       volume = {379},
       number = {2188},
          eid = {20190566},
        pages = {20190566},
          doi = {10.1098/rsta.2019.0566},
archivePrefix = {arXiv},
       eprint = {2007.15794},
 primaryClass = {astro-ph.IM},
       adsurl = {https://ui.adsabs.harvard.edu/abs/2021RSPTA.37990566C}
}

@ARTICLE{2021MNRAS.500.5195E,
       author = {{Ewall-Wice}, Aaron and {Kern}, Nicholas and {Dillon}, Joshua S. and {Liu}, Adrian and {Parsons}, Aaron and {Singh}, Saurabh and {Lanman}, Adam and {La Plante}, Paul and {Fagnoni}, Nicolas and {Acedo}, Eloy de Lera and {DeBoer}, David R. and {Nunhokee}, Chuneeta and {Bull}, Philip and {Chang}, Tzu-Ching and {Lazio}, T. Joseph W. and {Aguirre}, James and {Weinberg}, Sean},
        title = "{DAYENU: a simple filter of smooth foregrounds for intensity mapping power spectra}",
      journal = {\mnras},
         year = 2021,
        month = jan,
       volume = {500},
       number = {4},
        pages = {5195-5213},
          doi = {10.1093/mnras/staa3293},
archivePrefix = {arXiv},
       eprint = {2004.11397},
 primaryClass = {astro-ph.CO},
       adsurl = {https://ui.adsabs.harvard.edu/abs/2021MNRAS.500.5195E}
}

@ARTICLE{2020MNRAS.498.4178M,
       author = {{Mondal}, R. and {Fialkov}, A. and {Fling}, C. and {Iliev}, I.~T. and {Barkana}, R. and {Ciardi}, B. and {Mellema}, G. and {Zaroubi}, S. and {Koopmans}, L.~V.~E. and {Mertens}, F.~G. and {Gehlot}, B.~K. and {Ghara}, R. and {Ghosh}, A. and {Giri}, S.~K. and {Offringa}, A. and {Pandey}, V.~N.},
        title = "{Tight constraints on the excess radio background at z = 9.1 from LOFAR}",
      journal = {\mnras},
         year = 2020,
        month = nov,
       volume = {498},
       number = {3},
        pages = {4178-4191},
          doi = {10.1093/mnras/staa2422},
archivePrefix = {arXiv},
       eprint = {2004.00678},
 primaryClass = {astro-ph.CO},
       adsurl = {https://ui.adsabs.harvard.edu/abs/2020MNRAS.498.4178M}
}

@ARTICLE{2020A&A...641A...6P,
       author = {{Planck Collaboration} and {Aghanim}, N. and {Akrami}, Y. and {Ashdown}, M. and {Aumont}, J. and {Baccigalupi}, C. and {Ballardini}, M. and {Banday}, A.~J. and {Barreiro}, R.~B. and {Bartolo}, N. and {Basak}, S. and {Battye}, R. and {Benabed}, K. and {Bernard}, J.-P. and {Bersanelli}, M. and {Bielewicz}, P. and {Bock}, J.~J. and {Bond}, J.~R. and {Borrill}, J. and {Bouchet}, F.~R. and {Boulanger}, F. and {Bucher}, M. and {Burigana}, C. and {Butler}, R.~C. and {Calabrese}, E. and {Cardoso}, J.-F. and {Carron}, J. and {Challinor}, A. and {Chiang}, H.~C. and {Chluba}, J. and {Colombo}, L.~P.~L. and {Combet}, C. and {Contreras}, D. and {Crill}, B.~P. and {Cuttaia}, F. and {de Bernardis}, P. and {de Zotti}, G. and {Delabrouille}, J. and {Delouis}, J.-M. and {Di Valentino}, E. and {Diego}, J.~M. and {Dor{\'e}}, O. and {Douspis}, M. and {Ducout}, A. and {Dupac}, X. and {Dusini}, S. and {Efstathiou}, G. and {Elsner}, F. and {En{\ss}lin}, T.~A. and {Eriksen}, H.~K. and {Fantaye}, Y. and {Farhang}, M. and {Fergusson}, J. and {Fernandez-Cobos}, R. and {Finelli}, F. and {Forastieri}, F. and {Frailis}, M. and {Fraisse}, A.~A. and {Franceschi}, E. and {Frolov}, A. and {Galeotta}, S. and {Galli}, S. and {Ganga}, K. and {G{\'e}nova-Santos}, R.~T. and {Gerbino}, M. and {Ghosh}, T. and {Gonz{\'a}lez-Nuevo}, J. and {G{\'o}rski}, K.~M. and {Gratton}, S. and {Gruppuso}, A. and {Gudmundsson}, J.~E. and {Hamann}, J. and {Handley}, W. and {Hansen}, F.~K. and {Herranz}, D. and {Hildebrandt}, S.~R. and {Hivon}, E. and {Huang}, Z. and {Jaffe}, A.~H. and {Jones}, W.~C. and {Karakci}, A. and {Keih{\"a}nen}, E. and {Keskitalo}, R. and {Kiiveri}, K. and {Kim}, J. and {Kisner}, T.~S. and {Knox}, L. and {Krachmalnicoff}, N. and {Kunz}, M. and {Kurki-Suonio}, H. and {Lagache}, G. and {Lamarre}, J.-M. and {Lasenby}, A. and {Lattanzi}, M. and {Lawrence}, C.~R. and {Le Jeune}, M. and {Lemos}, P. and {Lesgourgues}, J. and {Levrier}, F. and {Lewis}, A. and {Liguori}, M. and {Lilje}, P.~B. and {Lilley}, M. and {Lindholm}, V. and {L{\'o}pez-Caniego}, M. and {Lubin}, P.~M. and {Ma}, Y.-Z. and {Mac{\'\i}as-P{\'e}rez}, J.~F. and {Maggio}, G. and {Maino}, D. and {Mandolesi}, N. and {Mangilli}, A. and {Marcos-Caballero}, A. and {Maris}, M. and {Martin}, P.~G. and {Martinelli}, M. and {Mart{\'\i}nez-Gonz{\'a}lez}, E. and {Matarrese}, S. and {Mauri}, N. and {McEwen}, J.~D. and {Meinhold}, P.~R. and {Melchiorri}, A. and {Mennella}, A. and {Migliaccio}, M. and {Millea}, M. and {Mitra}, S. and {Miville-Desch{\^e}nes}, M.-A. and {Molinari}, D. and {Montier}, L. and {Morgante}, G. and {Moss}, A. and {Natoli}, P. and {N{\o}rgaard-Nielsen}, H.~U. and {Pagano}, L. and {Paoletti}, D. and {Partridge}, B. and {Patanchon}, G. and {Peiris}, H.~V. and {Perrotta}, F. and {Pettorino}, V. and {Piacentini}, F. and {Polastri}, L. and {Polenta}, G. and {Puget}, J.-L. and {Rachen}, J.~P. and {Reinecke}, M. and {Remazeilles}, M. and {Renzi}, A. and {Rocha}, G. and {Rosset}, C. and {Roudier}, G. and {Rubi{\~n}o-Mart{\'\i}n}, J.~A. and {Ruiz-Granados}, B. and {Salvati}, L. and {Sandri}, M. and {Savelainen}, M. and {Scott}, D. and {Shellard}, E.~P.~S. and {Sirignano}, C. and {Sirri}, G. and {Spencer}, L.~D. and {Sunyaev}, R. and {Suur-Uski}, A.-S. and {Tauber}, J.~A. and {Tavagnacco}, D. and {Tenti}, M. and {Toffolatti}, L. and {Tomasi}, M. and {Trombetti}, T. and {Valenziano}, L. and {Valiviita}, J. and {Van Tent}, B. and {Vibert}, L. and {Vielva}, P. and {Villa}, F. and {Vittorio}, N. and {Wandelt}, B.~D. and {Wehus}, I.~K. and {White}, M. and {White}, S.~D.~M. and {Zacchei}, A. and {Zonca}, A.},
        title = "{Planck 2018 results. VI. Cosmological parameters}",
      journal = {\aap},
         year = 2020,
        month = sep,
       volume = {641},
          eid = {A6},
        pages = {A6},
          doi = {10.1051/0004-6361/201833910},
archivePrefix = {arXiv},
       eprint = {1807.06209},
 primaryClass = {astro-ph.CO},
       adsurl = {https://ui.adsabs.harvard.edu/abs/2020A&A...641A...6P}
}

@ARTICLE{2020ApJ...897..175T,
       author = {{Tauscher}, Keith and {Rapetti}, David and {Burns}, Jack O.},
        title = "{Global 21 cm Signal Extraction from Foreground and Instrumental Effects. III. Utilizing Drift-scan Time Dependence and Full Stokes Measurements}",
      journal = {\apj},
         year = 2020,
        month = jul,
       volume = {897},
       number = {2},
          eid = {175},
        pages = {175},
          doi = {10.3847/1538-4357/ab9b2a},
archivePrefix = {arXiv},
       eprint = {2003.05452},
 primaryClass = {astro-ph.IM},
       adsurl = {https://ui.adsabs.harvard.edu/abs/2020ApJ...897..175T}
}

@ARTICLE{2020MNRAS.492.6086E,
       author = {{Ewall-Wice}, Aaron and {Chang}, Tzu-Ching and {Lazio}, T. Joseph W.},
        title = "{The Radio Scream from black holes at Cosmic Dawn: a semi-analytic model for the impact of radio-loud black holes on the 21 cm global signal}",
      journal = {\mnras},
         year = 2020,
        month = mar,
       volume = {492},
       number = {4},
        pages = {6086-6104},
          doi = {10.1093/mnras/stz3501},
archivePrefix = {arXiv},
       eprint = {1903.06788},
 primaryClass = {astro-ph.GA},
       adsurl = {https://ui.adsabs.harvard.edu/abs/2020MNRAS.492.6086E}
}

@ARTICLE{2020MNRAS.493.4728G,
       author = {{Ghara}, R. and {Giri}, S.~K. and {Mellema}, G. and {Ciardi}, B. and {Zaroubi}, S. and {Iliev}, I.~T. and {Koopmans}, L.~V.~E. and {Chapman}, E. and {Gazagnes}, S. and {Gehlot}, B.~K. and {Ghosh}, A. and {Jeli{\'c}}, V. and {Mertens}, F.~G. and {Mondal}, R. and {Schaye}, J. and {Silva}, M.~B. and {Asad}, K.~M.~B. and {Kooistra}, R. and {Mevius}, M. and {Offringa}, A.~R. and {Pandey}, V.~N. and {Yatawatta}, S.},
        title = "{Constraining the intergalactic medium at z {\ensuremath{\approx}} 9.1 using LOFAR Epoch of Reionization observations}",
      journal = {\mnras},
         year = 2020,
        month = feb,
       volume = {493},
       number = {4},
        pages = {4728-4747},
          doi = {10.1093/mnras/staa487},
archivePrefix = {arXiv},
       eprint = {2002.07195},
 primaryClass = {astro-ph.CO},
       adsurl = {https://ui.adsabs.harvard.edu/abs/2020MNRAS.493.4728G}
}

@ARTICLE{2020MNRAS.492...22S,
       author = {{Sims}, Peter H. and {Pober}, Jonathan C.},
        title = "{Testing for calibration systematics in the EDGES low-band data using Bayesian model selection}",
      journal = {\mnras},
         year = 2020,
        month = feb,
       volume = {492},
       number = {1},
        pages = {22-38},
          doi = {10.1093/mnras/stz3388},
archivePrefix = {arXiv},
       eprint = {1910.03165},
 primaryClass = {astro-ph.CO},
       adsurl = {https://ui.adsabs.harvard.edu/abs/2020MNRAS.492...22S}
}

@ARTICLE{2019ApJ...884..105K,
       author = {{Kern}, Nicholas S. and {Parsons}, Aaron R. and {Dillon}, Joshua S. and {Lanman}, Adam E. and {Fagnoni}, Nicolas and {de Lera Acedo}, Eloy},
        title = "{Mitigating Internal Instrument Coupling for 21 cm Cosmology. I. Temporal and Spectral Modeling in Simulations}",
      journal = {\apj},
         year = 2019,
        month = oct,
       volume = {884},
       number = {2},
          eid = {105},
        pages = {105},
          doi = {10.3847/1538-4357/ab3e73},
       adsurl = {https://ui.adsabs.harvard.edu/abs/2019ApJ...884..105K}
}

@INPROCEEDINGS{2019BAAS...51g.255H,
       author = {{Hallinan}, Gregg and {Ravi}, V. and {Weinreb}, S. and {Kocz}, J. and {Huang}, Y. and {Woody}, D.~P. and {Lamb}, J. and {D'Addario}, L. and {Catha}, M. and {Law}, C. and {Kulkarni}, S.~R. and {Phinney}, E.~S. and {Eastwood}, M.~W. and {Bouman}, K. and {McLaughlin}, M. and {Ransom}, S. and {Siemens}, X. and {Cordes}, J. and {Lynch}, R. and {Kaplan}, D. and {Brazier}, A. and {Bhatnagar}, S. and {Myers}, S. and {Walter}, F. and {Gaensler}, B.},
        title = "{The DSA-2000 {\textemdash} A Radio Survey Camera}",
    booktitle = {Bulletin of the American Astronomical Society},
         year = 2019,
       volume = {51},
        month = sep,
          eid = {255},
        pages = {255},
          doi = {10.48550/arXiv.1907.07648},
archivePrefix = {arXiv},
       eprint = {1907.07648},
 primaryClass = {astro-ph.IM},
       adsurl = {https://ui.adsabs.harvard.edu/abs/2019BAAS...51g.255H}
}

@ARTICLE{2019AJ....158...84E,
       author = {{Eastwood}, Michael W. and {Anderson}, Marin M. and {Monroe}, Ryan M. and {Hallinan}, Gregg and {Catha}, Morgan and {Dowell}, Jayce and {Garsden}, Hugh and {Greenhill}, Lincoln J. and {Hicks}, Brian C. and {Kocz}, Jonathon and {Price}, Danny C. and {Schinzel}, Frank K. and {Vedantham}, Harish and {Wang}, Yuankun},
        title = "{The 21 cm Power Spectrum from the Cosmic Dawn: First Results from the OVRO-LWA}",
      journal = {\aj},
         year = 2019,
        month = aug,
       volume = {158},
       number = {2},
          eid = {84},
        pages = {84},
          doi = {10.3847/1538-3881/ab2629},
archivePrefix = {arXiv},
       eprint = {1906.08943},
 primaryClass = {astro-ph.CO},
       adsurl = {https://ui.adsabs.harvard.edu/abs/2019AJ....158...84E}
}

@ARTICLE{2019ApJ...880...26S,
       author = {{Singh}, Saurabh and {Subrahmanyan}, Ravi},
        title = "{The Redshifted 21 cm Signal in the EDGES Low-band Spectrum}",
      journal = {\apj},
         year = 2019,
        month = jul,
       volume = {880},
       number = {1},
          eid = {26},
        pages = {26},
          doi = {10.3847/1538-4357/ab2879},
archivePrefix = {arXiv},
       eprint = {1903.04540},
 primaryClass = {astro-ph.CO},
       adsurl = {https://ui.adsabs.harvard.edu/abs/2019ApJ...880...26S}
}

@ARTICLE{2019MNRAS.485L..24K,
       author = {{Kulkarni}, Girish and {Keating}, Laura C. and {Haehnelt}, Martin G. and {Bosman}, Sarah E.~I. and {Puchwein}, Ewald and {Chardin}, Jonathan and {Aubert}, Dominique},
        title = "{Large Ly {\ensuremath{\alpha}} opacity fluctuations and low CMB {\ensuremath{\tau}} in models of late reionization with large islands of neutral hydrogen extending to z < 5.5}",
      journal = {\mnras},
         year = 2019,
        month = may,
       volume = {485},
       number = {1},
        pages = {L24-L28},
          doi = {10.1093/mnrasl/slz025},
archivePrefix = {arXiv},
       eprint = {1809.06374},
 primaryClass = {astro-ph.CO},
       adsurl = {https://ui.adsabs.harvard.edu/abs/2019MNRAS.485L..24K}
}

@ARTICLE{2019ApJ...874..153B,
       author = {{Bradley}, Richard F. and {Tauscher}, Keith and {Rapetti}, David and {Burns}, Jack O.},
        title = "{A Ground Plane Artifact that Induces an Absorption Profile in Averaged Spectra from Global 21 cm Measurements, with Possible Application to EDGES}",
      journal = {\apj},
         year = 2019,
        month = apr,
       volume = {874},
       number = {2},
          eid = {153},
        pages = {153},
          doi = {10.3847/1538-4357/ab0d8b},
archivePrefix = {arXiv},
       eprint = {1810.09015},
 primaryClass = {astro-ph.IM},
       adsurl = {https://ui.adsabs.harvard.edu/abs/2019ApJ...874..153B}
}

@ARTICLE{2019PDU....24..289S,
       author = {{Sikivie}, Pierre},
        title = "{Axion dark matter and the 21-cm signal}",
      journal = {Physics of the Dark Universe},
         year = 2019,
        month = mar,
       volume = {24},
          eid = {100289},
        pages = {100289},
          doi = {10.1016/j.dark.2019.100289},
archivePrefix = {arXiv},
       eprint = {1805.05577},
 primaryClass = {astro-ph.CO},
       adsurl = {https://ui.adsabs.harvard.edu/abs/2019PDU....24..289S}
}

@ARTICLE{2019MNRAS.483.1980M,
       author = {{Mirocha}, Jordan and {Furlanetto}, Steven R.},
        title = "{What does the first highly redshifted 21-cm detection tell us about early galaxies?}",
      journal = {\mnras},
         year = 2019,
        month = feb,
       volume = {483},
       number = {2},
        pages = {1980-1992},
          doi = {10.1093/mnras/sty3260},
archivePrefix = {arXiv},
       eprint = {1803.03272},
 primaryClass = {astro-ph.GA},
       adsurl = {https://ui.adsabs.harvard.edu/abs/2019MNRAS.483.1980M}
}

@ARTICLE{2019JAI.....850004P,
       author = {{Philip}, L. and {Abdurashidova}, Z. and {Chiang}, H.~C. and {Ghazi}, N. and {Gumba}, A. and {Heilgendorff}, H.~M. and {J{\'a}uregui-Garc{\'\i}a}, J.~M. and {Malepe}, K. and {Nunhokee}, C.~D. and {Peterson}, J. and {Sievers}, J.~L. and {Simes}, V. and {Spann}, R.},
        title = "{Probing Radio Intensity at High-Z from Marion: 2017 Instrument}",
      journal = {Journal of Astronomical Instrumentation},
         year = 2019,
        month = jan,
       volume = {8},
       number = {2},
          eid = {1950004},
        pages = {1950004},
          doi = {10.1142/S2251171719500041},
archivePrefix = {arXiv},
       eprint = {1806.09531},
 primaryClass = {astro-ph.IM},
       adsurl = {https://ui.adsabs.harvard.edu/abs/2019JAI.....850004P}
}

@ARTICLE{2018Natur.564E..32H,
       author = {{Hills}, Richard and {Kulkarni}, Girish and {Meerburg}, P. Daniel and {Puchwein}, Ewald},
        title = "{Concerns about modelling of the EDGES data}",
      journal = {\nat},
         year = 2018,
        month = dec,
       volume = {564},
       number = {7736},
        pages = {E32-E34},
          doi = {10.1038/s41586-018-0796-5},
archivePrefix = {arXiv},
       eprint = {1805.01421},
 primaryClass = {astro-ph.CO},
       adsurl = {https://ui.adsabs.harvard.edu/abs/2018Natur.564E..32H}
}

@ARTICLE{2018PhRvD..98j3503Y,
       author = {{Yang}, Yupeng},
        title = "{Contributions of dark matter annihilation to the global 21 cm spectrum observed by the EDGES experiment}",
      journal = {\prd},
         year = 2018,
        month = nov,
       volume = {98},
       number = {10},
          eid = {103503},
        pages = {103503},
          doi = {10.1103/PhysRevD.98.103503},
archivePrefix = {arXiv},
       eprint = {1803.05803},
 primaryClass = {astro-ph.CO},
       adsurl = {https://ui.adsabs.harvard.edu/abs/2018PhRvD..98j3503Y}
}

@ARTICLE{2018PASA...35...33W,
       author = {{Wayth}, Randall B. and {Tingay}, Steven J. and {Trott}, Cathryn M. and {Emrich}, David and {Johnston-Hollitt}, Melanie and {McKinley}, Ben and {Gaensler}, B.~M. and {Beardsley}, A.~P. and {Booler}, T. and {Crosse}, B. and {Franzen}, T.~M.~O. and {Horsley}, L. and {Kaplan}, D.~L. and {Kenney}, D. and {Morales}, M.~F. and {Pallot}, D. and {Sleap}, G. and {Steele}, K. and {Walker}, M. and {Williams}, A. and {Wu}, C. and {Cairns}, Iver. H. and {Filipovic}, M.~D. and {Johnston}, S. and {Murphy}, T. and {Quinn}, P. and {Staveley-Smith}, L. and {Webster}, R. and {Wyithe}, J.~S.~B.},
        title = "{The Phase II Murchison Widefield Array: Design overview}",
      journal = {\pasa},
         year = 2018,
        month = nov,
       volume = {35},
          eid = {e033},
        pages = {e033},
          doi = {10.1017/pasa.2018.37},
archivePrefix = {arXiv},
       eprint = {1809.06466},
 primaryClass = {astro-ph.IM},
       adsurl = {https://ui.adsabs.harvard.edu/abs/2018PASA...35...33W}
}

@ARTICLE{2018ApJ...868...63E,
       author = {{Ewall-Wice}, A. and {Chang}, T.-C. and {Lazio}, J. and {Dor{\'e}}, O. and {Seiffert}, M. and {Monsalve}, R.~A.},
        title = "{Modeling the Radio Background from the First Black Holes at Cosmic Dawn: Implications for the 21 cm Absorption Amplitude}",
      journal = {\apj},
         year = 2018,
        month = nov,
       volume = {868},
       number = {1},
          eid = {63},
        pages = {63},
          doi = {10.3847/1538-4357/aae51d},
archivePrefix = {arXiv},
       eprint = {1803.01815},
 primaryClass = {astro-ph.CO},
       adsurl = {https://ui.adsabs.harvard.edu/abs/2018ApJ...868...63E}
}

@ARTICLE{2018PhLB..785..159F,
       author = {{Fraser}, Sean and {Hektor}, Andi and {H{\"u}tsi}, Gert and {Kannike}, Kristjan and {Marzo}, Carlo and {Marzola}, Luca and {Racioppi}, Antonio and {Raidal}, Martti and {Spethmann}, Christian and {Vaskonen}, Ville and {Veerm{\"a}e}, Hardi},
        title = "{The EDGES 21 cm anomaly and properties of dark matter}",
      journal = {Physics Letters B},
         year = 2018,
        month = oct,
       volume = {785},
        pages = {159-164},
          doi = {10.1016/j.physletb.2018.08.035},
archivePrefix = {arXiv},
       eprint = {1803.03245},
 primaryClass = {hep-ph},
       adsurl = {https://ui.adsabs.harvard.edu/abs/2018PhLB..785..159F}
}

@ARTICLE{2018ApJ...864...53E,
       author = {{Eilers}, Anna-Christina and {Davies}, Frederick B. and {Hennawi}, Joseph F.},
        title = "{The Opacity of the Intergalactic Medium Measured along Quasar Sightlines at z {\ensuremath{\sim}} 6}",
      journal = {\apj},
         year = 2018,
        month = sep,
       volume = {864},
       number = {1},
          eid = {53},
        pages = {53},
          doi = {10.3847/1538-4357/aad4fd},
archivePrefix = {arXiv},
       eprint = {1807.04229},
 primaryClass = {astro-ph.GA},
       adsurl = {https://ui.adsabs.harvard.edu/abs/2018ApJ...864...53E}
}

@ARTICLE{2018MNRAS.479.1055B,
       author = {{Bosman}, Sarah E.~I. and {Fan}, Xiaohui and {Jiang}, Linhua and {Reed}, Sophie and {Matsuoka}, Yoshiki and {Becker}, George and {Haehnelt}, Martin},
        title = "{New constraints on Lyman-{\ensuremath{\alpha}} opacity with a sample of 62 quasarsat z > 5.7}",
      journal = {\mnras},
         year = 2018,
        month = sep,
       volume = {479},
       number = {1},
        pages = {1055-1076},
          doi = {10.1093/mnras/sty1344},
archivePrefix = {arXiv},
       eprint = {1802.08177},
 primaryClass = {astro-ph.GA},
       adsurl = {https://ui.adsabs.harvard.edu/abs/2018MNRAS.479.1055B}
}

@ARTICLE{2018MNRAS.478.4193P,
       author = {{Price}, D.~C. and {Greenhill}, L.~J. and {Fialkov}, A. and {Bernardi}, G. and {Garsden}, H. and {Barsdell}, B.~R. and {Kocz}, J. and {Anderson}, M.~M. and {Bourke}, S.~A. and {Craig}, J. and {Dexter}, M.~R. and {Dowell}, J. and {Eastwood}, M.~W. and {Eftekhari}, T. and {Ellingson}, S.~W. and {Hallinan}, G. and {Hartman}, J.~M. and {Kimberk}, R. and {Lazio}, T. Joseph W. and {Leiker}, S. and {MacMahon}, D. and {Monroe}, R. and {Schinzel}, F. and {Taylor}, G.~B. and {Tong}, E. and {Werthimer}, D. and {Woody}, D.~P.},
        title = "{Design and characterization of the Large-aperture Experiment to Detect the Dark Age (LEDA) radiometer systems}",
      journal = {\mnras},
         year = 2018,
        month = aug,
       volume = {478},
       number = {3},
        pages = {4193-4213},
          doi = {10.1093/mnras/sty1244},
archivePrefix = {arXiv},
       eprint = {1709.09313},
 primaryClass = {astro-ph.IM},
       adsurl = {https://ui.adsabs.harvard.edu/abs/2018MNRAS.478.4193P}
}

@ARTICLE{2018PhRvL.121c1103P,
       author = {{Pospelov}, Maxim and {Pradler}, Josef and {Ruderman}, Joshua T. and {Urbano}, Alfredo},
        title = "{Room for New Physics in the Rayleigh-Jeans Tail of the Cosmic Microwave Background}",
      journal = {\prl},
         year = 2018,
        month = jul,
       volume = {121},
       number = {3},
          eid = {031103},
        pages = {031103},
          doi = {10.1103/PhysRevLett.121.031103},
archivePrefix = {arXiv},
       eprint = {1803.07048},
 primaryClass = {hep-ph},
       adsurl = {https://ui.adsabs.harvard.edu/abs/2018PhRvL.121c1103P}
}

@ARTICLE{2018PhRvD..98b3013S,
       author = {{Slatyer}, Tracy R. and {Wu}, Chih-Liang},
        title = "{Early-Universe constraints on dark matter-baryon scattering and their implications for a global 21 cm signal}",
      journal = {\prd},
         year = 2018,
        month = jul,
       volume = {98},
       number = {2},
          eid = {023013},
        pages = {023013},
          doi = {10.1103/PhysRevD.98.023013},
archivePrefix = {arXiv},
       eprint = {1803.09734},
 primaryClass = {astro-ph.CO},
       adsurl = {https://ui.adsabs.harvard.edu/abs/2018PhRvD..98b3013S}
}

@ARTICLE{2018ApJ...858L..17F,
       author = {{Feng}, Chang and {Holder}, Gilbert},
        title = "{Enhanced Global Signal of Neutral Hydrogen Due to Excess Radiation at Cosmic Dawn}",
      journal = {\apjl},
         year = 2018,
        month = may,
       volume = {858},
       number = {2},
          eid = {L17},
        pages = {L17},
          doi = {10.3847/2041-8213/aac0fe},
archivePrefix = {arXiv},
       eprint = {1802.07432},
 primaryClass = {astro-ph.CO},
       adsurl = {https://ui.adsabs.harvard.edu/abs/2018ApJ...858L..17F}
}

@ARTICLE{2018Natur.557..684M,
       author = {{Mu{\~n}oz}, Julian B. and {Loeb}, Abraham},
        title = "{A small amount of mini-charged dark matter could cool the baryons in the early Universe}",
      journal = {\nat},
         year = 2018,
        month = may,
       volume = {557},
       number = {7707},
        pages = {684-686},
          doi = {10.1038/s41586-018-0151-x},
archivePrefix = {arXiv},
       eprint = {1802.10094},
 primaryClass = {astro-ph.CO},
       adsurl = {https://ui.adsabs.harvard.edu/abs/2018Natur.557..684M}
}

@ARTICLE{2018Natur.555...67B,
       author = {{Bowman}, Judd D. and {Rogers}, Alan E.~E. and {Monsalve}, Raul A. and {Mozdzen}, Thomas J. and {Mahesh}, Nivedita},
        title = "{An absorption profile centred at 78 megahertz in the sky-averaged spectrum}",
      journal = {\nat},
         year = 2018,
        month = mar,
       volume = {555},
       number = {7694},
        pages = {67-70},
          doi = {10.1038/nature25792},
archivePrefix = {arXiv},
       eprint = {1810.05912},
 primaryClass = {astro-ph.CO},
       adsurl = {https://ui.adsabs.harvard.edu/abs/2018Natur.555...67B}
}

@ARTICLE{2018Natur.555...71B,
       author = {{Barkana}, Rennan},
        title = "{Possible interaction between baryons and dark-matter particles revealed by the first stars}",
      journal = {\nat},
         year = 2018,
        month = mar,
       volume = {555},
       number = {7694},
        pages = {71-74},
          doi = {10.1038/nature25791},
archivePrefix = {arXiv},
       eprint = {1803.06698},
 primaryClass = {astro-ph.CO},
       adsurl = {https://ui.adsabs.harvard.edu/abs/2018Natur.555...71B}
}

@ARTICLE{2017ApJ...847...64M,
       author = {{Monsalve}, Raul A. and {Rogers}, Alan E.~E. and {Bowman}, Judd D. and {Mozdzen}, Thomas J.},
        title = "{Results from EDGES High-band. I. Constraints on Phenomenological Models for the Global 21 cm Signal}",
      journal = {\apj},
         year = 2017,
        month = sep,
       volume = {847},
       number = {1},
          eid = {64},
        pages = {64},
          doi = {10.3847/1538-4357/aa88d1},
archivePrefix = {arXiv},
       eprint = {1708.05817},
 primaryClass = {astro-ph.CO},
       adsurl = {https://ui.adsabs.harvard.edu/abs/2017ApJ...847...64M}
}

@ARTICLE{2017ApJ...844...85O,
       author = {{Ota}, Kazuaki and {Iye}, Masanori and {Kashikawa}, Nobunari and {Konno}, Akira and {Nakata}, Fumiaki and {Totani}, Tomonori and {Kobayashi}, Masakazu A.~R. and {Fudamoto}, Yoshinobu and {Seko}, Akifumi and {Toshikawa}, Jun and {Ichikawa}, Akie and {Shibuya}, Takatoshi and {Onoue}, Masafusa},
        title = "{A New Constraint on Reionization from the Evolution of the Ly{\ensuremath{\alpha}} Luminosity Function at z {\ensuremath{\sim}} 6-7 Probed by a Deep Census of z = 7.0 Ly{\ensuremath{\alpha}} Emitter Candidates to 0.3L $^{*}$}",
      journal = {\apj},
         year = 2017,
        month = jul,
       volume = {844},
       number = {1},
          eid = {85},
        pages = {85},
          doi = {10.3847/1538-4357/aa7a0a},
archivePrefix = {arXiv},
       eprint = {1703.02501},
 primaryClass = {astro-ph.GA},
       adsurl = {https://ui.adsabs.harvard.edu/abs/2017ApJ...844...85O}
}

@ARTICLE{2017ApJ...840...33S,
       author = {{Sathyanarayana Rao}, Mayuri and {Subrahmanyan}, Ravi and {Udaya Shankar}, N. and {Chluba}, Jens},
        title = "{Modeling the Radio Foreground for Detection of CMB Spectral Distortions from the Cosmic Dawn and the Epoch of Reionization}",
      journal = {\apj},
         year = 2017,
        month = may,
       volume = {840},
       number = {1},
          eid = {33},
        pages = {33},
          doi = {10.3847/1538-4357/aa69bd},
archivePrefix = {arXiv},
       eprint = {1611.04602},
 primaryClass = {astro-ph.CO},
       adsurl = {https://ui.adsabs.harvard.edu/abs/2017ApJ...840...33S}
}

@ARTICLE{2017ApJS..230....7P,
       author = {{Perley}, R.~A. and {Butler}, B.~J.},
        title = "{An Accurate Flux Density Scale from 50 MHz to 50 GHz}",
      journal = {\apjs},
         year = 2017,
        month = may,
       volume = {230},
       number = {1},
          eid = {7},
        pages = {7},
          doi = {10.3847/1538-4365/aa6df9},
archivePrefix = {arXiv},
       eprint = {1609.05940},
 primaryClass = {astro-ph.IM},
       adsurl = {https://ui.adsabs.harvard.edu/abs/2017ApJS..230....7P}
}

@ARTICLE{2017PASP..129d5001D,
       author = {{DeBoer}, David R. and {Parsons}, Aaron R. and {Aguirre}, James E. and {Alexander}, Paul and {Ali}, Zaki S. and {Beardsley}, Adam P. and {Bernardi}, Gianni and {Bowman}, Judd D. and {Bradley}, Richard F. and {Carilli}, Chris L. and {Cheng}, Carina and {de Lera Acedo}, Eloy and {Dillon}, Joshua S. and {Ewall-Wice}, Aaron and {Fadana}, Gcobisa and {Fagnoni}, Nicolas and {Fritz}, Randall and {Furlanetto}, Steve R. and {Glendenning}, Brian and {Greig}, Bradley and {Grobbelaar}, Jasper and {Hazelton}, Bryna J. and {Hewitt}, Jacqueline N. and {Hickish}, Jack and {Jacobs}, Daniel C. and {Julius}, Austin and {Kariseb}, MacCalvin and {Kohn}, Saul A. and {Lekalake}, Telalo and {Liu}, Adrian and {Loots}, Anita and {MacMahon}, David and {Malan}, Lourence and {Malgas}, Cresshim and {Maree}, Matthys and {Martinot}, Zachary and {Mathison}, Nathan and {Matsetela}, Eunice and {Mesinger}, Andrei and {Morales}, Miguel F. and {Neben}, Abraham R. and {Patra}, Nipanjana and {Pieterse}, Samantha and {Pober}, Jonathan C. and {Razavi-Ghods}, Nima and {Ringuette}, Jon and {Robnett}, James and {Rosie}, Kathryn and {Sell}, Raddwine and {Smith}, Craig and {Syce}, Angelo and {Tegmark}, Max and {Thyagarajan}, Nithyanandan and {Williams}, Peter K.~G. and {Zheng}, Haoxuan},
        title = "{Hydrogen Epoch of Reionization Array (HERA)}",
      journal = {\pasp},
         year = 2017,
        month = apr,
       volume = {129},
       number = {974},
        pages = {045001},
          doi = {10.1088/1538-3873/129/974/045001},
archivePrefix = {arXiv},
       eprint = {1606.07473},
 primaryClass = {astro-ph.IM},
       adsurl = {https://ui.adsabs.harvard.edu/abs/2017PASP..129d5001D}
}

@ARTICLE{2017MNRAS.464.1365M,
       author = {{Mirocha}, Jordan and {Furlanetto}, Steven R. and {Sun}, Guochao},
        title = "{The global 21-cm signal in the context of the high- z galaxy luminosity function}",
      journal = {\mnras},
         year = 2017,
        month = jan,
       volume = {464},
       number = {2},
        pages = {1365-1379},
          doi = {10.1093/mnras/stw2412},
archivePrefix = {arXiv},
       eprint = {1607.00386},
 primaryClass = {astro-ph.GA},
       adsurl = {https://ui.adsabs.harvard.edu/abs/2017MNRAS.464.1365M}
}

@ARTICLE{2017ApJ...835...49M,
       author = {{Monsalve}, Raul A. and {Rogers}, Alan E.~E. and {Bowman}, Judd D. and {Mozdzen}, Thomas J.},
        title = "{Calibration of the EDGES High-band Receiver to Observe the Global 21 cm Signature from the Epoch of Reionization}",
      journal = {\apj},
         year = 2017,
        month = jan,
       volume = {835},
       number = {1},
          eid = {49},
        pages = {49},
          doi = {10.3847/1538-4357/835/1/49},
archivePrefix = {arXiv},
       eprint = {1602.08065},
 primaryClass = {astro-ph.IM},
       adsurl = {https://ui.adsabs.harvard.edu/abs/2017ApJ...835...49M}
}

@ARTICLE{2017MNRAS.464.3486Z,
       author = {{Zheng}, H. and {Tegmark}, M. and {Dillon}, J.~S. and {Kim}, D.~A. and {Liu}, A. and {Neben}, A.~R. and {Jonas}, J. and {Reich}, P. and {Reich}, W.},
        title = "{An improved model of diffuse galactic radio emission from 10 MHz to 5 THz}",
      journal = {\mnras},
         year = 2017,
        month = jan,
       volume = {464},
       number = {3},
        pages = {3486-3497},
          doi = {10.1093/mnras/stw2525},
archivePrefix = {arXiv},
       eprint = {1605.04920},
 primaryClass = {astro-ph.CO},
       adsurl = {https://ui.adsabs.harvard.edu/abs/2017MNRAS.464.3486Z}
}

@ARTICLE{2017AJ....153...26S,
       author = {{Sathyanarayana Rao}, Mayuri and {Subrahmanyan}, Ravi and {Udaya Shankar}, N. and {Chluba}, Jens},
        title = "{GMOSS: All-sky Model of Spectral Radio Brightness Based on Physical Components and Associated Radiative Processes}",
      journal = {\aj},
         year = 2017,
        month = jan,
       volume = {153},
       number = {1},
          eid = {26},
        pages = {26},
          doi = {10.3847/1538-3881/153/1/26},
archivePrefix = {arXiv},
       eprint = {1607.07453},
 primaryClass = {astro-ph.CO},
       adsurl = {https://ui.adsabs.harvard.edu/abs/2017AJ....153...26S}
}

@ARTICLE{2016JAI.....541001H,
       author = {{Hickish}, Jack and {Abdurashidova}, Zuhra and {Ali}, Zaki and {Buch}, Kaushal D. and {Chaudhari}, Sandeep C. and {Chen}, Hong and {Dexter}, Matthew and {Domagalski}, Rachel Simone and {Ford}, John and {Foster}, Griffin and {George}, David and {Greenberg}, Joe and {Greenhill}, Lincoln and {Isaacson}, Adam and {Jiang}, Homin and {Jones}, Glenn and {Kapp}, Francois and {Kriel}, Henno and {Lacasse}, Rich and {Lutomirski}, Andrew and {MacMahon}, David and {Manley}, Jason and {Martens}, Andrew and {McCullough}, Randy and {Muley}, Mekhala V. and {New}, Wesley and {Parsons}, Aaron and {Price}, Daniel C. and {Primiani}, Rurik A. and {Ray}, Jason and {Siemion}, Andrew and {van Tonder}, Verees{\'e} and {Vertatschitsch}, Laura and {Wagner}, Mark and {Weintroub}, Jonathan and {Werthimer}, Dan},
        title = "{A Decade of Developing Radio-Astronomy Instrumentation using CASPER Open-Source Technology}",
      journal = {Journal of Astronomical Instrumentation},
         year = 2016,
        month = dec,
       volume = {5},
       number = {4},
          eid = {1641001-12},
        pages = {1641001-12},
          doi = {10.1142/S2251171716410014},
archivePrefix = {arXiv},
       eprint = {1611.01826},
 primaryClass = {astro-ph.IM},
       adsurl = {https://ui.adsabs.harvard.edu/abs/2016JAI.....541001H}
}

@ARTICLE{2016ApJ...831....6D,
       author = {{Datta}, Abhirup and {Bradley}, Richard and {Burns}, Jack O. and {Harker}, Geraint and {Komjathy}, Attila and {Lazio}, T. Joseph W.},
        title = "{The Effects of the Ionosphere on Ground-based Detection of the Global 21 cm Signal from the Cosmic Dawn and the Dark Ages}",
      journal = {\apj},
         year = 2016,
        month = nov,
       volume = {831},
       number = {1},
          eid = {6},
        pages = {6},
          doi = {10.3847/0004-637X/831/1/6},
       adsurl = {https://ui.adsabs.harvard.edu/abs/2016ApJ...831....6D}
}

@ARTICLE{2016MNRAS.462..235L,
       author = {{Liu}, Chuanwu and {Mutch}, Simon J. and {Angel}, P.~W. and {Duffy}, Alan R. and {Geil}, Paul M. and {Poole}, Gregory B. and {Mesinger}, Andrei and {Wyithe}, J. Stuart B.},
        title = "{Dark-ages reionization and galaxy formation simulation - IV. UV luminosity functions of high-redshift galaxies}",
      journal = {\mnras},
         year = 2016,
        month = oct,
       volume = {462},
       number = {1},
        pages = {235-249},
          doi = {10.1093/mnras/stw1015},
archivePrefix = {arXiv},
       eprint = {1512.00563},
 primaryClass = {astro-ph.GA},
       adsurl = {https://ui.adsabs.harvard.edu/abs/2016MNRAS.462..235L}
}

@ARTICLE{2016MNRAS.460.4320E,
       author = {{Ewall-Wice}, A. and {Dillon}, Joshua S. and {Hewitt}, J.~N. and {Loeb}, A. and {Mesinger}, A. and {Neben}, A.~R. and {Offringa}, A.~R. and {Tegmark}, M. and {Barry}, N. and {Beardsley}, A.~P. and {Bernardi}, G. and {Bowman}, Judd D. and {Briggs}, F. and {Cappallo}, R.~J. and {Carroll}, P. and {Corey}, B.~E. and {de Oliveira-Costa}, A. and {Emrich}, D. and {Feng}, L. and {Gaensler}, B.~M. and {Goeke}, R. and {Greenhill}, L.~J. and {Hazelton}, B.~J. and {Hurley-Walker}, N. and {Johnston-Hollitt}, M. and {Jacobs}, Daniel C. and {Kaplan}, D.~L. and {Kasper}, J.~C. and {Kim}, HS and {Kratzenberg}, E. and {Lenc}, E. and {Line}, J. and {Lonsdale}, C.~J. and {Lynch}, M.~J. and {McKinley}, B. and {McWhirter}, S.~R. and {Mitchell}, D.~A. and {Morales}, M.~F. and {Morgan}, E. and {Thyagarajan}, Nithyanandan and {Oberoi}, D. and {Ord}, S.~M. and {Paul}, S. and {Pindor}, B. and {Pober}, J.~C. and {Prabu}, T. and {Procopio}, P. and {Riding}, J. and {Rogers}, A.~E.~E. and {Roshi}, A. and {Shankar}, N. Udaya and {Sethi}, Shiv K. and {Srivani}, K.~S. and {Subrahmanyan}, R. and {Sullivan}, I.~S. and {Tingay}, S.~J. and {Trott}, C.~M. and {Waterson}, M. and {Wayth}, R.~B. and {Webster}, R.~L. and {Whitney}, A.~R. and {Williams}, A. and {Williams}, C.~L. and {Wu}, C. and {Wyithe}, J.~S.~B.},
        title = "{First limits on the 21 cm power spectrum during the Epoch of X-ray heating}",
      journal = {\mnras},
         year = 2016,
        month = aug,
       volume = {460},
       number = {4},
        pages = {4320-4347},
          doi = {10.1093/mnras/stw1022},
archivePrefix = {arXiv},
       eprint = {1605.00016},
 primaryClass = {astro-ph.CO},
       adsurl = {https://ui.adsabs.harvard.edu/abs/2016MNRAS.460.4320E}
}

@ARTICLE{2016PhR...645....1B,
       author = {{Barkana}, Rennan},
        title = "{The rise of the first stars: Supersonic streaming, radiative feedback, and 21-cm cosmology}",
      journal = {\physrep},
         year = 2016,
        month = jul,
       volume = {645},
        pages = {1-59},
          doi = {10.1016/j.physrep.2016.06.006},
       adsurl = {https://ui.adsabs.harvard.edu/abs/2016PhR...645....1B}
}

@ARTICLE{2016MNRAS.457.1864L,
       author = {{Liu}, Adrian and {Parsons}, Aaron R.},
        title = "{Constraining cosmology and ionization history with combined 21 cm power spectrum and global signal measurements}",
      journal = {\mnras},
         year = 2016,
        month = apr,
       volume = {457},
       number = {2},
        pages = {1864-1877},
          doi = {10.1093/mnras/stw071},
archivePrefix = {arXiv},
       eprint = {1510.08815},
 primaryClass = {astro-ph.CO},
       adsurl = {https://ui.adsabs.harvard.edu/abs/2016MNRAS.457.1864L}
}

@ARTICLE{2016ApJ...821...59F,
       author = {{Fialkov}, Anastasia and {Loeb}, Abraham},
        title = "{Precise Measurement of the Reionization Optical Depth from the Global 21 cm Signal Accounting for Cosmic Heating}",
      journal = {\apj},
         year = 2016,
        month = apr,
       volume = {821},
       number = {1},
          eid = {59},
        pages = {59},
          doi = {10.3847/0004-637X/821/1/59},
archivePrefix = {arXiv},
       eprint = {1601.03058},
 primaryClass = {astro-ph.CO},
       adsurl = {https://ui.adsabs.harvard.edu/abs/2016ApJ...821...59F}
}

@ARTICLE{2016MNRAS.455.3890M,
       author = {{Mozdzen}, T.~J. and {Bowman}, J.~D. and {Monsalve}, R.~A. and {Rogers}, A.~E.~E.},
        title = "{Limits on foreground subtraction from chromatic beam effects in global redshifted 21 cm measurements}",
      journal = {\mnras},
         year = 2016,
        month = feb,
       volume = {455},
       number = {4},
        pages = {3890-3900},
          doi = {10.1093/mnras/stv2601},
archivePrefix = {arXiv},
       eprint = {1510.03477},
 primaryClass = {astro-ph.IM},
       adsurl = {https://ui.adsabs.harvard.edu/abs/2016MNRAS.455.3890M}
}

@ARTICLE{2016PhRvD..93d3013L,
       author = {{Liu}, Adrian and {Pritchard}, Jonathan R. and {Allison}, Rupert and {Parsons}, Aaron R. and {Seljak}, Uro{\v{s}} and {Sherwin}, Blake D.},
        title = "{Eliminating the optical depth nuisance from the CMB with 21 cm cosmology}",
      journal = {\prd},
         year = 2016,
        month = feb,
       volume = {93},
       number = {4},
          eid = {043013},
        pages = {043013},
          doi = {10.1103/PhysRevD.93.043013},
archivePrefix = {arXiv},
       eprint = {1509.08463},
 primaryClass = {astro-ph.CO},
       adsurl = {https://ui.adsabs.harvard.edu/abs/2016PhRvD..93d3013L}
}

@ARTICLE{2015ITAP...63.5433S,
       author = {{Sutinjo}, A.~T. and {Colegate}, T.~M. and {Wayth}, R.~B. and {Hall}, P.~J. and {de Lera Acedo}, E. and {Booler}, T. and {Faulkner}, A.~J. and {Feng}, L. and {Hurley-Walker}, N. and {Juswardy}, B. and {Padhi}, S.~K. and {Razavi-Ghods}, N. and {Sokolowski}, M. and {Tingay}, S.~J. and {Bij de Vaate}, J.~G.},
        title = "{Characterization of a Low-Frequency Radio Astronomy Prototype Array in Western Australia}",
      journal = {IEEE Transactions on Antennas and Propagation},
         year = 2015,
        month = dec,
       volume = {63},
       number = {12},
        pages = {5433-5442},
          doi = {10.1109/TAP.2015.2487504},
archivePrefix = {arXiv},
       eprint = {1510.01515},
 primaryClass = {astro-ph.IM},
       adsurl = {https://ui.adsabs.harvard.edu/abs/2015ITAP...63.5433S}
}

@ARTICLE{2015ApJ...813...21M,
       author = {{Mason}, Charlotte A. and {Trenti}, Michele and {Treu}, Tommaso},
        title = "{The Galaxy UV Luminosity Function before the Epoch of Reionization}",
      journal = {\apj},
         year = 2015,
        month = nov,
       volume = {813},
       number = {1},
          eid = {21},
        pages = {21},
          doi = {10.1088/0004-637X/813/1/21},
archivePrefix = {arXiv},
       eprint = {1508.01204},
 primaryClass = {astro-ph.GA},
       adsurl = {https://ui.adsabs.harvard.edu/abs/2015ApJ...813...21M}
}

@ARTICLE{2015MNRAS.450.3032M,
       author = {{McLeod}, D.~J. and {McLure}, R.~J. and {Dunlop}, J.~S. and {Robertson}, B.~E. and {Ellis}, R.~S. and {Targett}, T.~A.},
        title = "{New redshift z $\simeq$ 9 galaxies in the Hubble Frontier Fields: implications for early evolution of the UV luminosity density}",
      journal = {\mnras},
         year = 2015,
        month = jul,
       volume = {450},
       number = {3},
        pages = {3032-3044},
          doi = {10.1093/mnras/stv780},
archivePrefix = {arXiv},
       eprint = {1412.1472},
 primaryClass = {astro-ph.GA},
       adsurl = {https://ui.adsabs.harvard.edu/abs/2015MNRAS.450.3032M}
}

@ARTICLE{2015MNRAS.447.3402B,
       author = {{Becker}, George D. and {Bolton}, James S. and {Madau}, Piero and {Pettini}, Max and {Ryan-Weber}, Emma V. and {Venemans}, Bram P.},
        title = "{Evidence of patchy hydrogen reionization from an extreme Ly{\ensuremath{\alpha}} trough below redshift six}",
      journal = {\mnras},
         year = 2015,
        month = mar,
       volume = {447},
       number = {4},
        pages = {3402-3419},
          doi = {10.1093/mnras/stu2646},
archivePrefix = {arXiv},
       eprint = {1407.4850},
 primaryClass = {astro-ph.CO},
       adsurl = {https://ui.adsabs.harvard.edu/abs/2015MNRAS.447.3402B}
}

@ARTICLE{2015ApJ...799...90B,
       author = {{Bernardi}, G. and {McQuinn}, M. and {Greenhill}, L.~J.},
        title = "{Foreground Model and Antenna Calibration Errors in the Measurement of the Sky-averaged {\ensuremath{\lambda}}21 cm Signal at z\raisebox{-0.5ex}\textasciitilde 20}",
      journal = {\apj},
         year = 2015,
        month = jan,
       volume = {799},
       number = {1},
          eid = {90},
        pages = {90},
          doi = {10.1088/0004-637X/799/1/90},
archivePrefix = {arXiv},
       eprint = {1404.0887},
 primaryClass = {astro-ph.CO},
       adsurl = {https://ui.adsabs.harvard.edu/abs/2015ApJ...799...90B}
}

@ARTICLE{2014ApJ...793..102S,
       author = {{Switzer}, Eric R. and {Liu}, Adrian},
        title = "{Erasing the Variable: Empirical Foreground Discovery for Global 21 cm Spectrum Experiments}",
      journal = {\apj},
         year = 2014,
        month = oct,
       volume = {793},
       number = {2},
          eid = {102},
        pages = {102},
          doi = {10.1088/0004-637X/793/2/102},
archivePrefix = {arXiv},
       eprint = {1404.7561},
 primaryClass = {astro-ph.CO},
       adsurl = {https://ui.adsabs.harvard.edu/abs/2014ApJ...793..102S}
}

@ARTICLE{2014RaSc...49..157H,
       author = {{Helmboldt}, J.~F. and {Ellingson}, S.~W. and {Hartman}, J.~M. and {Lazio}, T.~J.~W. and {Taylor}, G.~B. and {Wilson}, T.~L. and {Wolfe}, C.~N.},
        title = "{All-sky imaging of meteor trails at 55.25 MHz with the first station of the Long Wavelength Array}",
      journal = {Radio Science},
         year = 2014,
        month = mar,
       volume = {49},
       number = {3},
        pages = {157-180},
          doi = {10.1002/2013RS005220},
       adsurl = {https://ui.adsabs.harvard.edu/abs/2014RaSc...49..157H}
}

@ARTICLE{2014MNRAS.437.1056V,
       author = {{Vedantham}, H.~K. and {Koopmans}, L.~V.~E. and {de Bruyn}, A.~G. and {Wijnholds}, S.~J. and {Ciardi}, B. and {Brentjens}, M.~A.},
        title = "{Chromatic effects in the 21 cm global signal from the cosmic dawn}",
      journal = {\mnras},
         year = 2014,
        month = jan,
       volume = {437},
       number = {2},
        pages = {1056-1069},
          doi = {10.1093/mnras/stt1878},
archivePrefix = {arXiv},
       eprint = {1306.2172},
 primaryClass = {astro-ph.IM},
       adsurl = {https://ui.adsabs.harvard.edu/abs/2014MNRAS.437.1056V}
}

@ARTICLE{2013A&A...556A...2V,
       author = {{\VAN{Harleem}{van}{van} Harleem}, M.~P. and {Wise}, M.~W. and {Gunst}, A.~W. and {Heald}, G. and {McKean}, J.~P. and {Hessels}, J.~W.~T. and {de Bruyn}, A.~G. and {Nijboer}, R. and {Swinbank}, J. and {Fallows}, R. and {Brentjens}, M. and {Nelles}, A. and {Beck}, R. and {Falcke}, H. and {Fender}, R. and {H{\"o}randel}, J. and {Koopmans}, L.~V.~E. and {Mann}, G. and {Miley}, G. and {R{\"o}ttgering}, H. and {Stappers}, B.~W. and {Wijers}, R.~A.~M.~J. and {Zaroubi}, S. and {van den Akker}, M. and {Alexov}, A. and {Anderson}, J. and {Anderson}, K. and {van Ardenne}, A. and {Arts}, M. and {Asgekar}, A. and {Avruch}, I.~M. and {Batejat}, F. and {B{\"a}hren}, L. and {Bell}, M.~E. and {Bell}, M.~R. and {van Bemmel}, I. and {Bennema}, P. and {Bentum}, M.~J. and {Bernardi}, G. and {Best}, P. and {B{\^\i}rzan}, L. and {Bonafede}, A. and {Boonstra}, A.-J. and {Braun}, R. and {Bregman}, J. and {Breitling}, F. and {van de Brink}, R.~H. and {Broderick}, J. and {Broekema}, P.~C. and {Brouw}, W.~N. and {Br{\"u}ggen}, M. and {Butcher}, H.~R. and {van Cappellen}, W. and {Ciardi}, B. and {Coenen}, T. and {Conway}, J. and {Coolen}, A. and {Corstanje}, A. and {Damstra}, S. and {Davies}, O. and {Deller}, A.~T. and {Dettmar}, R.-J. and {van Diepen}, G. and {Dijkstra}, K. and {Donker}, P. and {Doorduin}, A. and {Dromer}, J. and {Drost}, M. and {van Duin}, A. and {Eisl{\"o}ffel}, J. and {van Enst}, J. and {Ferrari}, C. and {Frieswijk}, W. and {Gankema}, H. and {Garrett}, M.~A. and {de Gasperin}, F. and {Gerbers}, M. and {de Geus}, E. and {Grie{\ss}meier}, J.-M. and {Grit}, T. and {Gruppen}, P. and {Hamaker}, J.~P. and {Hassall}, T. and {Hoeft}, M. and {Holties}, H.~A. and {Horneffer}, A. and {van der Horst}, A. and {van Houwelingen}, A. and {Huijgen}, A. and {Iacobelli}, M. and {Intema}, H. and {Jackson}, N. and {Jelic}, V. and {de Jong}, A. and {Juette}, E. and {Kant}, D. and {Karastergiou}, A. and {Koers}, A. and {Kollen}, H. and {Kondratiev}, V.~I. and {Kooistra}, E. and {Koopman}, Y. and {Koster}, A. and {Kuniyoshi}, M. and {Kramer}, M. and {Kuper}, G. and {Lambropoulos}, P. and {Law}, C. and {van Leeuwen}, J. and {Lemaitre}, J. and {Loose}, M. and {Maat}, P. and {Macario}, G. and {Markoff}, S. and {Masters}, J. and {McFadden}, R.~A. and {McKay-Bukowski}, D. and {Meijering}, H. and {Meulman}, H. and {Mevius}, M. and {Middelberg}, E. and {Millenaar}, R. and {Miller-Jones}, J.~C.~A. and {Mohan}, R.~N. and {Mol}, J.~D. and {Morawietz}, J. and {Morganti}, R. and {Mulcahy}, D.~D. and {Mulder}, E. and {Munk}, H. and {Nieuwenhuis}, L. and {van Nieuwpoort}, R. and {Noordam}, J.~E. and {Norden}, M. and {Noutsos}, A. and {Offringa}, A.~R. and {Olofsson}, H. and {Omar}, A. and {Orr{\'u}}, E. and {Overeem}, R. and {Paas}, H. and {Pandey-Pommier}, M. and {Pandey}, V.~N. and {Pizzo}, R. and {Polatidis}, A. and {Rafferty}, D. and {Rawlings}, S. and {Reich}, W. and {de Reijer}, J.-P. and {Reitsma}, J. and {Renting}, G.~A. and {Riemers}, P. and {Rol}, E. and {Romein}, J.~W. and {Roosjen}, J. and {Ruiter}, M. and {Scaife}, A. and {van der Schaaf}, K. and {Scheers}, B. and {Schellart}, P. and {Schoenmakers}, A. and {Schoonderbeek}, G. and {Serylak}, M. and {Shulevski}, A. and {Sluman}, J. and {Smirnov}, O. and {Sobey}, C. and {Spreeuw}, H. and {Steinmetz}, M. and {Sterks}, C.~G.~M. and {Stiepel}, H.-J. and {Stuurwold}, K. and {Tagger}, M. and {Tang}, Y. and {Tasse}, C. and {Thomas}, I. and {Thoudam}, S. and {Toribio}, M.~C. and {van der Tol}, B. and {Usov}, O. and {van Veelen}, M. and {van der Veen}, A.-J. and {ter Veen}, S. and {Verbiest}, J.~P.~W. and {Vermeulen}, R. and {Vermaas}, N. and {Vocks}, C. and {Vogt}, C. and {de Vos}, M. and {van der Wal}, E. and {van Weeren}, R. and {Weggemans}, H. and {Weltevrede}, P. and {White}, S. and {Wijnholds}, S.~J. and {Wilhelmsson}, T. and {Wucknitz}, O. and {Yatawatta}, S. and {Zarka}, P. and {Zensus}, A.},
        title = "{LOFAR: The LOw-Frequency ARray}",
      journal = {\aap},
         year = 2013,
        month = aug,
       volume = {556},
          eid = {A2},
        pages = {A2},
          doi = {10.1051/0004-6361/201220873},
archivePrefix = {arXiv},
       eprint = {1305.3550},
 primaryClass = {astro-ph.IM},
       adsurl = {https://ui.adsabs.harvard.edu/abs/2013A&A...556A...2V}
}

@ARTICLE{2013PhRvD..87d3002L,
       author = {{Liu}, Adrian and {Pritchard}, Jonathan R. and {Tegmark}, Max and {Loeb}, Abraham},
        title = "{Global 21 cm signal experiments: A designer's guide}",
      journal = {\prd},
         year = 2013,
        month = feb,
       volume = {87},
       number = {4},
          eid = {043002},
        pages = {043002},
          doi = {10.1103/PhysRevD.87.043002},
       adsurl = {https://ui.adsabs.harvard.edu/abs/2013PhRvD..87d3002L}
}

@ARTICLE{2013PASA...30....7T,
       author = {{Tingay}, S.~J. and {Goeke}, R. and {Bowman}, J.~D. and {Emrich}, D. and {Ord}, S.~M. and {Mitchell}, D.~A. and {Morales}, M.~F. and {Booler}, T. and {Crosse}, B. and {Wayth}, R.~B. and {Lonsdale}, C.~J. and {Tremblay}, S. and {Pallot}, D. and {Colegate}, T. and {Wicenec}, A. and {Kudryavtseva}, N. and {Arcus}, W. and {Barnes}, D. and {Bernardi}, G. and {Briggs}, F. and {Burns}, S. and {Bunton}, J.~D. and {Cappallo}, R.~J. and {Corey}, B.~E. and {Deshpande}, A. and {Desouza}, L. and {Gaensler}, B.~M. and {Greenhill}, L.~J. and {Hall}, P.~J. and {Hazelton}, B.~J. and {Herne}, D. and {Hewitt}, J.~N. and {Johnston-Hollitt}, M. and {Kaplan}, D.~L. and {Kasper}, J.~C. and {Kincaid}, B.~B. and {Koenig}, R. and {Kratzenberg}, E. and {Lynch}, M.~J. and {Mckinley}, B. and {Mcwhirter}, S.~R. and {Morgan}, E. and {Oberoi}, D. and {Pathikulangara}, J. and {Prabu}, T. and {Remillard}, R.~A. and {Rogers}, A.~E.~E. and {Roshi}, A. and {Salah}, J.~E. and {Sault}, R.~J. and {Udaya-Shankar}, N. and {Schlagenhaufer}, F. and {Srivani}, K.~S. and {Stevens}, J. and {Subrahmanyan}, R. and {Waterson}, M. and {Webster}, R.~L. and {Whitney}, A.~R. and {Williams}, A. and {Williams}, C.~L. and {Wyithe}, J.~S.~B.},
        title = "{The Murchison Widefield Array: The Square Kilometre Array Precursor at Low Radio Frequencies}",
      journal = {\pasa},
         year = 2013,
        month = jan,
       volume = {30},
          eid = {e007},
        pages = {e007},
          doi = {10.1017/pasa.2012.007},
archivePrefix = {arXiv},
       eprint = {1206.6945},
 primaryClass = {astro-ph.IM},
       adsurl = {https://ui.adsabs.harvard.edu/abs/2013PASA...30....7T}
}

@ARTICLE{2012RaSc...47.0K06R,
       author = {{Rogers}, Alan E.~E. and {Bowman}, Judd D.},
        title = "{Absolute calibration of a wideband antenna and spectrometer for accurate sky noise temperature measurements}",
      journal = {Radio Science},
         year = 2012,
        month = aug,
       volume = {47},
       number = {4},
          eid = {RS0K06},
        pages = {RS0K06},
          doi = {10.1029/2011RS004962},
archivePrefix = {arXiv},
       eprint = {1209.1106},
 primaryClass = {astro-ph.IM},
       adsurl = {https://ui.adsabs.harvard.edu/abs/2012RaSc...47.0K06R}
}

@ARTICLE{2010PhRvD..82b3006P,
       author = {{Pritchard}, Jonathan R. and {Loeb}, Abraham},
        title = "{Constraining the unexplored period between the dark ages and reionization with observations of the global 21 cm signal}",
      journal = {\prd},
         year = 2010,
        month = jul,
       volume = {82},
       number = {2},
          eid = {023006},
        pages = {023006},
          doi = {10.1103/PhysRevD.82.023006},
archivePrefix = {arXiv},
       eprint = {1005.4057},
 primaryClass = {astro-ph.CO},
       adsurl = {https://ui.adsabs.harvard.edu/abs/2010PhRvD..82b3006P}
}

@ARTICLE{2010AJ....139.1468P,
       author = {{Parsons}, Aaron R. and {Backer}, Donald C. and {Foster}, Griffin S. and {Wright}, Melvyn C.~H. and {Bradley}, Richard F. and {Gugliucci}, Nicole E. and {Parashare}, Chaitali R. and {Benoit}, Erin E. and {Aguirre}, James E. and {Jacobs}, Daniel C. and {Carilli}, Chris L. and {Herne}, David and {Lynch}, Mervyn J. and {Manley}, Jason R. and {Werthimer}, Daniel J.},
        title = "{The Precision Array for Probing the Epoch of Re-ionization: Eight Station Results}",
      journal = {\aj},
         year = 2010,
        month = apr,
       volume = {139},
       number = {4},
        pages = {1468-1480},
          doi = {10.1088/0004-6256/139/4/1468},
archivePrefix = {arXiv},
       eprint = {0904.2334},
 primaryClass = {astro-ph.CO},
       adsurl = {https://ui.adsabs.harvard.edu/abs/2010AJ....139.1468P}
}

@ARTICLE{2009IEEEP..97.1482D,
       author = {{Dewdney}, P.~E. and {Hall}, P.~J. and {Schilizzi}, R.~T. and {Lazio}, T.~J.~L.~W.},
        title = "{The Square Kilometre Array}",
      journal = {IEEE Proceedings},
         year = 2009,
        month = aug,
       volume = {97},
       number = {8},
        pages = {1482-1496},
          doi = {10.1109/JPROC.2009.2021005},
       adsurl = {https://ui.adsabs.harvard.edu/abs/2009IEEEP..97.1482D}
}

@ARTICLE{2008MNRAS.388..247D,
       author = {{de Oliveira-Costa}, Ang{\'e}lica and {Tegmark}, Max and {Gaensler}, B.~M. and {Jonas}, Justin and {Landecker}, T.~L. and {Reich}, Patricia},
        title = "{A model of diffuse Galactic radio emission from 10 MHz to 100 GHz}",
      journal = {\mnras},
         year = 2008,
        month = jul,
       volume = {388},
       number = {1},
        pages = {247-260},
          doi = {10.1111/j.1365-2966.2008.13376.x},
archivePrefix = {arXiv},
       eprint = {0802.1525},
 primaryClass = {astro-ph},
       adsurl = {https://ui.adsabs.harvard.edu/abs/2008MNRAS.388..247D}
}

@ARTICLE{2007MNRAS.381...75M,
       author = {{McQuinn}, Matthew and {Hernquist}, Lars and {Zaldarriaga}, Matias and {Dutta}, Suvendra},
        title = "{Studying reionization with Ly{\ensuremath{\alpha}} emitters}",
      journal = {\mnras},
         year = 2007,
        month = oct,
       volume = {381},
       number = {1},
        pages = {75-96},
          doi = {10.1111/j.1365-2966.2007.12085.x},
archivePrefix = {arXiv},
       eprint = {0704.2239},
 primaryClass = {astro-ph},
       adsurl = {https://ui.adsabs.harvard.edu/abs/2007MNRAS.381...75M}
}

@ARTICLE{2006PhR...433..181F,
       author = {{Furlanetto}, Steven R. and {Oh}, S. Peng and {Briggs}, Frank H.},
        title = "{Cosmology at low frequencies: The 21 cm transition and the high-redshift Universe}",
      journal = {\physrep},
         year = 2006,
        month = oct,
       volume = {433},
       number = {4-6},
        pages = {181-301},
          doi = {10.1016/j.physrep.2006.08.002},
archivePrefix = {arXiv},
       eprint = {astro-ph/0608032},
 primaryClass = {astro-ph},
       adsurl = {https://ui.adsabs.harvard.edu/abs/2006PhR...433..181F}
}

@ARTICLE{2006AJ....132..117F,
       author = {{Fan}, Xiaohui and {Strauss}, Michael A. and {Becker}, Robert H. and {White}, Richard L. and {Gunn}, James E. and {Knapp}, Gillian R. and {Richards}, Gordon T. and {Schneider}, Donald P. and {Brinkmann}, J. and {Fukugita}, Masataka},
        title = "{Constraining the Evolution of the Ionizing Background and the Epoch of Reionization with z\raisebox{-0.5ex}\textasciitilde6 Quasars. II. A Sample of 19 Quasars}",
      journal = {\aj},
         year = 2006,
        month = jul,
       volume = {132},
       number = {1},
        pages = {117-136},
          doi = {10.1086/504836},
archivePrefix = {arXiv},
       eprint = {astro-ph/0512082},
 primaryClass = {astro-ph},
       adsurl = {https://ui.adsabs.harvard.edu/abs/2006AJ....132..117F}
}

@ARTICLE{2004NSGeo...2..151F,
       author = {{Fechner}, T. and {B{\"o}rner}, F.~D. and {Richter}, T. and {Yaramanci}, U. and {Weihnacht}, B.},
        title = "{Lithological interpretation of the spectral dielectric properties of limestone}",
      journal = {Near Surface Geophysics},
         year = 2004,
        month = aug,
       volume = {2},
       number = {3},
        pages = {151-159},
          doi = {10.3997/1873-0604.2004013},
       adsurl = {https://ui.adsabs.harvard.edu/abs/2004NSGeo...2..151F}
}

@ARTICLE{2003MNRAS.346..871O,
       author = {{Oh}, S. Peng and {Mack}, Katherine J.},
        title = "{Foregrounds for 21-cm observations of neutral gas at high redshift}",
      journal = {\mnras},
         year = 2003,
        month = dec,
       volume = {346},
       number = {3},
        pages = {871-877},
          doi = {10.1111/j.1365-2966.2003.07133.x},
archivePrefix = {arXiv},
       eprint = {astro-ph/0302099},
 primaryClass = {astro-ph},
       adsurl = {https://ui.adsabs.harvard.edu/abs/2003MNRAS.346..871O}
}

@ARTICLE{2002ApJ...564..576D,
       author = {{Di Matteo}, Tiziana and {Perna}, Rosalba and {Abel}, Tom and {Rees}, Martin J.},
        title = "{Radio Foregrounds for the 21 Centimeter Tomography of the Neutral Intergalactic Medium at High Redshifts}",
      journal = {\apj},
         year = 2002,
        month = jan,
       volume = {564},
       number = {2},
        pages = {576-580},
          doi = {10.1086/324293},
archivePrefix = {arXiv},
       eprint = {astro-ph/0109241},
 primaryClass = {astro-ph},
       adsurl = {https://ui.adsabs.harvard.edu/abs/2002ApJ...564..576D}
}

@ARTICLE{2000ApJ...528..597T,
       author = {{Tozzi}, Paolo and {Madau}, Piero and {Meiksin}, Avery and {Rees}, Martin J.},
        title = "{Radio Signatures of H I at High Redshift: Mapping the End of the ``Dark Ages''}",
      journal = {\apj},
         year = 2000,
        month = jan,
       volume = {528},
       number = {2},
        pages = {597-606},
          doi = {10.1086/308196},
archivePrefix = {arXiv},
       eprint = {astro-ph/9903139},
 primaryClass = {astro-ph},
       adsurl = {https://ui.adsabs.harvard.edu/abs/2000ApJ...528..597T}
}

@ARTICLE{1999A&A...345..380S,
       author = {{Shaver}, P.~A. and {Windhorst}, R.~A. and {Madau}, P. and {de Bruyn}, A.~G.},
        title = "{Can the reionization epoch be detected as a global signature in the cosmic background?}",
      journal = {\aap},
         year = 1999,
        month = may,
       volume = {345},
        pages = {380-390},
          doi = {10.48550/arXiv.astro-ph/9901320},
archivePrefix = {arXiv},
       eprint = {astro-ph/9901320},
 primaryClass = {astro-ph},
       adsurl = {https://ui.adsabs.harvard.edu/abs/1999A&A...345..380S}
}

@ARTICLE{1997ApJ...475..429M,
       author = {{Madau}, Piero and {Meiksin}, Avery and {Rees}, Martin J.},
        title = "{21 Centimeter Tomography of the Intergalactic Medium at High Redshift}",
      journal = {\apj},
         year = 1997,
        month = feb,
       volume = {475},
       number = {2},
        pages = {429-444},
          doi = {10.1086/303549},
archivePrefix = {arXiv},
       eprint = {astro-ph/9608010},
 primaryClass = {astro-ph},
       adsurl = {https://ui.adsabs.harvard.edu/abs/1997ApJ...475..429M}
}

@book{1982nyhr.book.....B,
  author    = {Balanis, C. A.},
  title     = {Antenna Theory: Analysis and Design},
  year      = {1982},
  publisher = {John Wiley \& Sons},
  address   = {New York},
  edition   = {1st}
}

@ARTICLE{1978ATTTJ..57.1371S,
       author = {{Slepian}, D.},
        title = "{Prolate spheroidal wave functions, Fourier analysis, and uncertainty. V - The discrete case}",
      journal = {AT\&T Technical Journal},
         year = 1978,
        month = jun,
       volume = {57},
        pages = {1371-1430},
       adsurl = {https://ui.adsabs.harvard.edu/abs/1978ATTTJ..57.1371S}
}

@ARTICLE{1978ITMTT..26...34M,
       author = {{Meys}, R.~P.},
        title = "{A Wave Approach to the Noise Properties of Linear Microwave Devices}",
      journal = {IEEE Transactions on Microwave Theory and Techniques},
         year = 1978,
        month = jan,
       volume = {26},
       number = {1},
        pages = {34-37},
          doi = {10.1109/TMTT.1978.1129303},
       adsurl = {https://ui.adsabs.harvard.edu/abs/1978ITMTT..26...34M}
}

@ARTICLE{1977A&A....61...99B,
       author = {{Baars}, J.~W.~M. and {Genzel}, R. and {Pauliny-Toth}, I.~I.~K. and {Witzel}, A.},
        title = "{The absolute spectrum of Cas A: an accurate flux density scale and a set of secondary calibrators.}",
      journal = {\aap},
         year = 1977,
        month = oct,
       volume = {61},
        pages = {99-106},
       adsurl = {https://ui.adsabs.harvard.edu/abs/1977A&A....61...99B}
}

@ARTICLE{1948Natur.161..596E,
       author = {{Ellyett}, C.~D. and {Davies}, J.~G.},
        title = "{Velocity of Meteors Measured by Diffraction of Radio Waves from Trails during Formation}",
      journal = {\nat},
         year = 1948,
        month = apr,
       volume = {161},
       number = {4094},
        pages = {596-597},
          doi = {10.1038/161596a0},
       adsurl = {https://ui.adsabs.harvard.edu/abs/1948Natur.161..596E}
}

@ARTICLE{1947MNRAS.107..176H,
       author = {{Hey}, J.~S. and {Parsons}, S.~J. and {Stewart}, G.~S.},
        title = "{Radar observations of the Giacobinids meteor shower, 1946}",
      journal = {\mnras},
         year = 1947,
        month = jan,
       volume = {107},
        pages = {176},
          doi = {10.1093/mnras/107.2.176},
       adsurl = {https://ui.adsabs.harvard.edu/abs/1947MNRAS.107..176H}
}

@article{lestari2005adaptive,
author = {Lestari, Andrian Andaya and Yarovoy, Alexander G. and Ligthart, Leo P.},
title = {Adaptive Wire Bow-Tie Antenna for {GPR} Applications},
journal = {IEEE Transactions on Antennas and Propagation},
volume = {53},
number = {5},
pages = {1745--1754},
month = may,
year = {2005},
doi = {10.1109/TAP.2005.846726}
}

@article{lestari2010modified,
author = {Lestari, Andrian Andaya and Bharata, Endon and Suksmono, Andriyan Bayu and Kurniawan,
Adit and Yarovoy, Alexander G. and Ligthart, Leo P.},
title = {A Modified Bow-Tie Antenna for Improved Pulse Radiation},
journal = {IEEE Transactions on Antennas and Propagation},
volume = {58},
number = {7},
pages = {2184--2192},
month = jul,
year = {2010},
doi = {10.1109/TAP.2010.2048853}
}

@article{yang2020radiation,
author = {Yang, Guangyao and Ye, Shengbo and Ji, Yicai and Zhang, Xiaojuan and Fang, Guangyou},
title = {Radiation Enhancement of an Ultrawideband Unidirectional Folded Bowtie Antenna for {GPR}
Applications},
journal = {IEEE Access},
volume = {8},
pages = {182218--182228},
year = {2020},
doi = {10.1109/ACCESS.2020.3029050}
}

@misc{jax2018github,
  author = {James Bradbury and Roy Frostig and Peter Hawkins and Matthew James Johnson and Yash Katariya and Chris Leary and Dougal Maclaurin and George Necula and Adam Paszke and Jake Vander{P}las and Skye Wanderman-{M}ilne and Qiao Zhang},
  title = {{JAX}: composable transformations of {P}ython+{N}um{P}y programs},
  url = {http://github.com/jax-ml/jax},
  version = {0.3.13},
  year = {2018},
}

@inproceedings{Josh_2016,
  author       = {Josh, Matthew and Clennell, Ben and Cauchefert, Matthieu},
  title        = {Dielectric Permittivity and Anisotropy of Intact Multi-Saturated Organic Shales},
  booktitle    = {SPWLA 57th Annual Logging Symposium},
  year         = {2016},
  month        = jun,
  address      = {Reykjavik, Iceland},
  note         = {Paper presented at the SPWLA 57th Annual Logging Symposium},
}

@misc{Hintze_1974,
  author       = {Hintze, L. F.},
  title        = {Preliminary geologic map of the {Notch Peak} quadrangle, {Millard County}, {Utah}},
  year         = {1974},
  publisher    = {U.S. Geological Survey},
  address      = {Reston, VA},
  howpublished = {U.S. Geological Survey},
  note         = {Map}
}

@article{Parasnis_1956,
author = {Parasnis, D. S.},
title = {THE ELECTRICAL RESISTIVITY OF SOME SULPHIDE AND OXIDE MINERALS AND THEIR ORES},
journal = {Geophysical Prospecting},
volume = {4},
number = {3},
pages = {249-278},
doi = {https://doi.org/10.1111/j.1365-2478.1956.tb01409.x},
url = {https://onlinelibrary.wiley.com/doi/abs/10.1111/j.1365-2478.1956.tb01409.x},
eprint = {https://onlinelibrary.wiley.com/doi/pdf/10.1111/j.1365-2478.1956.tb01409.x},
year = {1956}
}








\bsp	
\label{lastpage}
\end{document}